\documentclass[twocolumn,superscriptaddress,amsmath,amssymb]{revtex4}
\pdfoutput=1
\usepackage[dvipsnames]{xcolor}
\usepackage{graphicx}
\usepackage{dcolumn,xcolor}
\usepackage{amsmath} 
\usepackage{amssymb}
\usepackage{amsfonts}
\usepackage{bm}
\usepackage{csvsimple}
\usepackage{ifsym,stmaryrd}
\usepackage{wasysym}
\usepackage{pifont}
\usepackage{overpic}
\usepackage{array}
\usepackage{hyperref}
\usepackage{scalerel}
\usepackage{tikz}
\usepackage[titletoc]{appendix}
           
\usepackage[latin1]{inputenc}
\usepackage{mathrsfs}

\DeclareFontEncoding{LS1}{}{}
\DeclareFontSubstitution{LS1}{stix}{m}{n}
\DeclareSymbolFont{symbols4}{LS1}{stixbb}{m}{it}
\DeclareMathSymbol{\varhexagonblack}{\mathord}{symbols4}{"DD}

\usepackage[latin1]{inputenc}

\begin{document}

\title{Radical scaling: beyond our feet and fingers}

\author{M.A.~Fardin}
\altaffiliation[Corresponding author ]{}
\email{marc-antoine.fardin@ijm.fr}
\affiliation{Universit\'{e} Paris Cit{\'e}, CNRS, Institut Jacques Monod, F-75013 Paris, France}
\affiliation{The Academy of Bradylogists}
\author{M.~Hautefeuille}
\affiliation{Institut de Biologie Paris Seine, UMR 7622, Sorbonne Universit\'{e}, 7 quai Saint Bernard, 75005 Paris, France}
\author{V.~Sharma}
\affiliation{Department  of Chemical  Engineering, University of Illinois at Chicago, Chicago, Illinois 60608, United States}
\affiliation{The Academy of Bradylogists}

\date{February 5, 2025}

\begin{abstract}
Scaling laws arise and are eulogized across disciplines from natural to social sciences for providing pithy, quantitative, `scale-free', and `universal' power law relationships between two variables. On a log-log plot, the power laws display as straight lines, with a slope set by the exponent of the scaling law. In practice, a scaling relationship works only for a limited range, bookended by crossovers to other scaling laws. Leading with Taylor's oft-cited scaling law for the blast radius of an explosion against time, and by collating an unprecedented amount of datasets for laser-induced, chemical and nuclear explosions, we show distinct kinematics arise at the early and late stages. We illustrate that picking objective scales for the two axes using the transitions between regimes leads to the collapse of the data for the two regimes and their crossover, but the third regime is typically not mapped to the master curve. The objective scales permit us to abandon the arbitrarily chosen anthropocentric units of measurement, like feet for length and heart-beat for time, but the decimal system with ten digits (fingers) is still part of the picture. We show a remarkable collapse of all three regimes onto a common master curve occurs if we replace the base 10 by a dimensionless radix that combines the scales from the two crossovers. We also illustrate this approach of radical scaling for capillarity-driven pinching, coalescence and spreading of drops and bubbles, expecting such generalizations will be made for datasets across many disciplines.
\end{abstract}

\maketitle

\textbf{\textit{Significance Statement:}} \textit{Scaling laws expressed in arbitrary units often fail when observations span a broader range. Transitions between regimes reveal objective units, allowing to capture these regimes and their crossovers. Beyond units, we must reconsider the numerical base (radix) we use. Decimal, derived from our ten fingers, dominates, but natural phenomena operate independently of human conventions. By analyzing transitions between successive scaling regimes, we propose using a number derived from the system itself as the base instead of 10. This approach captures universal behavior across regimes, creating new opportunities to revisit examples from diverse disciplines. Such a framework challenges anthropocentric standards, offering deeper insight into how numbers and units emerge directly from physical phenomena.}


What we see, perceive, and measure in the natural world is the combination of what \textit{is}, and the angle, perspective, or reference frame we have chosen. Finding out what this elusive thing \textit{is} then requires an array of viewpoints to overlap. For the practitioners of scaling analysis this overlap is understood quite literally. One may initially start with a messy set of intersecting data sets, and then strive to find the ``right scale'' with which the data almost magically overlap. Usually this quest stops when one finds judicious units for both axes of the plot. In this contribution, we show that a more complete scaling analysis should seek not solely to find units, but also to renormalize the very base we use for counting. Conventionally the base is 10, like the number of our fingers. However, fingers are no more legitimate than feet to count and measure. 

The starting point of a scaling analysis is usually a scaling law or power law, i.e. a relation of proportionality between one variable and some power of another, $y=K x^\alpha$~\cite{Barenblatt2003,Santiago2019}. Relationships of this kind are found everywhere. The periods of rotation of a planet is proportional to its distance to the Sun raised to a power $\alpha=\frac{3}{2}$ (Kepler law). The mean square displacement of a diffusive particle is proportional to the square root of the time, so with $\alpha=\frac{1}{2}$ (Einstein-Smoluchowski law). The metabolic rate of many animals is proportional to their mass to the power $\alpha=\frac{3}{4}$ (Kleiber law). The forces of gravity or of electrostatics are inversely proportional to the square of the distance, $\alpha=-2$ (Newton and Coulomb laws). The power radiated by a black body scales with the fourth power of the temperature, $\alpha=4$ (Stefan-Boltzmann law). The frequency of occurrence of a word is inversely proportional to its rank, $\alpha=-1$ (Zipf-Mandelbrot law). The list goes on and on. Scaling laws are ubiquitous but they are usually studied in isolation. Their apparent simplicity is often a consequence of the narrowness of the observational range. When data are gathered more broadly any scaling law is bound to meet its demise. This statement has been verified time and time again by experiments. What we will show is that the eventual breakdown of a power law is actually a necessity if the associated phenomenon is to be independent of our human imprint. We will show that two intersecting power laws are needed to find objective units, and three to find an objective base.

This article is accompanied by \href{www.youtube/@naturesnumbers}{video lectures on Youtube}, and a \href{www.numbersnature.org/explosions/data/trinity}{website} making the data freely available. Details on how to use this material are provided in Supporting Information~(SI). 

\section*{Scaling laws}
\begin{figure*}
\centering
\includegraphics[width=17cm,clip]{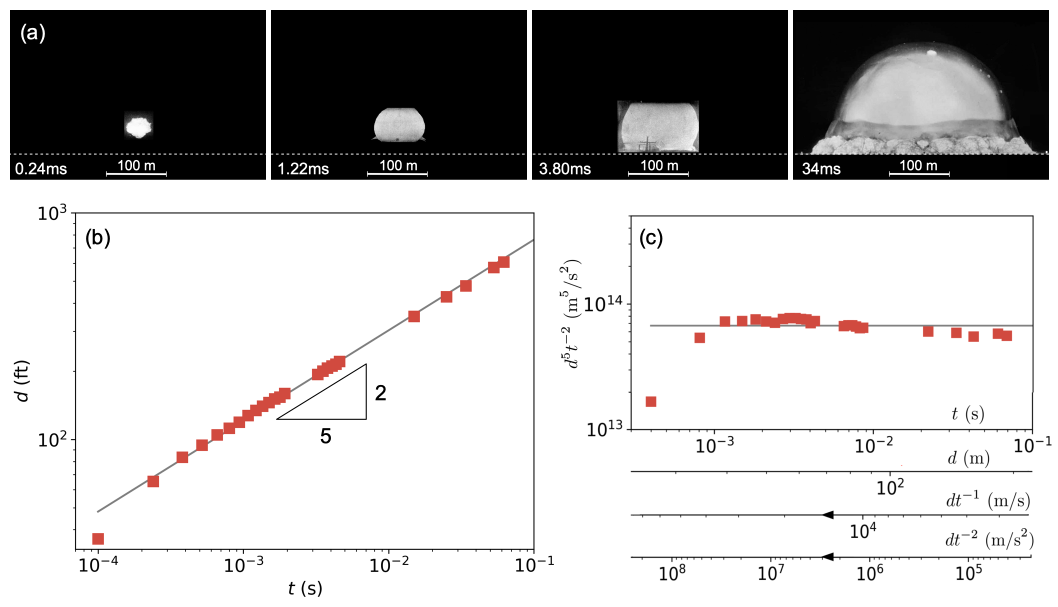}
\caption{Trinity: a paragon of scaling law. (a) Pictures of the nuclear test, taken by the optical team led by J.E. Mack~\cite{Mack1946,Mack1947}, and used by Taylor to analyze the kinematics of the explosion blast. Details and additional images are provided in SI and on a \protect\href{www.numbersnature.org/explosions/data/trinity}{webpage} we created for the greater dispersion of these data (Images courtesy of the Los Alamos National Laboratory). (b) Radius of the nuclear blast over time based on the images available to Taylor in 1950~\cite{Taylor1950b}. The grey line is $d=Kt^\frac{2}{5}$, with $K=1913$~ft.s$^{-\frac{2}{5}}$. (c) Replotting the data as an explosivity $\tilde{X}\equiv d^5/t^2$ versus time $t$, distance $d$, speed $d/t$, or acceleration $d/t^2$--examples of the fact that the horizontal variable is indeterminate. The horizontal grey line is $X\equiv K^5$. 
\label{fig1}}
\end{figure*} 
Scaling laws have been studied in a wide variety of contexts and the arguments we shall develop in this article can be generally applied. We will however restrict our examples to particularly visual scaling laws tracking the evolution of a size or distance $d$ over time $t$: 
\begin{equation}
d\simeq K t^\alpha
\label{powerlaw}
\end{equation}
Note that we use an approximate rather than strict equality, because actual data sets rarely perfectly match a power law. 

In Eq.~\ref{powerlaw} if $\alpha=1$ then motion is uniform and $K$ is a speed. If $\alpha=2$, motion is uniformly accelerated and $K$ is an acceleration. If $\alpha=\frac{1}{2}$, $K$ may be called a `sorptivity'~\cite{Philip1957}, but one usually writes $d=(Dt)^\frac{1}{2}$, defining a `coefficient of diffusion' or `diffusivity' $D\equiv K^2$. This definition allows to deal with a kinematic quantity with integer exponents, since $[D]=\mathcal{L}^2.\mathcal{T}^{-1}$, but $[K]=\mathcal{L}.\mathcal{T}^{-\frac{1}{2}}$ (brackets are used to give the dimensions of the enclosed quantity). Generally $K$ is a kinematic quantity, i.e. depending solely on space and time, with $[K]=\mathcal{L}.\mathcal{T}^{-\alpha}$. 

One example of the kind of power law defined in Eq.~\ref{powerlaw} is found in many textbooks on scaling and dimensional analysis: $d=K t^\frac{2}{5}$, which describes the extension of a blast wave of radius $d$, a time $t$ after detonation~\cite{Barenblatt2003,Santiago2019}. In that case, $K$ has no standard name, but for future reference we may call $X\equiv K^5$ an `explosivity'~\cite{Fardin2024}, with $[X]=\mathcal{L}^5.\mathcal{T}^{-2}$, defined in such a way as to have integer exponents (like the diffusivity when $\alpha=\frac{1}{2}$). This $\frac{2}{5}$ scaling law was famously derived by G.I. Taylor and used to analyze the footage of Trinity, the first atomic test~\cite{Taylor1950a,Taylor1950b}. 

Fig.~\ref{fig1}a provides pictures of the Trinity test in the first few milliseconds after detonation. A number of these images had been declassified in a report published in 1947 by J.E. Mack~\cite{Mack1947}, the head of the optical team for the Trinity test. Taylor also got access to a few more pictures through his connection to the British Ministry of Supply, including the picture at $t=1.22$~ms in Fig.~\ref{fig1}a~\cite{Taylor1950b}. During World War 2 Taylor had been involved in the Tube Alloys program, the secret British nuclear weapon project. After the Quebec Agreement on August 19th 1943, the British program was subsumed into the Manhattan Project, its American counterpart, and Taylor continued to play a major role~\cite{Bainbridge1976}. In fact, Taylor was one of only two foreigners (the other being James Chadwick) in a very short list of ten ``Distinguished Visitors'' to be officially invited to the Trinity test in New Mexico~\cite{Bainbridge1976}. In 1941 Taylor had predicted that the motion of a nuclear blast wave should follow a power law of the form $d=K t^\frac{2}{5}$, and his prediction was confirmed by the Trinity test on July 16th 1945~\cite{Taylor1950a}. In 1950 Taylor was cleared to publish his own account in a couple of papers, a first paper on the theory behind such prediction~\cite{Taylor1950a}, and a second on the agreement between the prediction and the data from the Trinity test~\cite{Taylor1950b}. 

Fig.~\ref{fig1}b gives a logarithmic plot of the growth of the blast radius measured by Taylor on the pictures of the test, replotted from Taylor's second paper~\cite{Taylor1950b}. The agreement between the data and Taylor's scaling is indeed quite remarkable. On a logarithmic plot a power law appears as a straight line with a slope given by the exponent, here $\alpha=\frac{2}{5}$. The prefactor $K$ sets the position of the line. In the case of Trinity, $K\simeq 1913$~ft.s$^{-\frac{2}{5}}$. Taylor's impressive achievement was to connect this value to the yield of the explosion and to the density of the ambient medium (air in that case), providing a rational for this exotic exponent of $\frac{2}{5}$~\cite{Taylor1950a,Taylor1950b,Barenblatt2003,Fardin2024}. However, the underlying dimensional analysis is not the focus of this article. Our approach here is essentially phenomenological. We do not ask why the dynamics occur but how their existence shape our point of view. 

Power laws, like Taylor's $\frac{2}{5}$ scaling, are generally understood to be `scale-free'~\cite{Barabasi2009}, since no preferred units of space nor time stand out. A power law is also said to be `self-similar'~\cite{Barenblatt2003}, the dynamics following the same law regardless of scale, i.e. no matter how small or large the time $t$ or radius $d$ are. This nomenclature was introduced in the 1960s, in the wake of Benoit Mandelbrot's work on fractals~\cite{Mandelbrot1977}, and has remained popular in the literature on dimensional analysis~\cite{Barenblatt2003}. However, both of these terms, `scale-free' and `self-similar', can be slightly misleading. 

When space is measured in feet and time in seconds, then the value of $K$ is around 1913. If we change the units, the value changes accordingly, for instance $K\simeq 583$~m.s$^{-\frac{2}{5}}$. The fact that a power law is `scale-free' does not mean that all units are equivalent, but that there is an infinity of equally good units. In this context, feet, meters and second are not ``good units''. As illustrated in the animated figures in SI (c.f. SI section VII), if the coordinates $(t_i,d_i)$ of any point along the power law are used as units, then the value of $K$ in these units becomes trivial: $K\simeq 1~d_i/t_i^\alpha$. These units are ``good units''. Assuming $d$ and $t$ to be initially measured in any arbitrary units, this prevalence of some choices of units can be written in the following way: 
\begin{equation}
\frac{d}{d_i}\simeq  \Big(\frac{t}{t_i}\Big)^\alpha
\label{powerlaw2}
\end{equation}
Note that $d/d_i$ and $t/t_i$ can be read as $d$ and $t$ ``in units of'' $d_i$ and $t_i$ respectively. What remains arbitrary is the choice of point along the power law, i.e. the choice of a pair $(t_i,d_i)$. 

\section*{Units for a single axis}
\begin{figure*}
\centering
\includegraphics[width=17cm,clip]{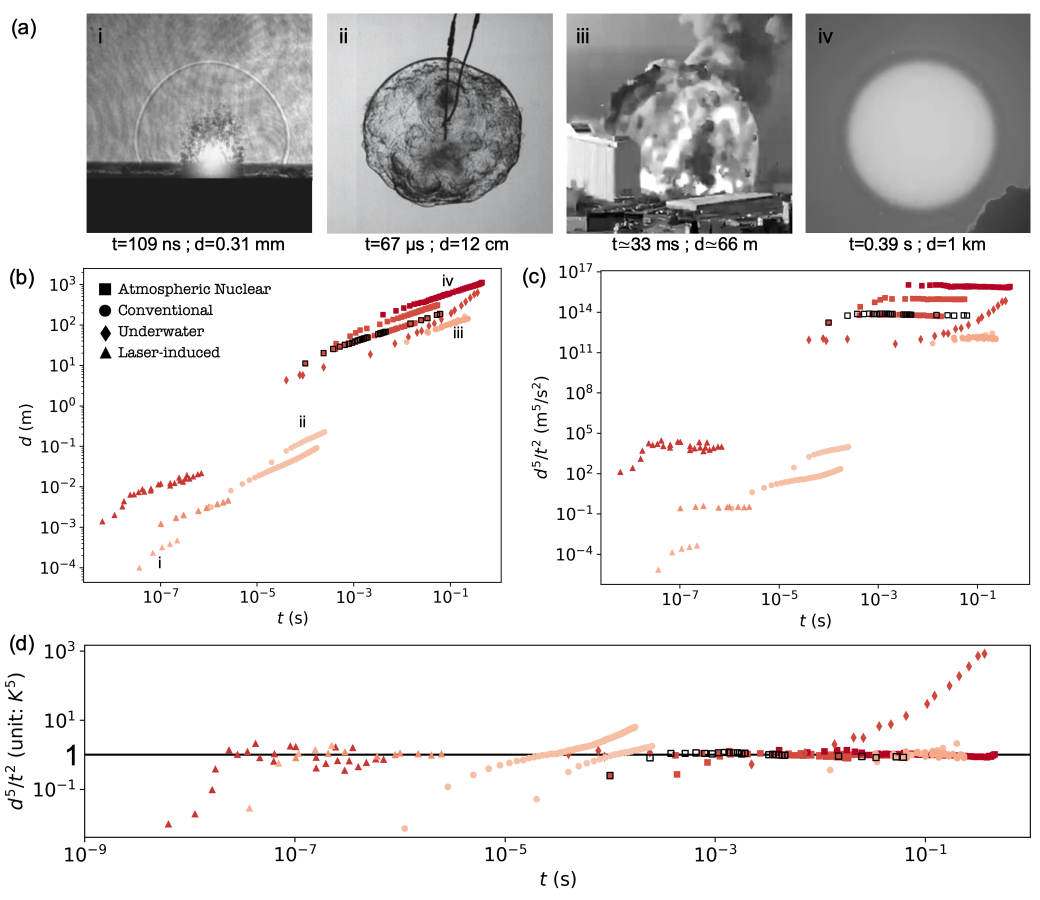}
\caption{Explosions across scales. (a) Pictures of explosions across scales. i) Laser-induced explosion~\cite{Porneala2006}.  ii) Explosion of a 1~gram charge of  pentaerythritol tetranitrate~\cite{Hargather2007}. iii) 2020 Beirut explosion caused by 2.75 kilotons of ammonium nitrate~\cite{Aouad2021,Rigby2021}. iv) \protect\href{https://youtu.be/IZZ_IsyE_iE?si=ggmL8Z4s3XJloGUv}{Dominic Housatonic nuclear test} (personal communication with G. Spriggs - Lawrence Livermore National Laboratory ). Below each image the time since detonation $t$ and the blast radius $d$ are specified. (b) Blast radius over time for a number of nuclear ($\square$: Trinity~\cite{Mack1947,Taylor1950b}, $\blacksquare$:~\cite{OConnell1957,Schmitt2016,Nguyen2017}) and conventional ($\bullet$:~\cite{Kingery1962,Aouad2021,Hargather2007,Kleine2010}) explosions in air, and underwater ($\blacklozenge$:~\cite{Porzel1957}), together with laser-induced explosions ($\blacktriangle$:~\cite{Porneala2006,Gatti1988,Grun1991}). (c) Replotting the data as an explosivity $\tilde{X}\equiv d^5/t^2$ versus time. Most data sets show a plateau extending over a significant time range. The ordinate of each plateau gives the value of $K^5$ for that data set. (d) Using $X\equiv K^5$ as unit of explosivity, all plateaus align on $\tilde{X}/X\simeq 1$. A guide to the data and additional images are provided in SI and on this \protect\href{https://www.numbersnature.org/explosions/2-beyond-trinity}{webpage}. An animated version of panel b illustrating all data sets is given in SI (Fig2b.gif). The color code is explained later in the paper and quantified in Fig.~\ref{fig5}. Note that the data sets on conventional explosions in pale pink only follow Taylor's regime over a very narrow time range, as will be explained later in the article. 
\label{fig2}}
\end{figure*} 
Technically, Eq.~\ref{powerlaw} is often said to be scale-free or self-similar because $d/t^\alpha\simeq K$ is constant~\cite{Barenblatt2003,Barabasi2009}. In the case of Taylor's scaling, $d/t^\frac{2}{5}\simeq K$. We are of course free to raise both sides to an identical power, in particular to a power of 5, and get $d^5/t^2\simeq K^5$. The right-hand side is constant and is what we called an ``explosivity''~\cite{Fardin2024}. Thus, the combination of variables in the left-hand side must also be a constant explosivity. Indeed, as shown in Fig.~\ref{fig1}c, if $d^5/t^2$ is plotted against any other combination of variables, the data appear flat, and the value of the plateau is set by $X\equiv K^5$. 

If $d$ is plotted against $t$ there is indeed no preferred units of space and time, any pair $(t_i,d_i)$ is equally valid, and even any other units if we are allowing $K$ to take non-trivial values, like $K\simeq 583$~m.s$^{-\frac{2}{5}}$. However, the same dynamics can also be tracked using different variables. For instance, one might follow the speed of the shock front over time, or at various distances from ground zero. Measurements performed from different perspectives should be consistent, so in particular we should have $v\simeq K t^{-\frac{3}{5}}$ and $v\simeq K^\frac{5}{2} d^{-\frac{3}{2}}$~\cite{Fardin2024}, where $v\simeq d/t$ is the front speed (numerical factors are omitted; see SI section III.A for details). The speed $v$, just like the size $d$ and time $t$ has no preferred scale. However, if instead we use the new variable $\tilde{X}\equiv d^5/t^2$, then this quantity is constant and it has a preferred scale, the unit of explosivity $X\equiv K^5$. 

Power laws are apparently scale-free. The axes of the two primitive variables do not have preferred units. Nevertheless, there is always a way to combine the initial variables in such a way as to obtain a preferred unit for one axis, while the other axis remains indeterminate~\cite{Barenblatt2003,Fardin2024}. Basically, if $y=K x^\alpha$, $x$ and $y$ do not have preferred units, but $(y/x^\alpha)^\gamma$ has units $K^\gamma$, for any value of the free exponent $\gamma$. Such switch in perspective may seem a bit extravagant when performed on a single instance of a power law, but it becomes quite useful when comparing multiple examples. For instance, Fig.~\ref{fig2}b gives the blast radii of a number of other atmospheric nuclear explosions~\cite{Mack1947,Taylor1950b,OConnell1957,Schmitt2016,Nguyen2017}, conventional explosions~\cite{Kingery1962,Aouad2021,Hargather2007,Kleine2010}, an underwater explosion~\cite{Porzel1957}, and laser-induced explosions~\cite{Porneala2006,Gatti1988,Grun1991}. A guide to the data is provided in SI. Some of these explosions are pictured in Fig.~\ref{fig2}a. When these data are represented as the radius $d$ versus the time $t$ in conventional units, then the explosions appear quite different. The blasts go from microscopic to terrifying. 

Since the scales of the explosions in Fig.~\ref{fig2}b vary so much, it can be hard to believe that all these dynamics essentially display the same $\frac{2}{5}$ scaling derived by Taylor. Yet, as shown in Fig.~\ref{fig2}c, if instead we plot the explosivity $\tilde{X}\equiv d^5/t^2$ for each explosion we indeed see that a portion of the data fall on plateaus, the ordinates of the plateaus are set by the values of $K$, i.e. the value of the explosivity scale $X\equiv K^5$ in each case. In Fig.~\ref{fig2}c the explosivities of all examples are still measured in conventional units (m$^5$/s$^2$). If instead we use the particular values of $X$ as units in each case, all curves lie on the same horizontal (unity) plateau, as shown in Fig.~\ref{fig2}d. Effectively, we have constructed a dimensionless number $N_1\equiv \tilde{X}/X\equiv d^5/(t^2 K^5)$, equal to unity as long as the dynamics follow Taylor's scaling (this number does not have a standard name, but we have recently proposed it be called the Taylor-Sedov number~\cite{Fardin2024}; see SI section III.A.1 for details). However, the horizontal axis in Fig.~\ref{fig2}d still awaits a proper scale, and some part of the data depart from the plateaus, where they also cease to overlap. As we shall see now these two issues are connected and can be resolved.  

\section*{Units for both axes}
\begin{figure}
\centering
\includegraphics[width=8.5cm,clip]{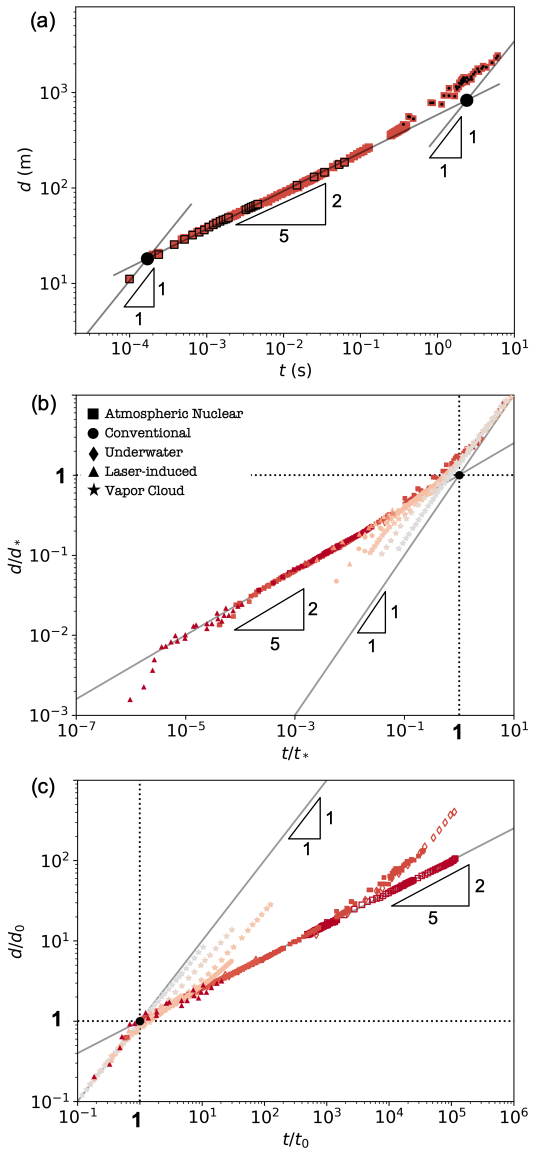}
\caption{Representing the dynamics of explosions with objective units. (a) Representation in standard units of an extended data set on the Trinity nuclear test, declassified in the 1980s. As a comparison, the black squares are the data used by Taylor~\cite{Taylor1950b}. Red squares are from optical measurement collected by Mack's team~\cite{Mack1946}, except those marked with a black dot, where the blast radius was inferred from pressure measurements~\cite{Bainbridge1976}. (b) Data from Fig.~\ref{fig2} are replotted in Hopkinson-Cranz units, $(t_*,d_*)$ defined in Eq.~\ref{HC1} and~\ref{HC2}. Also included are data on vapor cloud explosions ($\star$), which were only provided in scaled form in the original paper~\cite{Tang1999}. (c) Objective units from the early dynamics of explosions, $(t_0,d_0)$, defined in Eq.~\ref{early1} and~\ref{early2}.  The grey lines are $d=Kt^\frac{2}{5}$, $d=c_0 t$, and $d=c_s t$. For data sets with hollow symbols the initial speed of the explosion is not directly measured but estimated from mechanical considerations (see SI section II.B for details). The color code is explained later in the paper and quantified in Fig.~\ref{fig5}. Animated versions of panels b and c are given in SI to highlight each data set (Fig3b.gif, Fig3c.gif). The files SI3b.gif and SI3c.gif provide animated transitions between Fig.~\ref{fig2}b and Fig.~\ref{Sfig3}b and c respectively, as explained in SI section VII.  \label{fig3}}
\end{figure} 
In his 1950 papers, Taylor considered the intermediate stage of the dynamics of a large explosion~\cite{Taylor1950a,Taylor1950b}. The pictures he used were just a subset of the ones taken by Mack's team~\cite{Mack1946}. Fig.~\ref{fig3}a gives a more complete account of the dynamics, using data declassified in the 1980s~\cite{Mack1946} (we highly recommend reading this full report of immense historical value). The hollow squares are optical measurements and the stars were obtained from pressure measurements~\cite{Bainbridge1976}, allowing to track the shock front beyond the point where it becomes transparent~\cite{Mack1946,Mack1947}. Although the shock front follows Taylor's scaling from a fraction of a millisecond to about 0.1~s after detonation, it eventually departs from it, decelerating progressively until it reaches the speed of sound, $c_s\simeq 344$~m/s, denoted by the continuous line of slope 1 in Fig.~\ref{fig3}a. This ultimate weakening of shocks had been understood since the beginning of the 20th century, notably thanks to studies by Bertram Hopkinson~\cite{Hopkinson1915} and Carl Cranz~\cite{Cranz1926,Fuller2005}, the two men generally credited for understanding this transition~\cite{Sachs1944,Westine2012,Wei2021}. 

Hopkinson and Cranz realized that the blast of different explosions could be superposed if 
distance and time were measured in scaled units, which can be expressed using the speed of sound~\cite{Wei2021}. With the insight from Taylor~\cite{Taylor1950a,Taylor1950b}, we can understand the intermediate regime of the explosion as abiding to the scaling $d\simeq K t^\frac{2}{5}$. As we previously saw, the front speed decreases over time as $v\simeq K t^{-\frac{3}{5}}$. Eventually the speed of the front reaches the sound speed, $v(t_*)\simeq c_s$. The time $t_*$ and radius $d_*$ at which this transition occurs can be estimated by simply equating Taylor's regime, $d\simeq Kt^\frac{2}{5}$, with the late propagation at the speed of sound, $d\simeq c_s t$: 
\begin{align}
t_* &\simeq \Big(\frac{K}{c_s}\Big)^\frac{5}{3} \label{HC1}\\
d_* &\simeq \frac{K^\frac{5}{3}}{c_s^\frac{2}{3}} \label{HC2}
\end{align}
These units of space and time are often called the Hopkinson-Cranz units~\cite{Westine2012}. In contrast to the second and meter (or any absolute standards) these units depend solely on the characteristics of the dynamics. They are not set subjectively, but objectively, by the phenomenon at play. 

Because the data eventually depart from Taylor's scaling, we acquire units for both space and time. Single power laws, $d\simeq Kt^\frac{2}{5}$ or $d\simeq c_s t$, do not have preferred scales. More precisely, for each power law taken separately, any couple of coordinates $(t_i,d_i)$ on the power law provides equally valid units. However, when we now have two intersecting power laws, their point of intersection, here  $(t_*,d_*)$, provides a unique pair of units, a special point of view, common to both regimes.  

For any explosion depicted in Fig.~\ref{fig2}, we can compute the values of the associated Hopkinson-Cranz units, based on the measured explosivity and on the speed of sound in the medium (air, water, and some rarefied gases--see SI section VIII for details). For tiny laser-induced explosions, the Hopkinson-Cranz point may occur after just a few microseconds and for distances in the millimeter or centimeter range~\cite{Porneala2006,Gatti1988}. At the other end, a nuclear explosion like Dominic Housatonic has $t_*\simeq 20$~s and $d_*\simeq 5$~km (values of $t_*$ and $d_*$ are tabulated for all data sets in SI Table V). Although these explosions may seem to have very different scales when measured with absolute but subjective units, their similarity is manifest once represented in relative but objective units. As shown in Fig.~\ref{fig3}b, when the dynamics of small and large explosions are plotted in the Hopkinson-Cranz units, they largely overlap. 

Note that the scaled units $t/t_*$ and $d/d_*$ are dimensionless numbers, and they can be connected to $N_1$ and to $N_2\equiv (d/t)/c_s$, the Mach number associated with the late regime. As shown in SI~Fig.~4b, the data can then be plotted as $N_1$ vs $N_2$. Nevertheless, the same portions of the data overlap as in Fig.~\ref{fig3}b, the same portion does not, and it is now time to address this point.   

\section*{Competing units}
Taylor was well aware that the $\frac{2}{5}$ scaling could only apply to the intermediate range of explosions~\cite{Taylor1950a,Taylor1950b}. He knew that at later times the shock would weaken sufficiently as to be almost indistinguishable from a sound wave, and he also knew that at very short time his scaling would fail. Indeed, we have seen that if the blast radius follows $d\simeq K t^\frac{2}{5}$, then its speed follows $v\simeq K t^{-\frac{3}{5}}$. The speed would seem to diverge at the instant of detonation. This is, of course, not the case in practice because at very short time the dynamics are governed by the initial ejection speed of the explosion, which we may call $c_0$~\cite{Bethe1947}. Initially, the explosion front follows a third power law, $d\simeq c_0 t$. The various departures from Taylor's regime at short time seen in the figures capture this initial phase. In this initial phase, the impeding factor is usually the inertia of the ejected mass, but as for the previously discussed regimes, the mechanics are not in focus here (see SI section II.B.1 for details). 

In the same way that we computed the crossover between Taylor's regime and the regime of sound propagation, we can now obtain the point of intersection between Taylor's regime and the initial regime at constant ejection speed: 
\begin{align}
t_0 &\simeq \Big(\frac{K}{c_0}\Big)^\frac{5}{3}\label{early1}\\
d_0 &\simeq \frac{K^\frac{5}{3}}{c_0^\frac{2}{3}}\label{early2}
\end{align}
This transition seems to have first been studied in the context of nuclear weapons development, in particular by the team led by Hans Bethe~\cite{Bethe1947}, which included some famous members, like John von Neumann, and an infamous one, Klaus Fuchs (a notable atomic spy; see 2nd season of the \href{https://www.bbc.co.uk/programmes/p08llv8n}{BBC series ``The Bomb''}). This crossover is also discussed quite clearly in the literature on supernovae~\cite{Cioffi1988,Truelove1999}. In this context, the initial ejection regime can last for centuries.  

In Fig.~\ref{fig3}b we had chosen to rescale the dynamics in such a way as to overlap the intermediate and late stages of the dynamics. Data sets with enough time resolution to capture the initial stage would not overlap. Instead, we can use the units $(t_0,d_0)$ obtained from the early crossover to rescale the plot. Fig.~\ref{fig3}c gives the result of such approach. Note that as done in SI~Fig.~4b  for the late crossover, we could tilt this scaled plot by using $N_1$ and $N_0\equiv (d/t)/c_0$ (SI~Fig.~4a). Either way, now the initial and intermediate regimes overlap but the late regimes do not. We have three consecutive regimes but it seems that we cannot rescale all of them at once. It is like having a short blanket on a cold night: pull it over your head and your feet get cold, cover your feet and your neck gets cold! Let now see how to knit a blanket with the perfect size to cover the three regimes of the dynamics.  

\section*{A special number}
With $N_0$, $N_1$ and $N_2$ we introduced three dimensionless numbers, which we may call ``simple''~\cite{Fardin2024}. Each one of these numbers depends on a combination of the primitive variables, $d$ and $t$, and on the kinematic constant of one of the three regimes, respectively, the initial speed $c_0$, Taylor's prefactor $K$, and the speed of sound $c_s$. Scaled variables, $t/t_*$ and $d/d_*$, or $t/t_0$ and $d/d_0$ are more composite kinds of numbers, depending respectively on $N_1$ and $N_2$, or $N_0$ and $N_1$ (see SI section IV.A). These dimensionless numbers are still dependent on the variables $d$ and $t$. There is, however, a special kind of dimensionless number depending solely on the parameters: 
\begin{equation}
\mathcal{N}\equiv \frac{c_0}{c_s}
\end{equation}
This number may be called the `initial Mach number'~\cite{Kleine2010} (also called the `flame Mach number' for vapor cloud explosions~\cite{Tang1999}). For each explosion this number is a constant, and its value has been surreptitiously used as a color code in all figures. As seen in Fig.~\ref{fig3}c for the units $(t_0,d_0)$, and in Fig.~\ref{fig3}b for the units $(t_*,d_*)$, the non-overlapping part of the data produce a series of parallel curves with a darker shade of red the further they are from the origin (an illustration is provided in SI Fig. 6). Indeed, dark shades of red encode large values of $\mathcal{N}$, and we have $t_*/t_0\simeq \mathcal{N}^\frac{5}{3}$, and $d_*/d_0\simeq \mathcal{N}^\frac{2}{3}$. More broadly, the left-out regime in each system of units can be expressed using the number $\mathcal{N}$:  
 \begin{align}
d\simeq c_0 t \leftrightarrow \frac{d}{d_*} &\simeq \mathcal{N} \frac{t}{t_*} \\
d\simeq c_s t \leftrightarrow \frac{d}{d_0} &\simeq \mathcal{N}^{-1} \frac{t}{t_0}
\end{align}
Solely rescaling the units does not allow the overlap of the three consecutive regimes at once. The two regimes going through the point of unit coordinates will overlap, but not the third regime. However, the prefactors of this third regime are set by the value of $\mathcal{N}$, and a complete overlap can be achieved if the importance of this number is fully acknowledged. 

\section*{A new base for counting}
In physics the term ``order of magnitude'' is often thrown around a bit loosely. Implicitly it is usually assumed that an order of magnitude is a decade, so two quantities separated by an order of magnitude will roughly differ by a factor of 10. How many decades a power law extends over is routinely used as a criterion to assess its worth. One may, for instance, say that Taylor's scaling is quite strong because it extends over almost three decades in time and over one in space. Ten is almost universally accepted as the base for counting, and indeed all logs we have used in the figures so far where logs in base 10. Nevertheless, this 10 is simply a convention, just like the meter or the second.  

When we are dealing with a single power law, we have seen that the dynamics are scale-free, in the sense that no preferred units stand out. In the same way, a scaling law is also base-free. The choice of base for the log is completely arbitrary and so the meaning of an ``order of magnitude'' can be adjusted at will. This freedom is not lifted by the addition of a second intersecting power law, but by considering a third. In that case we have seen that a special kind of dimensionless number emerges, $\mathcal{N}$. The value of this number changes from one explosion to another, based on the values of the ejection and sound speeds. However, in every explosion, $\mathcal{N}$ plays the same role. It sets the coordinates of the crossover and the prefactors of the power laws. The number 10 was arbitrary, bound to our subjective choices. In contrast, $\mathcal{N}$ is an objective number set by the dynamics. The number $\mathcal{N}$ provides an objective base, or `radix'. Indeed, when this number is used as a base for our logs, we can finally overlap all three regimes, as shown in Fig.~\ref{fig5}a. In effect, Fig.~\ref{fig5}a represents $\log_\mathcal{N} (d/d_*)$ vs $\log_\mathcal{N} (t/t_*)$. SI Fig. 6 and 7 summarizes the whole route from subjective units and base to the more objective representation of Fig.~\ref{fig5}a. 

On a plot with objective units and base, choosing the origin to be at $(t_0,d_0)$ or $(t_*,d_*)$ does not affect the appearance of the plot, it solely shifts the coordinates (Compare SI Fig. 6c-i and ii, and SI Fig. 7c-i and ii). For all curves, the full time range of Taylor's regime is always equal to $\frac{5}{3}$ ``orders of magnitude'' in base $\mathcal{N}$, and $\frac{2}{3}$ orders of magnitude in space, since $t_*/t_0\simeq \mathcal{N}^\frac{5}{3}$, and $d_*/d_0\simeq \mathcal{N}^\frac{2}{3}$ (top and right scale in Fig.~\ref{fig5}a). If these fractions are unsettling, one may prefer to use the radix $R\equiv \mathcal{N}^\frac{1}{3}$ to get 2 orders of magnitude in space and 5 in time (bottom and left scale in Fig.~\ref{fig5}a). 
\begin{figure*}
\centering
\includegraphics[width=17cm,clip]{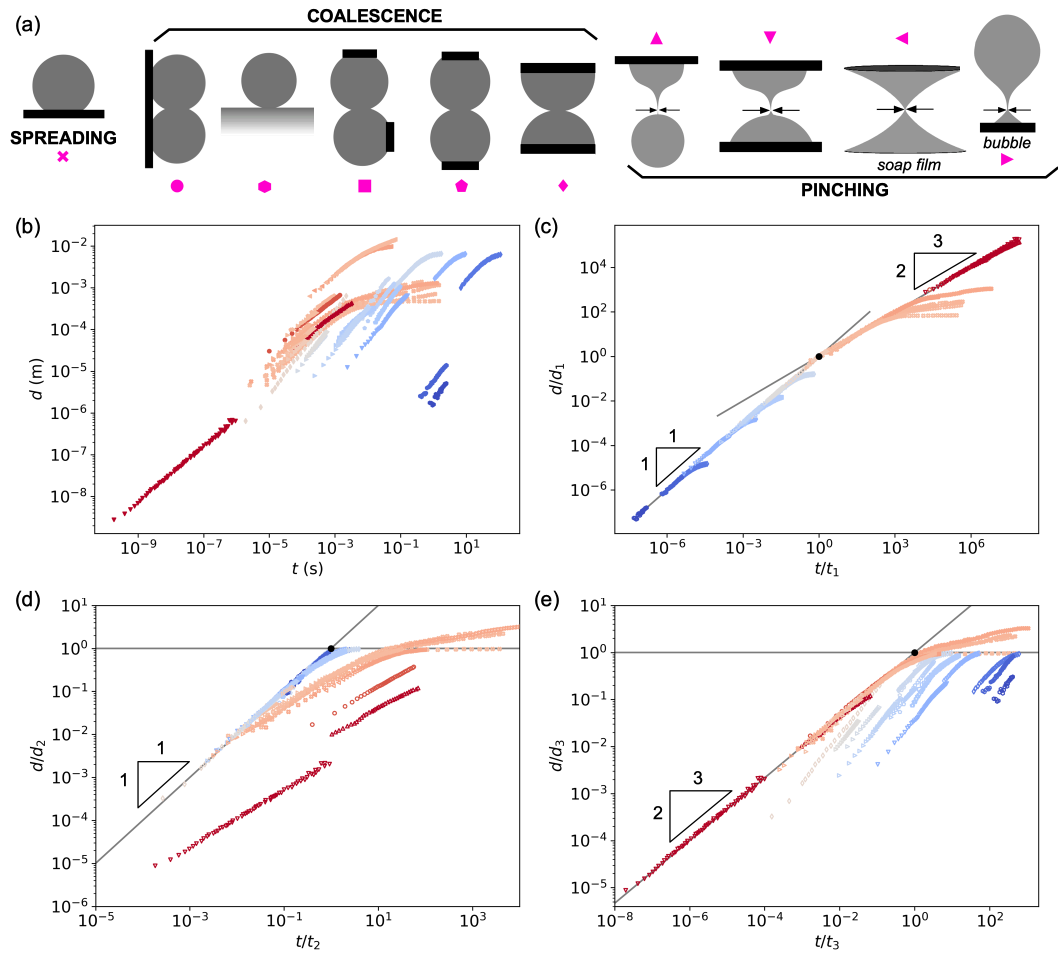}
\caption{Capillary dynamics of spreading~\cite{Eddi2013}, coalescing~\cite{Yao2005,Rahman2019,Aarts2005,Aarts2008,Paulsen2011}, and pinching droplets and bubbles~\cite{Chen1997,McKinley2000,Chen2002,Burton2004,Burton2005,Bolanos2009,Goldstein2010}, represented with standard units or with the objective units provided by the crossovers. (a) Sketches of the various spreading, coalescence and pinching setups present in the figure. Each setup is associated with a symbol used for the data sets in panels b to e. (b) The data are represented in conventional units (1~s, 1~m). (c) The data are represented in Ohnesorge units ($t_1$,$d_1$), defined in Eq.~\ref{Ohunits}. (d) The data are represented in visco-capillary units ($t_2$,$d_2$), defined in Eq.~\ref{VCunits}. (e) The data are represented in inertio-capillary units ($t_3$,$d_3$), defined in Eq.~\ref{ICunits}. The continuous grey lines are $d\simeq c_v t$, $d\simeq K_i t^\frac{2}{3}$, and $d\simeq D$.  The color code is explained later in the paper and quantified in Fig.~\ref{fig5}. A guide to the data is provided in SI. Animated versions of each panel illustrating all data sets are given in SI (Fig4b.gif, Fig4c.gif, Fig4d.gif, Fig4e.gif). The files SI4c.gif, SI4d.gif, and SI4e.gif provide animated transitions between panel b and panels c, d and e respectively, as explained in SI section VII. 
\label{fig4}}
\end{figure*} 

\section*{Duality}
All explosions we have discussed so far were detonations: the initial ejection speed was always greater than the sound speed, so $\mathcal{N}>1$. The converse is also possible. In that case, one speaks of deflagrations. As is apparent in Fig.~\ref{fig3}b or Fig.~\ref{fig3}c, the extent of Taylor's regime shrinks as $\mathcal{N}$ decreases toward unity. For deflagrations, when $\mathcal{N}<1$, this regime is not expected to be present. This progressive disappearance is quite clear in the data on conventional explosions~\cite{Hargather2007,Kleine2010}, or on vapor cloud explosions~\cite{Tang1999} (pale pink data sets in the figures). 

\begin{figure*}
\centering
\includegraphics[width=17cm,clip]{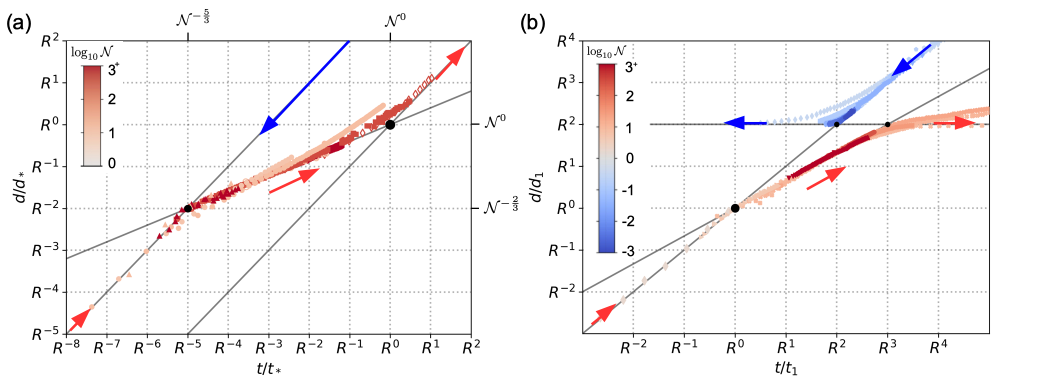}
\caption{Objective representations of the kinematics of explosions, and droplets and bubbles. (a) The explosion data introduced in Fig.~\ref{fig2} are plotted with objective units and an objective base. The units are $(t_*,d_*)$ defined in Eq.~\ref{HC1} and~\ref{HC2}. The base is the initial Mach number $\mathcal{N}\equiv c_0/c_s$ (top and right scales), or $R\equiv \mathcal{N}^\frac{1}{3}$ (left and bottom), a radix chosen such that the points of intersections between the regimes occur at integer coordinates. (b) Capillary dynamics of pinching, spreading and coalescing droplets and bubbles~\cite{Fardin2022} in objective units and base. The units are $(t_1,d_1)$ defined in Eq.~\ref{Ohunits}. The base is the inverse of the Ohnesorge number, $R\equiv \mathcal{N}\equiv c_v D^\frac{1}{2}/K_i^\frac{3}{2}$. In both plots, dynamics with $\mathcal{N}<1$ or $\mathcal{N}>1$ proceed in opposite directions, since $\mathcal{N}^{n+1}>\mathcal{N}^{n} \leftrightarrow \mathcal{N}>1$, a quite visual display of the duality of the dynamics. A guide to the data is provided in SI. Animated versions of each panel illustrating all data sets are given in SI (Fig5a.gif, Fig5b.gif). The files SI5a.gif and SI5b.gif provide animated transitions respectively between Fig.~\ref{fig3}b and panel a, and between Fig.~\ref{fig4}c and panel b, as explained in SI section VII. 
\label{fig5}}
\end{figure*}

The blue path in Fig.~\ref{fig5}a sketches what can be expected for deflagrations. The front proceeds at the constant ejection (or flame) speed. However, deflagrations are often affected by additional mechanisms (gravity, friction, etc), and so their dynamics may deviate from this template. We invite the reader to contact us to point us toward data sets on deflagration that could be included in a revised version of Fig.~\ref{fig5}a. 

Note that $\mathcal{N}=1$ is of course a singular case, which is quite obvious in a logarithmic representation in base $\mathcal{N}$, as in Fig.~\ref{fig5}a. For data sets with $\mathcal{N}$ just slightly above 1 the transitions from one regime to another tends to be stretched out and smoothed out (curves with pale shades). SI Fig.~10 shows the extent of such distortions in the case of the vapor cloud explosions~\cite{Tang1999}, which were included in Fig.~\ref{fig3}b and Fig.~\ref{fig3}c, but excluded from Fig.~\ref{fig5}a, since $\mathcal{N}\simeq 1^+$. This stretching effect and the case $\mathcal{N}=1$ go beyond the scope of this article.  

The duality between $\mathcal{N}<1$, and $\mathcal{N}>1$ can be sketched in the case of explosions, but it can be revealed more clearly for a different example, a second demonstration of the radical scaling approach we are promoting here. Whenever we have two intersecting power laws we acquire objective units. Whenever we have a third power law we can build an objective base or `radix', hence the term `radical' scaling. This procedure is absolutely general, so let us apply it to a different context: the dynamics of pinching~\cite{Chen1997,McKinley2000,Chen2002,Burton2004,Burton2005,Bolanos2009,Goldstein2010}, spreading~\cite{Eddi2013} and coalescing droplets and bubbles~\cite{Yao2005,Rahman2019,Aarts2005,Aarts2008,Paulsen2011}. 
(Data used in Fig.~\ref{fig4} and Fig.~\ref{fig5}b are from these cited references). We recently had the opportunity to review this field but we had not gone through the extra step of renormalizing the number base~\cite{Fardin2022}. 

Just as in the case of explosions, we are considering three consecutive power laws. For pinching, the neck of the drop is tracked as a function of the duration before pinch-off, for spreading, the contact radius is tracked since the instant of contact, and for coalescence, the radius of contact between the drops or bubbles is tracked since the instant they first touched~\cite{Fardin2022}. In these three setups, for drops and for bubbles, a number of experiments have progressively evidenced the possible existence of three consecutive regimes~\cite{Fardin2022} (other paths are possible~\cite{Xia2019,Eggers2024}, but they are not in focus here). At short time, the trajectory is linear, $d\simeq c_{v} t$, then a power law of the form $d\simeq K_i t^\frac{2}{3}$ is observed, until the variable size $d$ eventually reaches its maximum, set by the droplet or bubble size, $d\simeq D$. Experiments rarely have enough resolution to capture the three regimes, but their existence is inferred by piecing together multiple experiments~\cite{Fardin2022}. In the case of explosions, the initial and final regimes were linear, hence parallel in a logarithmic plot. Thus we only had two points of intersection, with coordinates $(t_0,d_0)$ and $(t_*,d_*)$. Since the three regimes now have different slopes ($\frac{2}{3}\neq 1 \neq 0$), we have three points of intersection:
 \begin{align}
d_1\simeq c_{v} t_1 \simeq K_i t_1^\frac{2}{3}  &\rightarrow t_1 \simeq\Big(\frac{K_i}{c_v}\Big)^3~;~~d_1\simeq \frac{K_i^3}{c_v^2} \label{Ohunits}\\
d_2\simeq D\simeq c_{v} t_2   &\rightarrow t_2 \simeq\frac{D}{c_v}~;~~d_2\simeq D \label{VCunits} \\
d_3\simeq D\simeq K_i t_3^\frac{2}{3}   &\rightarrow t_3 \simeq \Big(\frac{D}{K_i}\Big)^\frac{3}{2}~;~~d_3\simeq D  \label{ICunits}
\end{align}
The first pair of coordinates provides what we have called the Ohnesorge units~\cite{Fardin2022}. The two other pairs respectively correspond to what are usually called the visco-capillary and inertio-capillary units~\cite{Fardin2022}, due to the mechanical underpinning of the constants $K_i$ and $c_v$ (see SI section II.B.2 for details).   

As seen in Fig.~\ref{fig4}a, when plotted in standard units and with the traditional base 10, experiments on pinching, spreading, and coalescence crisscross each other in a tremendous mess. By choosing one of the three objective systems of units only two out of three regimes can be overlapped, as shown in Fig.~\ref{fig4}b, c and d. The full overlap is reached by rescaling the base. As in the case of explosions, the kinematic parameters can be combined to obtain a dimensionless number. Now we have:
\begin{equation}
 \mathcal{N}\equiv \frac{c_v D^\frac{1}{2}}{K_i^\frac{3}{2}} \label{Ndroplets}
\end{equation}
Note that this number (like any dimensionless number) is defined modulo an overall power~\cite{Fardin2024}, so for instance we could also use $\mathcal{N}^{-1}$, which is called the Ohnesorge number, or $\mathcal{N}^{2}$, which is called the Laplace number~\cite{McKinley2011}. We choose the inverse of the Ohnesorge number (i.e. the square root of the Laplace number) as defined in Eq.~\ref{Ndroplets}, in order to facilitate the comparison with the dynamics of explosions. The number $\mathcal{N}$ is a constant for each experiment, and just as with the initial Mach number for explosions, we can use $\mathcal{N}$ as our objective base. Fig.~\ref{fig5}b gives the objective plot for these dynamics. 

For these pinching, coalescence and spreading dynamics, the duality of the scaled plot is now clear. The succession of three regimes is only seen as long as $\mathcal{N}>1$. When $\mathcal{N}<1$ the linear regime at speed $c_v$ intersects the maximum size $D$ before reaching the $\frac{2}{3}$ regime, which is now inaccessible. Usually one speaks of inertial dynamics when $\mathcal{N}>1$ and of viscous dynamics when $\mathcal{N}<1$~\cite{Fardin2022}. Note that in Fig.~\ref{fig5}, dynamics with $\mathcal{N}<1$ or $\mathcal{N}>1$ proceed in opposite directions, since $\mathcal{N}^{n+1}>\mathcal{N}^{n} \leftrightarrow \mathcal{N}>1$, a quite visual display of the duality of the dynamics. 

We are currently studying how this duality manifests itself in other examples and we encourage the readers to reanalyze the data they may be familiar with under this new light. 

\section*{The pursuit of objectivity: radical scaling}
Plotting a single power law requires a choice of units and a choice of base, choices that are made subjectively, solely guided by arbitrary conventions. In a plot like Fig.~\ref{fig1}b the human presence is everywhere. Our feet are used to measure space, our hands to count, and the unit of time is barely less biased, a mixture of Egyptian and Babylonian fractions of the rotational period of our own planet. These choices are required to describe the phenomenon, but the phenomenon itself is expected to be independent of these choices. This independence can only be reconquered if the phenomenon is not as simple as initially thought, if the dynamics show at least three connected trends, rather than a single power law. 

This inherent connection between the wondrous diversity of nature and the objective description of phenomena reveals a key insight: simplicity, as appealing as it may seem, often obscures the underlying richness of the world. A phenomenon that can be described by a single power law is inherently tied to the subjective choices of the observer--choices that impose our human scales and perspectives onto the data. In contrast, when we encounter a phenomenon characterized by multiple, connected power laws, we are granted the opportunity to strip away this human imprint. The units and bases used in our descriptions become dictated by the data themselves, reflecting the true nature of the phenomenon rather than our conventions.


It is important to distinguish, however, between the objective diversity of nature and mere complexity. Consider Fig.~\ref{fig2}b or Fig.~\ref{fig4}a, where multiple datasets, gathered under varying conditions, are plotted subjectively. The result is a tangled web of data points that seems overwhelmingly complex. This apparent complexity, however, is often a reflection of our arbitrary choices in units and base rather than the phenomenon itself. When these units and bases are determined objectively, as in Fig.~\ref{fig5}, the data align in a more coherent and understandable pattern. By removing the subjective layers, we unveil a clearer, more interpretable representation of the underlying phenomena. What once appeared as a convoluted mess now reveals a pattern free from the distortions of human perspective. 

Once they are stripped from our footprints and fingerprints, the data are ready to be interpreted. This interpretation does not rely on our measuring and reckoning conventions anymore: instead it usually invokes dimensions beyond those of the variables. For instance, when power laws are kinematic, relating space and time $d(t)$, like those we used as examples, then the interpretation may be mechanical, involving forces, pressures, etc., quantities adding the dimension of mass to those of space and time. For instance, Taylor showed that the kinematic constant of the blast could be factorized as $K\simeq (E/\rho)^\frac{1}{5}$, where $E$ is the explosion yield and $\rho$ is the ambient density~\cite{Taylor1950a,Taylor1950b}. We recently reviewed this widespread decomposition of kinematic constants into pairs of mechanical parameters~\cite{Fardin2024}, and have been publishing \href{www.youtube/@naturesnumbers}{video lectures} on how to address objective units and bases from such a mechanical point of view. What this article reveals is that a pair of mechanical quantities and its associated regime can only tell an incomplete story. A more complete scaling analysis should make an effort to identify three connected power laws, not just one, and this would require a minimum number of four mechanical factors. We will address these questions of mechanical combinatorics in a future article. 

\subsubsection*{Acknowledgments}
M.H. thanks i-Bio funding. We thank N. Nikolova for her input on the manuscript. We thank G. McKinley, G. Spriggs, T. Wei and C. Aouad for enlightening discussions about drops, bubbles and explosions. We are grateful to A. Part (Atomic Heritage Foundation, National Museum of Nuclear Science \& History) and A. Carr (Los Alamos National Lab.) for their help in collecting material about the Trinity test. Finally, we acknowledge the joyful atmosphere of the Ladoux-M{\`e}ge lab, the stimulating haven that made this project possible.

\clearpage

\begin{widetext}

\begin{center}
\Huge Supplementary Information
\end{center}

\newcommand{\savedthefigure}{\thefigure}
\setcounter{figure}{0}  
\renewcommand{\thefigure}{S\arabic{figure}}
\renewcommand{\thetable}{S\arabic{table}}

\section{Experimental data summary}
Our study provides a meta-analysis of a number of experiments on explosions (nuclear, conventional, underwater and laser-induced), as well as experiments on spreading, coalescence and pinching of fluids of various properties. In this section we explain the protocol we followed to extract the data sets from the original articles and we give tables summarizing the properties of all experiments reproduced in the figures of the article. 

\subsection{Data extraction}
All data were extracted semi-manually from the figures of the original sources given in Table~\ref{table1} and \ref{table2}. For a given figure, the data points were identified manually on the free imagining software Fiji. The coordinates of the data points were stored and converted to standard units (seconds and meters). The precision is expected to be on the order of the size of the symbols used in the original graphs. When the sampling was high and multiple data points overlapped we only selected a subsets of the original data points. We omitted data points with large error bars, which usually corresponded to the first few measurements at the limit of the resolution of the experiment. All extracted data sets ($t$ and $d$) are given as two-columns text-files in the supplementary archive `\textbf{DataSets.zip}'.   

\begin{table}[h!]
\begin{tabular}{|l|c|c|l|}
\hline
 \textbf{Label} & \textbf{Type} & \textbf{Symbol} & \textbf{Reference} \\ \hline\hline
Taylor1950  & Nuclear & $\blacksquare$ & G.I. Taylor, \textit{Proc. R. Soc. Lond. A} \textbf{201}, 175-186 (1950) \\\hline
Mack1946  & Nuclear & $\blacksquare$ & J.E. Mack, LANL Technical Report LA-531 (1946) \\\hline
Bainbridge1976  & Nuclear & $\blacksquare$ & K.T. Bainbridge, LANL Technical Report (1976) \\\hline
OConnell1957  & Nuclear & $\blacksquare$ & P. O'Connell, EG{\&}G Technical Report (1957)\\\hline
Nguyen2017  & Nuclear & $\blacksquare$ & J.D. Nguyen and G.D. Spriggs, LLNL Technical Report (2017)\\\hline
Schmitt2016  & Nuclear & $\blacksquare$ & D.T. Schmitt, Thesis, Air Force Institute of Technology (2016)\\\hline
DominicHousatonic  & Nuclear & $\blacksquare$ &  Private communication with G.D. Spriggs (LLNL) \\\hline
Porzel1957  & Underwater nuclear & $\blacklozenge$ & F. Porzel, IIT Technical Report (1957) \\\hline
Kingery1962  & Conventional & $\bullet$ & C. Kingery \textit{et al.}, Army Ballistic Research Lab. Technical Report (1962) \\\hline
Aouad2021  & Conventional & $\bullet$ & C. Aouad \textit{et al.}, \textit{Shock Waves} \textbf{31}, 813-827 (2021) \\\hline
Hargather2007  & Conventional & $\bullet$ & M.J. Hargather and M.J. Settles, \textit{Shock Waves} \textbf{17}, 215--223 (2007)\\\hline
Kleine2010  & Conventional & $\bullet$ & H. Kleine, \textit{Eur. Phys. J. Spec. Top.} \textbf{182}, 3--34 (2010)\\\hline
Grun1991  & Laser-induced & $\blacktriangle$ & J. Grun  \textit{et al.}, \textit{Phys. Rev. Lett.} \textbf{66}, 2738 (1991)\\\hline
Porneala2006  & Laser-induced & $\blacktriangle$ & C. Porneala and D.A. Willis, \textit{Appl. Phys. Lett.} \textbf{89}, 211121 (2006)\\\hline
Gatti1988  & Laser-induced & $\blacktriangle$ & M. Gatti \textit{et al.}, \textit{Opt. Commun.} \textbf{69}, 141--146 (1988)\\\hline
Tang1999  & Vapor-cloud & $\bigstar$ & M.J Tang and Q.A. Baker, \textit{Process Saf. Prog.} \textbf{18}, 235--240 (1999)\\\hline
\end{tabular}
\caption{Origin of the data on explosions used in the article. Abbreviations: LANL (Los Alamos National Laboratory), LLNL (Lawrence Livermore National Laboratory), EG{\&}G (Edgerton, Germeshausen, and Grier, Inc.), IIT (Illinois Institute of Technology). All cited technical reports have been declassified and are freely available. 
\label{table1}}
\end{table}

\begin{table}[h!]
\begin{tabular}{|l|c|c|l|}
\hline
 \textbf{Label} & \textbf{Type} & \textbf{Symbol} & \textbf{Reference} \\ \hline\hline
Eddi2013  & Spreading & \ding{54} & A. Eddi \textit{et al.}, \textit{Phys. Fluids} \textbf{25}, 013102 (2013) \\\hline
Eddi2013b  & Coalescence & $\bullet$ & A. Eddi \textit{et al.}, \textit{Phys. Rev. Lett.} \textbf{111}, 144502 (2013) \\\hline
Yao2005  & Coalescence & $\blacklozenge$ & W. Yao \textit{et al.}, \textit{Phys. Rev. E} \textbf{71}, 016309 (2005)\\\hline
Aarts2005  & Coalescence & \begin{tikzpicture}
\filldraw[black] (90:0.1) -- (162:0.1) -- (234:0.1) -- (306:0.1) -- (18:0.1) -- cycle;
\end{tikzpicture}  & D. Aarts \textit{et al.}, \textit{Phys. Rev. Lett.} \textbf{95}, 164503 (2005) \\\hline
Aarts2008  & Coalescence & \begin{tikzpicture}
\filldraw[black] (90:0.1) -- (150:0.1) -- (210:0.1) -- (270:0.1) -- (330:0.1) -- (30:0.1) -- cycle;
\end{tikzpicture}   & D. Aarts and H. Lekkerkerker, \textit{J. Fluid Mech.} \textbf{71}, 275--294 (2008)\\\hline
Paulsen2011  & Coalescence & $\blacklozenge$ & J. Paulsen  \textit{et al.}, \textit{Phys. Rev. Lett.} \textbf{106}, 114501 (2011)\\\hline
Rahman2019  & Coalescence & $\blacksquare$ & M. Rahman  \textit{et al.}, \textit{Phys. Fluids} \textbf{31}, 012104 (2019)\\\hline
Chen1997  & Pinching & $\blacktriangleleft$ & Y.J. Chen and P.H. Steen, \textit{J. Fluid Mech.} \textbf{341}, 245--267 (1997)\\\hline
McKinley2000  & Pinching &  $\blacktriangledown$ & G.H. McKinley and A. Tripathi, \textit{J. Rheol.} \textbf{44}, 653--670 (2000)\\\hline
Chen2002  & Pinching &  $\blacktriangle$ & A.U. Chen \textit{et al.}, \textit{Phys. Rev. Lett.} \textbf{88}, 174501 (2002)\\\hline
Burton2004  & Pinching &  $\blacktriangledown$ & J.C. Burton \textit{et al.}, \textit{Phys. Rev. Lett.} \textbf{92}, 244505 (2004)\\\hline
Burton2005  & Pinching &  $\blacktriangleright$ & J.C. Burton \textit{et al.}, \textit{Phys. Rev. Lett.} \textbf{94}, 184502 (2005)\\\hline
Bolanos2009  & Pinching &  $\blacktriangleright$ & R. Bolanos-Jim{\'e}nez \textit{et al.}, \textit{Phys. Fluids} \textbf{21}, 072103 (2009)\\\hline
Goldstein2010  & Pinching &  $\blacktriangleleft$ & R.E. Goldstein \textit{et al.}, \textit{PNAS} \textbf{107}, 21979--21984 (2010)\\\hline
\end{tabular}
\caption{Origin of the data on spreading, pinching and coalescence used in the article.
\label{table2}}
\end{table}
 
\subsection{Data summary}
The values of the kinematic parameters associated to the regimes discussed in the article are given in in Table~\ref{table2a} for explosions, and in Table~\ref{table2b} for droplets and bubbles. Also included are the underlying mechanical parameters, which can be used to estimate the kinematic parameters, when the range of a particular experiment did not cover the regime (c.f. section~\ref{mechmod}). In such case the kinematic values appear between brackets. The data in Tables~\ref{table2a} and~\ref{table2b} can also be found in the supplementary files `\textbf{SummaryExplo.csv}' and `\textbf{SummaryDrop.csv}'. More details on each data set are given in section~\ref{datacomments}. 

\begin{table}[ht]
\centering
\csvreader[
    tabular=|l|c|c|c|c|c|c|c|c|, 
    respect all,                 
    table head=\hline 
    \textbf{Label} & $\bm\rho$ (kg.m$^{-3}$) & $\bm E$ (kg.m$^{2}$.s$^{-2}$) & 
    $\bm\Sigma$ (kg.m$^{-1}$.s$^{-2}$) & $\bm{m}$ (kg) & $\bm{\mathcal{N}}$ & 
    $\bm{c_{0}}$ (m/s) & $\bm{c_{s}}$ (m/s) & $\bm{K}$ (m/s$^\frac{2}{5}$) \\ \hline,
    late after line=\\\hline 
    ]%
    {./Figures/SummaryExplo.csv} 
    {
         Label=\Label,            
		Type=\Type,
		Symbol=\Symbol,
         Density=\Density, 
         Energy=\Energy, 
         BulkModulus=\BulkModulus, 
         EjectedMass=\EjectedMass, 
         InitialMach=\InitialMach, 
         co=\co, 
         cs=\cs, 
         K=\K
    }
    {\Label & \Density & \Energy & \BulkModulus & \EjectedMass & \InitialMach & \co & \cs & \K} 
\caption{Summary of the parameters of the explosions reproduced in the article: mechanical parameters (ambient density $\rho$, yield $E$, ambient bulk modulus $\Sigma$, ejected mass $m$), initial Mach number $\mathcal{N}\equiv c_0/c_s$ , kinematic parameters (initial speed $c_0$, ambient speed of sound $c_s$, and Taylor's prefactor $K$).  The content of this table is available in the supplementary file `\textbf{SummaryExplo.csv}'.   The first line of the table corresponds to the parameters associated with the Trinity test, corresponding to data from `Taylor1950', `Mack1946' and `Bainbridge1976', from the references given in Table~\ref{table1}.
}
\label{table2a}
\end{table}

\begin{table}
\csvreader[tabular=|l|c|c|c|c|c|c|c|,respect all,
    table head=\hline \textbf{Label} & $\bm\rho$ (kg.m$^{-3}$) & $\bm\eta$ (kg.m$^{-1}$.s$^{-1}$) & $\bm\Gamma$ (kg.s$^{-2}$)  & $\bm{\mathcal{N}}$ & 
    $\bm{c_{v}}$ (m/s) & $\bm{K_i}$ (m/s$^\frac{2}{3}$) & $\bm{D}$ (m)\\ \hline,
    late after line=\\\hline
  ]%
{./Figures/SummaryDrop.csv}{Label=\Label,Type=\Type,Density=\Density,Viscosity=\Viscosity, 5=\SufTens,7=\Size, 8=\Oh, 9=\Ki, 10=\cv}%
{\Label  &\Density &\Viscosity &\SufTens & \Oh & \cv & \Ki & \Size }
\caption{Summary of the parameters of the spreading, coalescence and pinching experiments reproduced in the article: mechanical parameters (density $\rho$, viscosity $\eta$, surface-tension $\Gamma$), inverse Ohnesorge number $\mathcal{N}\equiv (c_vD^\frac{1}{2})/K_i^\frac{3}{2}$ , kinematic parameters (visco-capillary speed $c_v$, inertio-capillary prefactor $K_i$, and droplet/bubble size $D$). The content of this table is available in the supplementary file `\textbf{SummaryDrop.csv}'.  
\label{table2b}}
\end{table}

\clearpage


\section{Systems of units}
In this section, we provide additional details on the two systems of units introduced in the case of explosions, and the three systems introduced for droplets and bubbles. These systems of units correspond to the crossovers between regimes. 

\subsection{Kinematic definitions}
\subsubsection{Explosions}
\begin{table}[ht]
\centering
\csvreader[
    tabular=|l|c|c|c|c|, 
    respect all,                 
    table head=\hline 
    \textbf{Label} &  $\bm{t_{*}}$ (s) & $\bm{d_{*}}$ (m) & $\bm{t_0}$ (s)  & $\bm{d_0}$ (m)\\ \hline,
    late after line=\\\hline 
    ]%
    {./Figures/SummaryExplo.csv} 
    {
         Label=\Label,            
		Type=\Type,
         12=\t1,
         13=\d1,
         14=\to,
         15=\do
    }
    {\Label & \t1 & \d1 & \to & \do} 
\caption{Values of the crossovers $(t_0,d_0)$ and $(t_*,d_*)$ between the three regimes of explosions discussed in the article. Values between brackets are estimated from the mechanical models described in section~\ref{mechexplo}, when the data sets did not extend enough to capture them. The content of this table is available in the supplementary file `\textbf{SummaryExplo.csv}'.
}\label{table3a}
\end{table}
For explosions we considered three regimes: 
 \begin{align}
\text{Regime 1:}\quad& d\simeq c_0 t\\
\text{Regime 2:}\quad& d\simeq K t^\frac{2}{5}\\
\text{Regime 3:}\quad& d\simeq c_s t
\end{align}
These three regimes lead to two points of intersection: 
\begin{align}
\text{Regime 1} \cap \text{Regime 2} \quad \rightarrow  t_0\simeq \Big(\frac{K}{c_0} \Big)^\frac{5}{3} \quad \& \quad d_0\simeq \frac{K^\frac{5}{3}}{c_0^\frac{2}{3}}\label{todo} \\
\text{Regime 2} \cap \text{Regime 3} \quad \rightarrow  t_*\simeq \Big(\frac{K}{c_s} \Big)^\frac{5}{3} \quad \& \quad d_*\simeq \frac{K^\frac{5}{3}}{c_s^\frac{2}{3}}\label{txdx}
\end{align}

\subsubsection{Droplets and bubbles}
\begin{table}
\csvreader[tabular=|l|c|c|c|c|c|,respect all,
    table head=\hline \textbf{Label} &  $\bm{t_{1}}$ (s) & $\bm{d_{1}}$ (m) & $\bm{t_2}$ (s) & $\bm{t_3}$ (s) & $\bm{D}$  (m)\\ \hline,
    late after line=\\\hline
  ]%
{./Figures/SummaryDrop.csv}{Label=\Label , 7=\Size , t1=\timo ,d1=\diso ,t2=\timv ,14=\timi}%
{\Label  & \timo & \diso & \timv & \timi & \Size  }
\caption{Values of the crossovers $(t_1,d_1)$, $(t_2,d_2)$, and $(t_3,d_3)$ between the three regimes of the dynamics of droplets and bubbles discussed in the article. Values between brackets are estimated from the mechanical models described in section~\ref{mechdrop}, when the data sets did not extend enough to capture them. The content of this table is available in the supplementary file `\textbf{SummaryDrop.csv}'. }
\label{table3b} 
\end{table}
For droplets and bubbles we considered three regimes: 
\begin{align}
\text{Regime 1:}\quad& d\simeq c_v t\\
\text{Regime 2:}\quad& d\simeq K_i t^\frac{2}{3}\\
\text{Regime 3:}\quad& d\simeq D
\end{align}
These three regimes lead to three points of intersection: 
\begin{align}
\text{Regime 1} \cap \text{Regime 2} \quad \rightarrow  t_1\simeq \Big(\frac{K_i}{c_v} \Big)^3 \quad \& \quad d_1\simeq \frac{K_i^3}{c_v^2} \label{t1d1}\\
\text{Regime 1} \cap \text{Regime 3} \quad \rightarrow  t_2\simeq \frac{D}{c_v} \quad \& \quad d_1\simeq D  \label{t2d2}\\
\text{Regime 2} \cap \text{Regime 3} \quad \rightarrow  t_3\simeq \Big(\frac{D}{K_i} \Big)^\frac{3}{2} \quad \& \quad d_3\simeq D  \label{t3d3}
\end{align}

\subsection{Mechanical models\label{mechmod}}
Data sets rarely extend over a large enough range to capture two consecutive regimes, let alone three. As shown in the article, knowledge on the various crossovers is then built piece by piece, by assembling data sets covering different parts of the master curves. For this assemblage to occur we need a way to predict the locations of the crossovers beyond any particular experimental range. We need mechanical models of the kinematics. Both explosions and capillary dynamics of droplets and bubbles have well established mechanical models that we can rely on. We recently published a review focusing on the mechanical models for spreading, pinching and coalescence~\cite{Fardin2022}. More generally, we discuss how mechanical models can be systematically tied to kinematic power laws in another publication~\cite{Fardin2024}. Finally, we dedicated a series of \href{https://www.youtube.com/playlist?list=PLbMiQs7eX-bbNTc-7HwdWzohUs8yPw300}{video lectures } to the mechanics of explosions, which provide additional information on this topic. Thus, we only here recall the essential aspects of the mechanical decomposition of the kinematic parameters of the power laws discussed in the article.

\subsubsection{Explosions\label{mechexplo}}
\begin{figure*}
\centering
\includegraphics[width=17cm,clip]{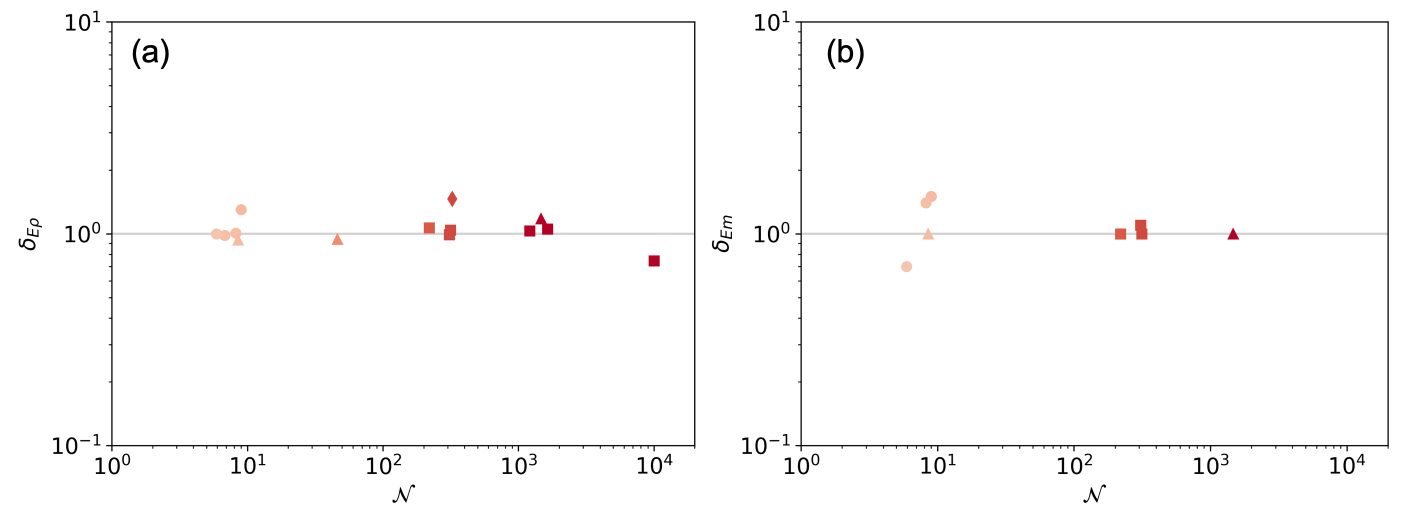}
\caption{(a) Values of the numerical correction $\delta_{E\rho}$ obtained by comparing the fitted value of the kinematic parameter $K$ and the value expected from the mechanical model, i.e.  $\delta_{E\rho}\equiv K  (E/\rho)^{-\frac{1}{5}}$. All data sets from Table~\ref{table1} are included. (b) Values of the numerical correction $\delta_{Em}\equiv c_0 (E/m)^{-\frac{1}{2}}$. Only a fraction of the data sets are included, those that captured the initial regime at constant speed, and so which could provide an experimental value for $c_0$: Taylor 1950, O'Connell 1957, Nguyen 2017, Kingery 1962, Hargather 2007, Kleine 2010, Grun 1991, and Porneala 2006.  
\label{Sfig1}}
\end{figure*} 
~\citet{Taylor1950a,Taylor1950b} famously established the mechanical scaling for the power law $d\simeq K t^\frac{2}{5}$. During this phase of the dynamics, the energy $E$ of the explosion is impelling the advancement of the blast, whereas the inertia of the surrounding air being swept-up by the shock is providing the impeding factor, through its density $\rho$. The dimensions of these two mechanical factors suggest that $K\simeq (E/\rho)^\frac{1}{5}$~\cite{Taylor1950a,Taylor1950b,Fardin2024}. This mechanical decomposition is actually quite effective in recovering the measured values of the kinematic parameter $K$ for a a range of explosions. We can introduce a numerical correction $\delta_{E\rho}$, such that $K=\delta_{E\rho} (E/\rho)^\frac{1}{5}$. As shown in Fig.~\ref{Sfig1}a the value of $\delta_{E\rho}$ for the explosions in Table~\ref{table2a} always remains close to 1 (For additional details see: \href{https://youtu.be/tJxJAh7_h3w?si=kSOn7KMudzGBHboq}{Explosions - Lecture 3}). 

Beyond Taylor's $\frac{2}{5}$ regime the speed of the shock has decreased to such an extent that its motion is almost indistinguishable from that of a sound wave of constant speed $c_s$~\cite{Hopkinson1915,Cranz1926,Fuller2005,Sachs1944,Westine2012,Wei2021}. This sound speed can be expressed mechanically, from the ratio between the bulk modulus $\Sigma$ of the ambient medium (air, water, etc.), and its density, i.e. $c_s\simeq (\Sigma/\rho)^\frac{1}{2}$ (c.f. \href{https://youtu.be/JcqVP6Q22PU?si=2wqqNRttbm1A8n39}{Explosions - Lecture 4}). 

Finally the initial regime of the explosion is often expressed by a ballistic model. The initial speed $c_0$ is impelled by the energy of the explosion and impeded by the ejected mass $m$, such that $c_0\simeq (E/m)^\frac{1}{2}$~\cite{Bethe1947}. The ejected mass is typically the mass of the bomb, and of solid materials in its immediate proximity. For instance, for Trinity $m\simeq 9$~tons, including the bomb and the cabin around it~\cite{Mack1946,Bainbridge1976}. A numerical correction $\delta_{Em}$ may eventually be necessary, $c_0=\delta_{Em} (E/m)^\frac{1}{2}$. Only a fraction of the data sets in Table~\ref{table1} extend to short enough times to capture the initial regime. The values of the correction $\delta_{Em}$ for these data sets are plotted in Fig.~\ref{Sfig1}b, showing that they always remain close to 1, attesting the validity of the mechanical model (c.f. \href{https://youtu.be/vSLiD-1FkcU?si=nDmSLpFqaxeAxNne}{Explosions - Lecture 5}).  

Overall the the three regimes described in the article can be expressed from three mechanical ratios involving the energy $E$, the ejected mass $m$, and the bulk modulus $\Sigma$ and density $\rho$ of the surrounding medium:
\begin{align}
\text{Regime 1:}\quad& c_0 \simeq \Big(\frac{E}{m}\Big)^\frac{1}{2} \label{comech}\\
\text{Regime 2:}\quad& K\simeq \Big(\frac{E}{\rho}\Big)^\frac{1}{5} \label{Kmech}\\
\text{Regime 3:}\quad& c_s\simeq  \Big(\frac{\Sigma}{\rho}\Big)^\frac{1}{2} \label{csmech}
\end{align}

The values of the four mechanical parameters for each explosion are given in Table~\ref{table2a}. Masses appearing between brackets correspond to data sets where the ejected mass could not be estimated based on the information provided in the associated reference. Two cases must be distinguished. When the value appears between square brackets, it is an upper-bound based on the earliest data point abiding to Taylor's regime. For instance, for `Aouad 2021' the first data point was collected at $t_i=33$~ms, and it still follows Taylor's regime (Regime 2), indicating that $t_i>t_0$, so $t_i>(K/c_0)^\frac{5}{3}$, i.e. $t_i\gtrsim(Km^\frac{1}{2}/E^\frac{1}{2})^\frac{5}{3}$, thus $m\lesssim t_i^\frac{6}{5} E/K^2$, and $m\lesssim 40$~tons. In the second case, within regular brackets, the data were collected early enough to capture the initial regime (Regime 1), and the mass was estimated using $m= E(\delta_{Em}/c_0)^2$, using $\delta_{Em}=1$. 

In Table~\ref{table2a}, the values of $c_0$ appearing between brackets are estimated using $c_0=\delta_{Em} (E/m)^\frac{1}{2}$, assuming $\delta_{Em}=1$. The values between square brackets use the estimated upper bound for the mass $m$. 

The coordinates of the crossovers can also be expressed mechanically. Using Eq.~\ref{todo}, \ref{comech} and \ref{Kmech}, the coordinates of the early crossover can be written in the following way: 
\begin{align}
t_0 &= \Big(\frac{\delta_{E\rho}}{\delta_{Em}}\Big)^\frac{5}{3} \Big(\frac{m^5}{E^3\rho^2}\Big)^\frac{1}{6}\simeq \Big(\frac{m^5}{E^3\rho^2}\Big)^\frac{1}{6}\label{Ohtime}\\
d_0 &=\Big(\frac{\delta_{E\rho}^5}{\delta_{Em}^2}\Big)^\frac{1}{3}  \Big(\frac{m}{\rho}\Big)^\frac{1}{3}\simeq \Big(\frac{m}{\rho}\Big)^\frac{1}{3}\label{Ohsize}\
\end{align}
In the right-hand side, the formulas neglect the numerical corrections $\delta_{Em}$ and $\delta_{E\rho}$. 

Using Eq.~\ref{txdx}, \ref{Kmech}, and  \ref{csmech}, the coordinates of the late crossover can be written in the following way: 
\begin{align}
t_* &= \delta_{E\rho}^\frac{5}{3} \Big(\frac{\rho^3 E^2}{\Sigma^5}\Big)^\frac{1}{6}\simeq\Big(\frac{\rho^3 E^2}{\Sigma^5}\Big)^\frac{1}{6}\\
d_* &=\delta_{E\rho}^\frac{5}{3}  \Big(\frac{E}{\Sigma}\Big)^\frac{1}{3}\simeq \Big(\frac{E}{\Sigma}\Big)^\frac{1}{3}
\end{align}
In the right-hand side, the formulas neglect the numerical correction $\delta_{E\rho}$. These formulas also assume that the speed of sound is exactly given by the formula $c_s=(\Sigma/\rho)^\frac{1}{2}$. 

\subsubsection{Droplets and bubbles\label{mechdrop}}
\begin{figure*}
\centering
\includegraphics[width=17cm,clip]{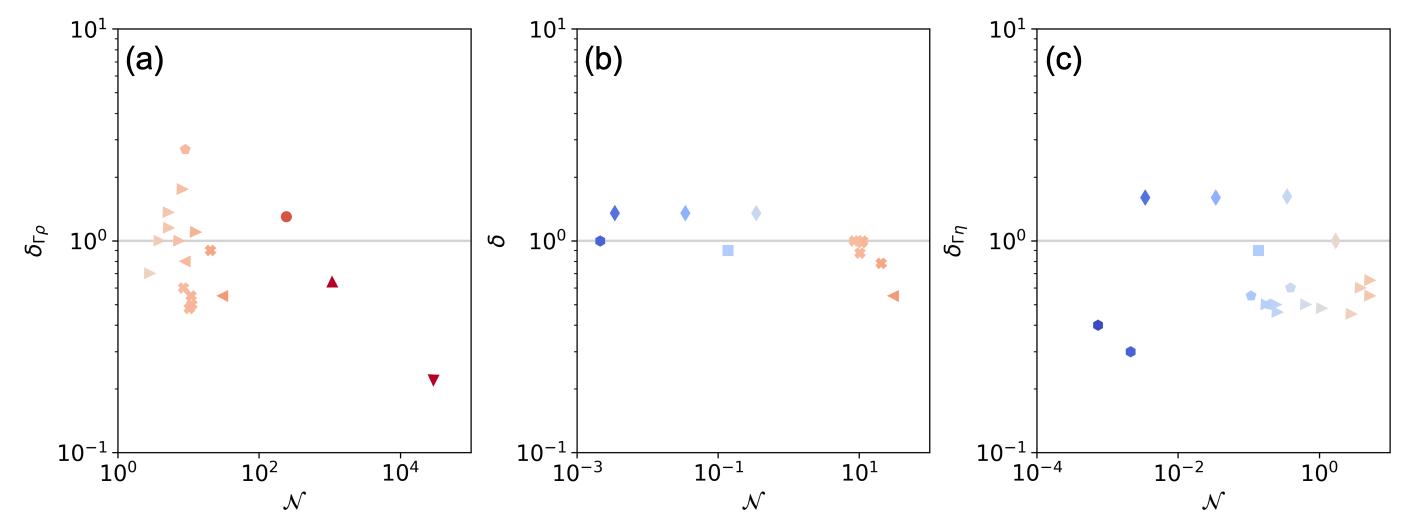}
\caption{(a) Values of the numerical correction $\delta_{\Gamma\rho}$ obtained by comparing the fitted value of the kinematic parameter $K_i$ and the value expected from the mechanical model, i.e.  $\delta_{\Gamma\rho}\equiv K_i  (\Gamma/\rho)^{-\frac{1}{3}}$. All values of $K_i$ appearing without brackets in Table~\ref{table2} are included. (b) Values of the numerical correction $\delta \equiv d_f/D$, where $d_f$ is the asymptotic value of the variable $d$. Only data sets capturing the final regime are included. Thus, data sets from Table~\ref{table2} where $t_1$ and $t_2$ both appear between brackets are excluded. (c) Values of the numerical correction $\delta_{\Gamma\eta} \equiv c_v (\Gamma/\eta)^{-1}$. Only data sets capturing the visco-capillary regime are included. Thus, data sets from Table~\ref{table2} where $c_v$ appear between brackets are excluded.
\label{Sfig2}}
\end{figure*} 
The mechanical scaling for the power law $d\simeq K_i t^\frac{2}{3}$ was established by~\citet{Keller1983} in the context of pinching, but the scaling had been identified earlier by~\citet{Thomson1871} and \citet{Rayleigh1890} in the context of capillary ripples~\cite{Fardin2024}. The kinematic parameter comes from the interplay between the surface-tension $\Gamma$ and the dominant density $\rho$, which is the density of the outer fluid for bubbles and the density of the inner fluid for droplets (when inner and outer fluids have similar densities, the parameter $\rho$ must combine both~\cite{Leppinen2003}). The dimensions of these two mechanical factors suggest that $K_i\simeq (\Gamma/\rho)^\frac{1}{3}$~\cite{Thomson1871,Keller1983,Fardin2024}. A more precise equation may introduce a numerical correction $\delta_{\Gamma\rho}$, such that $K_i=\delta_{\Gamma\rho} (\Gamma/\rho)^\frac{1}{3}$. As shown in Fig.~\ref{Sfig2}a the value of $\delta_{\Gamma\rho}$ for the pinching, spreading and coalescence experiments in Table~\ref{table2b} always remains close to 1 (For additional details see~\citet{Fardin2022}). 

Beyond the $\frac{2}{3}$ regime, the growth of the variable size $d$ (neck radius for coalescence and pinching, or contact radius for spreading) is typically bounded by the droplet or bubble size, such that $d\simeq D$. This limit is usually not written in mechanical terms (we show how it can be written mechanically in an upcoming publication; such mechanical decomposition is not required for the purpose of the present article). The actual final radius may include a numerical correction, $d= \delta D$. The values of $\delta$ obtained for the data sets including the late regime are given in Fig.~\ref{Sfig2}b, they are always close to 1. More precisely, the correction $\delta$ may slightly depend on the geometry. For instance, for the coalescence of two spherical droplets of initial radius $D$, the final droplet after merger will have a radius around $2^\frac{1}{3} D\simeq 1.26D$ (by volume conservation). For spreading droplets in total wetting conditions the contact radius does not actually stop but continues to grow at a very slow rate, following `Tanner's law'~\cite{Tanner1979,Leger1992}: 
\begin{equation}
d \simeq \Big(\frac{\Gamma}{\eta}\Big)^\frac{1}{10} D^\frac{9}{10} t^\frac{1}{10}
\label{Tanner}
\end{equation}
\noindent where $\eta$ is the viscosity of the fluid. When it exists this regime is represented by the dotted-dashed lines in the plots in section~\ref{datadrop}. We ignore this regime in the main text, and consider $\delta=1$ for these data sets. We analyzed Tanner's regime in more detail in a recent publication~\cite{Fardin2022}. 

Finally, the regime at constant speed, $d\simeq c_v t$, is often called `visco-capillary'~\cite{Fardin2022}, it is impelled by surface-tension and impeded by the dominant viscosity $\eta$ (inner fluid for droplets, outer fluid for bubbles), such that $c_v\simeq \Gamma/\eta$~\cite{McKinley2000}. Again, a numerical correction $\delta_{\Gamma\eta}$ may eventually be necessary, $c_v=\delta_{\Gamma\eta} \Gamma/\eta$. The values of the correction $\delta_{\Gamma\eta}$ for the data sets capturing this regime are plotted in Fig.~\ref{Sfig2}c. The values are close to 1, attesting the validity of the mechanical model.  

Except for the asymptotic regime $d\simeq D$, which is expressed purely geometrically, the other two regimes described in the article can be expressed from three mechanical parameters, the surface-tension $\Gamma$, the density $\rho$, and the viscosity $\eta$:
\begin{align}
\text{Regime 1:}\quad& c_v \simeq \frac{\Gamma}{\eta} \label{cvmech}\\
\text{Regime 2:}\quad& K_i \simeq \Big(\frac{\Gamma}{\rho}\Big)^\frac{1}{3} \label{Kimech}
\end{align}

The values of the geometric parameter $D$, and the three mechanical parameters for each experiment used in the article are given in Table~\ref{table2b}. The values of $c_v$ or $K_i$ appearing between brackets are estimated using the mechanical models, assuming $\delta_{\Gamma\eta}=1$ and $\delta_{\Gamma\rho}=1$.  

The coordinates of the crossovers can also be expressed mechanically. Using Eq.~\ref{t1d1}, \ref{cvmech} and \ref{Kimech}, the coordinates of the early crossover can be written in the following way: 
\begin{align}
t_1 &= \Big(\frac{\delta_{\Gamma\rho}}{\delta_{\Gamma\eta}}\Big)^3 \frac{\eta^3}{\Gamma^2\rho}\simeq \frac{\eta^3}{\Gamma^2\rho}\\
d_1 &=\frac{\delta_{\Gamma\rho}^3}{\delta_{\Gamma\eta}^2}  \frac{\eta^2}{\Gamma\rho}\simeq \frac{\eta^2}{\Gamma\rho}
\end{align}
In the right-hand side, the formulas neglect the numerical corrections $\delta_{\Gamma\eta}$ and $\delta_{\Gamma\rho}$. 

Using Eq.~\ref{t2d2} and \ref{cvmech}, the coordinates of the visco-capillary crossover can be written in the following way: 
\begin{align}
t_2 &= \frac{\delta}{\delta_{\Gamma\eta}} \frac{\eta D}{\Gamma} \simeq  \frac{\eta D}{\Gamma}  \\
d_2 &= \delta D \simeq D
\end{align}
In the right-hand side, the formulas neglect the numerical corrections $\delta_{\Gamma\eta}$ and $\delta$. 

Using Eq.~\ref{t3d3} and \ref{Kimech}, the coordinates of the inertio-capillary crossover can be written in the following way: 
\begin{align}
t_3 &= \Big(\frac{\delta}{\delta_{\Gamma\rho}}\Big)^\frac{3}{2} \Big(\frac{\rho D^3}{\Gamma}\Big)^\frac{1}{2} \simeq   \Big(\frac{\rho D^3}{\Gamma}\Big)^\frac{1}{2}  \\
d_3 &= \delta D \simeq D
\end{align}
In the right-hand side, the formulas neglect the numerical corrections $\delta_{\Gamma\rho}$ and $\delta$. 

\section{Simple dimensionless numbers}
In this section, we provide additional details on `simple dimensionless numbers', which provide a privileged perspective on a given power law. The concept was illustrated in the case of Taylor's $\frac{2}{5}$ regime, in Fig.~2 of the main text. We present here the form these numbers take for the three regimes of explosions and the three regimes of capillary dynamics discussed in the article.  

\subsection{Explosions\label{Nexplo}}
\subsubsection{The Taylor-Sedov number}
\begin{figure*}
\centering
\includegraphics[width=17cm,clip]{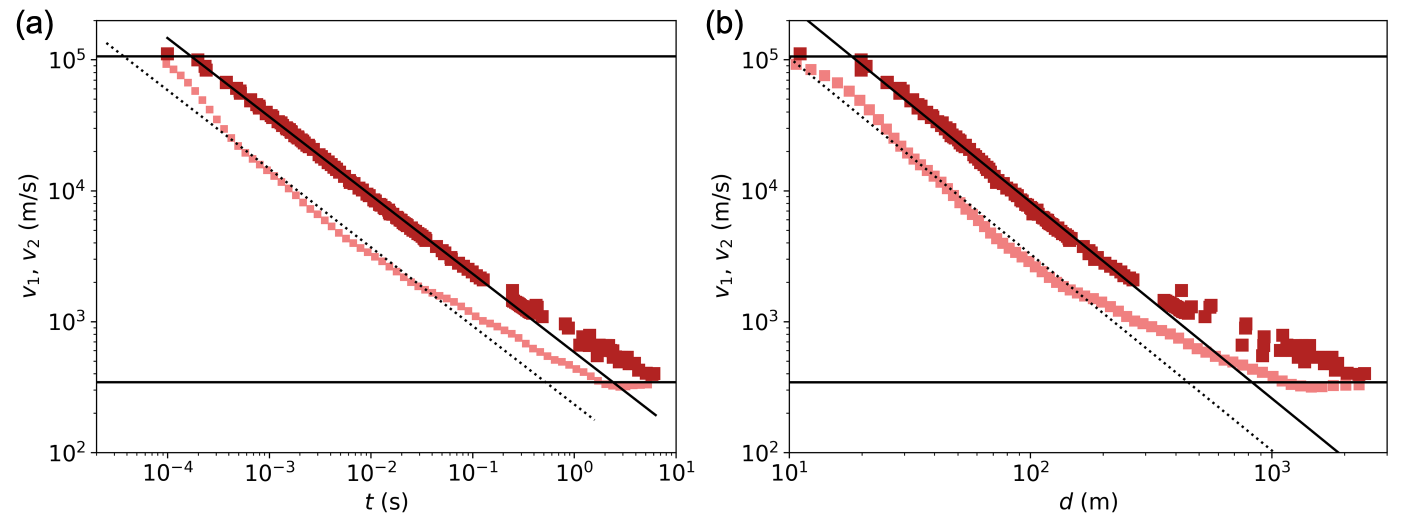}
\caption{Mean ($v_1$, dark red) and instantaneous speeds ($v_2$, light red) of the front of the Trinity explosion plotted against the time since detonation, $t$ (a), or the distance to ground zero, $d$ (b). To limit noise in the computation of the instantaneous speed, the initial data $d(t)$ were smoothed using a Savgol filter. In both plots the two horizontal lines are the initial and sound speeds ($c_0$ and $c_s$). In (a) the continuous diagonal line is $v_1 = K t^{-\frac{3}{5}}$, the dotted line is $v_2=(2/5) v_1$. In (b) the continuous diagonal line is $v_1 = K^\frac{5}{2} d^{-\frac{3}{2}}$, the dotted line is $v_2=(2/5) v_1$. (For additional details see: \href{https://youtu.be/bvVCvdB5Uzk?si=T7pXAZC4vu0DkvKY}{Explosions - Lecture 8}.)
\label{Sfig3}}
\end{figure*} 
Taylor's regime is traditionally expressed as a relationship between the radius of the explosion, $d$, and the time since detonation, $t$, as $d\simeq Kt^\frac{2}{5}$~\cite{Taylor1950a,Taylor1950b}. Nevertheless, other perspectives are possible to describe the same motions. For instance, as mentioned in the article, one may wish to represent the speed of the moving front over time, or at various distances from ground zero. One may consider two kinds of speeds: the integral (or `mean') speed $v_1\equiv d/t$, or the instantaneous speed $v_2\equiv \partial d/\partial t$. In the time range abiding to Taylor's scaling, these two speeds are related by a constant numerical factor: 
\begin{equation}
v_2\equiv \frac{\partial d}{\partial t} \simeq \frac{\partial (Kt^\frac{2}{5})}{\partial t} \simeq \frac{2}{5} \frac{d}{t} \simeq \frac{2}{5} v_1
\end{equation} 
Both speed are power laws of time with the same exponent, since $d/t \simeq K t^{-\frac{3}{5}}$. A comparison of the measurements of these two speeds in the case of the Trinity test are given in Fig.~\ref{Sfig3}a. 

The perspective is a mater of choice. One may prefer to represent a radius over time, $d(t)$, or a speed over time, $v_1(t)$ or $v_2(t)$, one may also prefer to represent the speed at various distances from ground zero $r$, when the explosion front reaches these distances, and so when $r=d$. Such choice may be motivated by the situation, for instance if speed detectors have been placed at various distances (this was not the case for the Trinity test~\cite{Bainbridge1976}). In the time range abiding to Taylor's scaling the power laws $v_1(d)$ and $v_2(d)$ can be obtained from $v_1(t)$ and $v_2(t)$ by using $t\simeq (d/K)^\frac{5}{2}$: 
\begin{equation}
v_2 \simeq \frac{2}{5} v_1 \simeq  \frac{2}{5} K^\frac{5}{2} d^{-\frac{3}{2}}
\end{equation} 
A comparison of the measurements of these two speeds in the case of the Trinity test are given in Fig.~\ref{Sfig3}b. 

Whether we choose to represent a distance versus a time ($d(t)$), a speed versus a time ($v(t)$), or a speed versus a distance ($v(d)$) the motions abiding to Taylor's regime are represented in log scale by a diagonal line, with a slope depending on the choice of perspective, i.e. on the choice of variables. However, for every regime there exist a special perspective, a special variable, which when plotted against any other variable will produce a constant plateau. We illustrated this for the case of Taylor's regime in Fig.~2c of the main text. In that case, since $d\simeq Kt^\frac{2}{5}$, we are free to define a new variable $\tilde{X}\equiv d^5/t^2$, such that $\tilde{X}\simeq X$, where $X\equiv K^5$ is a constant for all data points abiding to Taylor's regime. Note that any power of such variable will necessarily be constant as well, in particular $d^/t^\frac{2}{5}\simeq K$. The choice of overall exponent is usually such that the units of the new variable are the smallest possible integers~\cite{Fardin2024}. So $d^5/t^2$ is favored over $d^/t^\frac{2}{5}$, since the latter has a fractional time dimension. 

As seen in the main text, once a new `constant' variable has been defined, it can be `scaled' by the associated parameter. In the case of Taylor's regime, one may then define a simple dimensionless number $N_1\equiv \tilde{X}/X \equiv d^5/(t^2 K^5)$, which is equal to unity, $N_1\simeq 1$, for all data points abiding to Taylor's regime. As stressed in Fig.~2d of the main text, such dimensionless number can also be interpreted as giving the value of the variable $\tilde{X}$ \textit{in units of} $X\equiv K^5$. There is always a way to represent any apparently scale-free power law such that a single axis acquires a `natural' or `objective' unit (these terms do not have a broadly agreed meaning, both are used depending on context). 

The dimensionless number $N_1\equiv \tilde{X}/X \equiv d^5/(t^2 K^5)$ may not immediately resemble more well-known dimensionless numbers such as the Reynolds number, $\text{Re}\equiv \rho v d/\eta$. Whereas $N_1$ is expressed purely kinematically ($d$, $t$ and $K$ only depend on the dimensions of space and time), $\text{Re}$ combines kinematic ($d$, $v$) and mechanical ($\rho$, $\eta$) terms. The similarity between $N_1$ and $\text{Re}$ can be revealed if the kinematic constant $K$ is expressed from its underlying mechanical factors (Eq.~\ref{Kmech}): 
\begin{equation}
N_1 \equiv  \frac{\tilde{X}}{X} \equiv \frac{d^5}{t^2 K^5} \simeq  \frac{\rho d^5}{t^2 E} \simeq  \frac{\rho v^2 d^3}{E}\label{N1eq}
\end{equation}
The numerical correction ($\delta_{E\rho}$) is neglected for clarity. In the last equation we use $v\simeq d/t$ to express the dimensionless number from a distance $d$ and a speed $v$, the perspective usually chosen in hydrodynamics and for the Reynolds number in particular. 

Surprisingly, to the best of our knowledge the dimensionless number $N_1$ does not have a standard name. Since the `Taylor number' refers to something else already~\cite{Fardin2014}, we have recently taken the liberty of naming it the \textit{Taylor-Sedov number}~\cite{Fardin2024}, from the name of a Soviet physicist whom also contributed substantially to the understanding of explosion blasts~\cite{Sedov1993,Deakin2011}. One may use `$\text{Se}$' as its symbol (to mimic the standard hydrodynamic nomenclature), although as advocated in our recent review `$N_{E\rho}$' may be a more judicious symbol~\cite{Fardin2024}.  

\subsubsection{The Gurney number}
The initial regime of explosions, $d\simeq c_0 t$, involves a speed $c_0$, which is a more traditional kinematic parameter, more easily understood that the prefactor $K$ of Taylor's regime. The constant variable associated with this regime is simply the speed $v\equiv d/t$. The corresponding simple dimensionless number is: 
\begin{equation}
N_0 \equiv  \frac{d}{t c_0} \simeq  \frac{m^\frac{1}{2} d}{t E^\frac{1}{2}} \simeq \Big( \frac{m v^2}{E} \Big)^\frac{1}{2}\label{N0eq}
\end{equation}
The last two equations use the mechanical factors of the initial speed (Eq.~\ref{comech}). 

To our knowledge the dimensionless number $N_0$ does not have a standard name. To avoid fractional exponent on the mechanical factors one may prefer to use $N_0^2 \simeq mv^2/E$~\cite{Fardin2024}, but this dimensionless number does not seem to have a name either. Since it is directly connected to the concept of kinetic energy ($E\simeq m v^2$, neglecting numerical factors), founding figures such as Leibniz or {\'E}milie du Ch{\^a}telet may be associated with it. If we restrict ourselves to the context of explosions, $N_0$ (or $N_0^2$) may be called the \textit{Gurney number}, an homage to Ronald W. Gurney, who pioneered research on the initial phase of explosions~\cite{Gurney1943}. We would advocate that $N_0^2$ be called $N_{Em}$~\cite{Fardin2024}. 

\subsubsection{The Mach number}
The last regime of explosion is simply that of sound propagation, for which the corresponding simple dimensionless number is the well-known Mach number: 
\begin{equation}
N_2 \equiv  \frac{d}{t c_s} \simeq  \frac{\rho^\frac{1}{2} d}{t \Sigma^\frac{1}{2}} \simeq \Big( \frac{\rho v^2}{\Sigma} \Big)^\frac{1}{2}\label{N2eq}
\end{equation}
The traditional symbol used for $N_2$ is $\text{Ma}$, although we would advocate that $N_2^2$ be called $N_{\Sigma\rho}$~\cite{Fardin2024}. 

\subsection{Droplets and bubbles\label{Ndrop}}
\subsubsection{The Weber number}
For the regime $d\simeq K_i t^\frac{2}{3}$ one may use $d^3/t^2$ as constant variable, and the associated dimensionless number is the Weber number:  
\begin{equation}
N_{{\scaleto{\Gamma\rho}{4pt}}} \equiv  \frac{d^3}{t^2 K_i^3} \simeq  \frac{\rho d^3}{t^2 \Gamma} \simeq \frac{\rho v^2 d}{\Gamma}\label{N1beq}
\end{equation}
The traditional symbol used for the Weber number is $\text{We}$, although we advocate for the more explicit $N_{{\scaleto{\Gamma\rho}{4pt}}}$~\cite{Fardin2024}. 

\subsubsection{The Capillary number}
For the regime $d\simeq c_v t$ the constant variable is simply $d/t$, and the associated dimensionless number is the Capillary number:  
\begin{equation}
N_{{\scaleto{\Gamma\eta}{4pt}}} \equiv  \frac{d}{t c_v} \simeq  \frac{\eta d}{t \Gamma} \simeq \frac{\eta v}{\Gamma}\label{N0beq}
\end{equation}
The traditional symbol used for the Capillary number is $\text{Ca}$, although we advocate for the more explicit $N_{{\scaleto{\Gamma\eta}{4pt}}}$~\cite{Fardin2024}. 

\subsubsection{Third number}
Since the last regime is simply $d\simeq D$, the constant variable is simply $d$ and the simple dimensionless number is just $N\equiv d/D$. This number could be written in mechanical terms if the constant $D$ was written mechanically, but such decomposition goes beyond the scope of this article. (For additional details see: \href{https://youtu.be/wlckZEXOKJo?si=K6O5QtDV5MkujcYr}{Mechanics - Lecture 10}.) 

\section{Representations in terms of simple dimensionless numbers}
In the main text we favored a representation of the data using the primitive variables, $d$ and $t$, which we subsequently scaled with objective units. For instance, for explosions we represented $d/d_*$ vs $t/t_*$ in Fig.~3b of the main text, and $d/d_0$ vs $t/t_0$ in Fig.~3c. In this section we show how to connect these scaled variables to the simple dimensionless numbers associated to the different regimes.  

\subsection{Explosions\label{Nplotsexplo}}
\begin{figure*}
\centering
\includegraphics[width=17cm,clip]{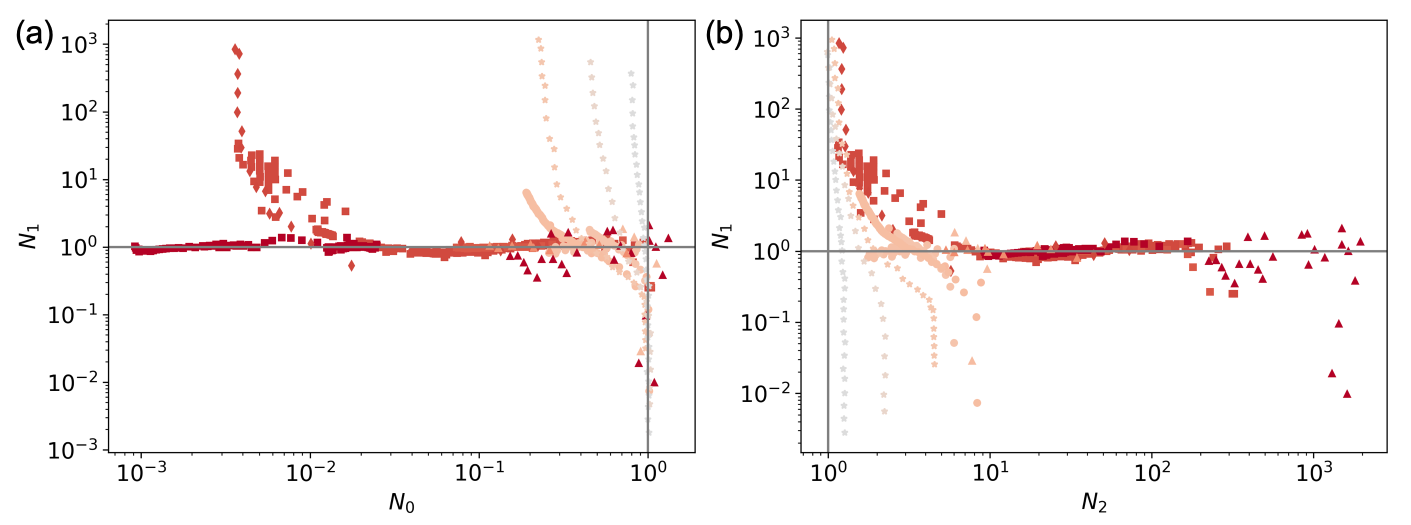}
\caption{The data sets on explosions used in the article are represented using the simple dimensionless numbers associated to the three regimes, $N_1$ vs $N_0$ (a), or $N_1$ vs $N_2$ (b). Note that in both plots time runs from the bottom right to the top left. $N_0\simeq 1$, $N_1\simeq 1$ and $N_2\simeq 1$ are respectively equivalent to $d\simeq c_0 t$, $d\simeq Kt^\frac{2}{5}$, and $d\simeq c_s t$.
\label{Sfig4}}
\end{figure*} 
Recalling Eq.~\ref{N1eq},~\ref{N0eq} and~\ref{N2eq} the initial and late units respectively defined in Eq.~\ref{todo} and~\ref{txdx} can be expressed from the simple dimensionless numbers: 
\begin{align}
\frac{t}{t_0} &= \simeq \Big(\frac{t^3 c_0^5}{K^5}\Big)^\frac{1}{3} \simeq \Big(\frac{d^5}{t^2 K^5} \frac{t^5 c_0^5}{d^5}\Big)^\frac{1}{3} \simeq \Big(\frac{N_1}{N_0^5}\Big)^\frac{1}{3}  \\
\frac{d}{d_0}  &\simeq \Big(\frac{d^3 c_0^2}{K^5}\Big)^\frac{1}{3} \simeq \Big(\frac{d^5}{t^2 K^5} \frac{t^2 c_0^2}{d^2}\Big)^\frac{1}{3} \simeq \Big(\frac{N_1}{N_0^2}\Big)^\frac{1}{3} \\
\frac{t}{t_*} &= \simeq \Big(\frac{t^3 c_s^5}{K^5}\Big)^\frac{1}{3} \simeq \Big(\frac{d^5}{t^2 K^5} \frac{t^5 c_s^5}{d^5}\Big)^\frac{1}{3} \simeq \Big(\frac{N_1}{N_2^5}\Big)^\frac{1}{3}  \\
\frac{d}{d_*}  &\simeq \Big(\frac{d^3 c_s^2}{K^5}\Big)^\frac{1}{3} \simeq \Big(\frac{d^5}{t^2 K^5} \frac{t^2 c_s^2}{d^2}\Big)^\frac{1}{3} \simeq \Big(\frac{N_1}{N_2^2}\Big)^\frac{1}{3} 
\end{align}

The representations using $d/d_0$ and $t/t_0$ (main text, Fig.~3c), or $d/d_*$ and $t/t_*$ (main text, Fig.~3b) have the advantage of being faithful to the initial variables $d$ and $t$, which facilitate their interpretation. However, one may instead choose to use the simple dimensionless numbers as scaled variables and represent the dynamics using $N_0$, $N_1$ and $N_2$ as axes. The representation using $N_0$ and $N_2$ is trivial, since $N_2\equiv N_0 \mathcal{N}$, where $\mathcal{N}\equiv c_0/c_s$ is the base. The two other representations are given in Fig.~\ref{Sfig4}. 

\subsection{Droplets and bubbles\label{Nplotsdrop}}
\begin{figure*}
\centering
\includegraphics[width=17cm,clip]{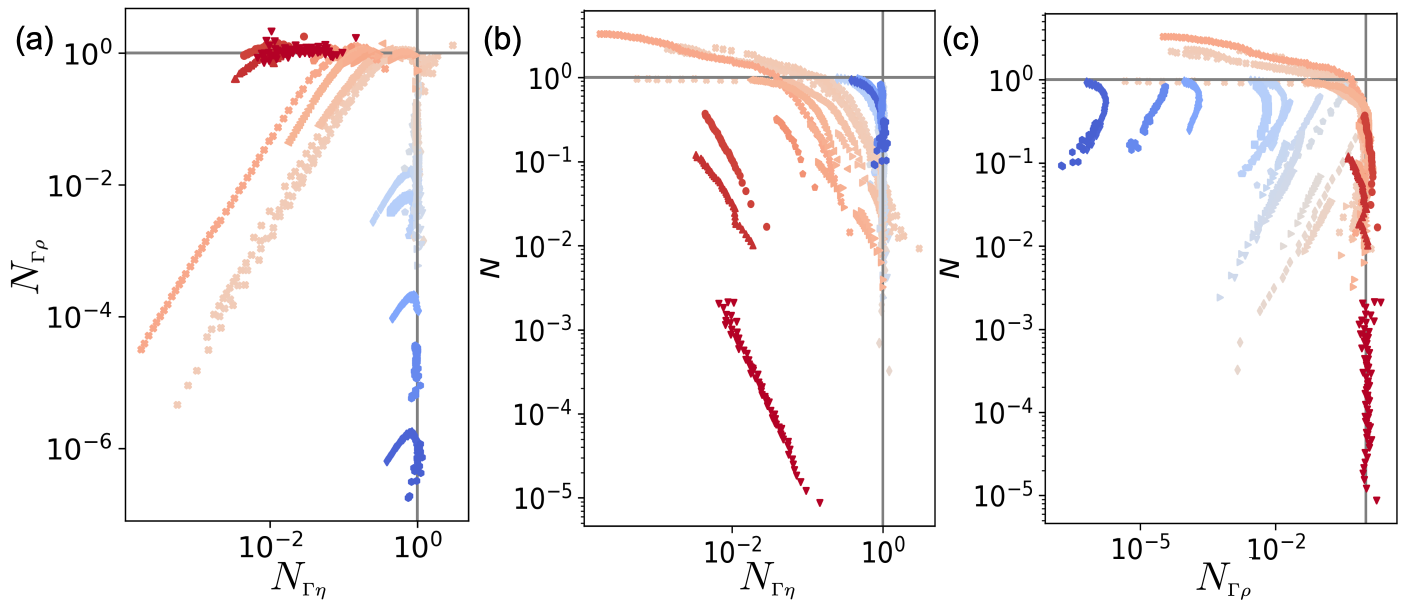}
\caption{The data sets on spreading, pinching and coalescence of droplets and bubbles used in the article are represented using the simple dimensionless numbers associated to the three regimes, $N_{{\scaleto{\Gamma\rho}{4pt}}}$ vs $N_{{\scaleto{\Gamma\eta}{4pt}}}$ (a), or $N$ vs $N_{{\scaleto{\Gamma\eta}{4pt}}}$ (b), or $N$ vs $N_{{\scaleto{\Gamma\rho}{4pt}}}$ (c). Note that in the three plots time runs from the bottom right to the top left. $N_{{\scaleto{\Gamma\eta}{4pt}}}\simeq 1$, $N_{{\scaleto{\Gamma\rho}{4pt}}}\simeq 1$ and $N\simeq 1$ are respectively equivalent to $d\simeq c_v t$, $d\simeq K_i t^\frac{2}{3}$, and $d\simeq D$. Note that the data in (b) and (c) with $N>1$ correspond to spreading experiments in total wetting conditions, where the contact radius continues to grow according to Tanner's law (Eq.~\ref{Tanner}). 
\label{Sfig5}}
\end{figure*} 
Recalling Eq.~\ref{N0beq} and~\ref{N1beq}, the three systems of units respectively defined in Eq.~\ref{t1d1}, \ref{t2d2} and~\ref{t3d3} can be expressed from the simple dimensionless numbers: 
\begin{align}
\frac{t}{t_1} &= \simeq  \frac{t c_v^3}{K_i^3}  \simeq \frac{t^3 c_v^3}{d^3} \frac{d^3}{t^2 K_i^3} \simeq \frac{N_{{\scaleto{\Gamma\rho}{4pt}}}}{N_{{\scaleto{\Gamma\eta}{4pt}}}^3}  \\
\frac{d}{d_1}  &\simeq \frac{d c_v^2}{K_i^3} \simeq \frac{d^3}{t^2 K_i^3} \frac{t^2 c_v^2}{d^2} \simeq \frac{N_{{\scaleto{\Gamma\rho}{4pt}}}}{N_{{\scaleto{\Gamma\eta}{4pt}}}^2} \\
\frac{t}{t_2} &= \simeq \frac{t c_v}{D}  \simeq \frac{d}{D} \frac{t c_v}{d} \simeq \frac{N}{N_{{\scaleto{\Gamma\eta}{4pt}}}}  \\
\frac{t}{t_3} &= \simeq \Big(\frac{t^2 K_i^3}{D^3}\Big)^\frac{1}{2}  \simeq \Big(\frac{t^2 K_i^3}{d^3} \frac{d^3}{D^3}\Big)^\frac{1}{2} \simeq \Big(\frac{N^3}{N_{{\scaleto{\Gamma\rho}{4pt}}}} \Big)^\frac{1}{2}
\end{align}
We recall that $N\equiv d/D$, where $D$ is the length scale associated to both $t_2$ and $t_3$. 

The representations using $d/d_1$ and $t/t_1$ (main text, Fig.~4b), $d/d_2$ and $t/t_2$ (main text, Fig.~4c), or $d/d_3$ and $t/t_3$ (main text, Fig.~4d) have the advantage of being faithful to the initial variables $d$ and $t$, which facilitate their interpretation. However, one may instead choose to use the simple dimensionless numbers as scaled variables and represent the dynamics using $N_{{\scaleto{\Gamma\eta}{4pt}}}$, $N_{{\scaleto{\Gamma\rho}{4pt}}}$ and $N$ as axes. Since $N=d/d_2=d/d_3$, the scaled distances in Fig.~4c and d are already a simple dimensionless number. The three possible plots using pairs of simple dimensionless numbers as axes are given in Fig.~\ref{Sfig5}.

\newpage

\begin{figure*}[h!]
\centering
\includegraphics[width=17cm,clip]{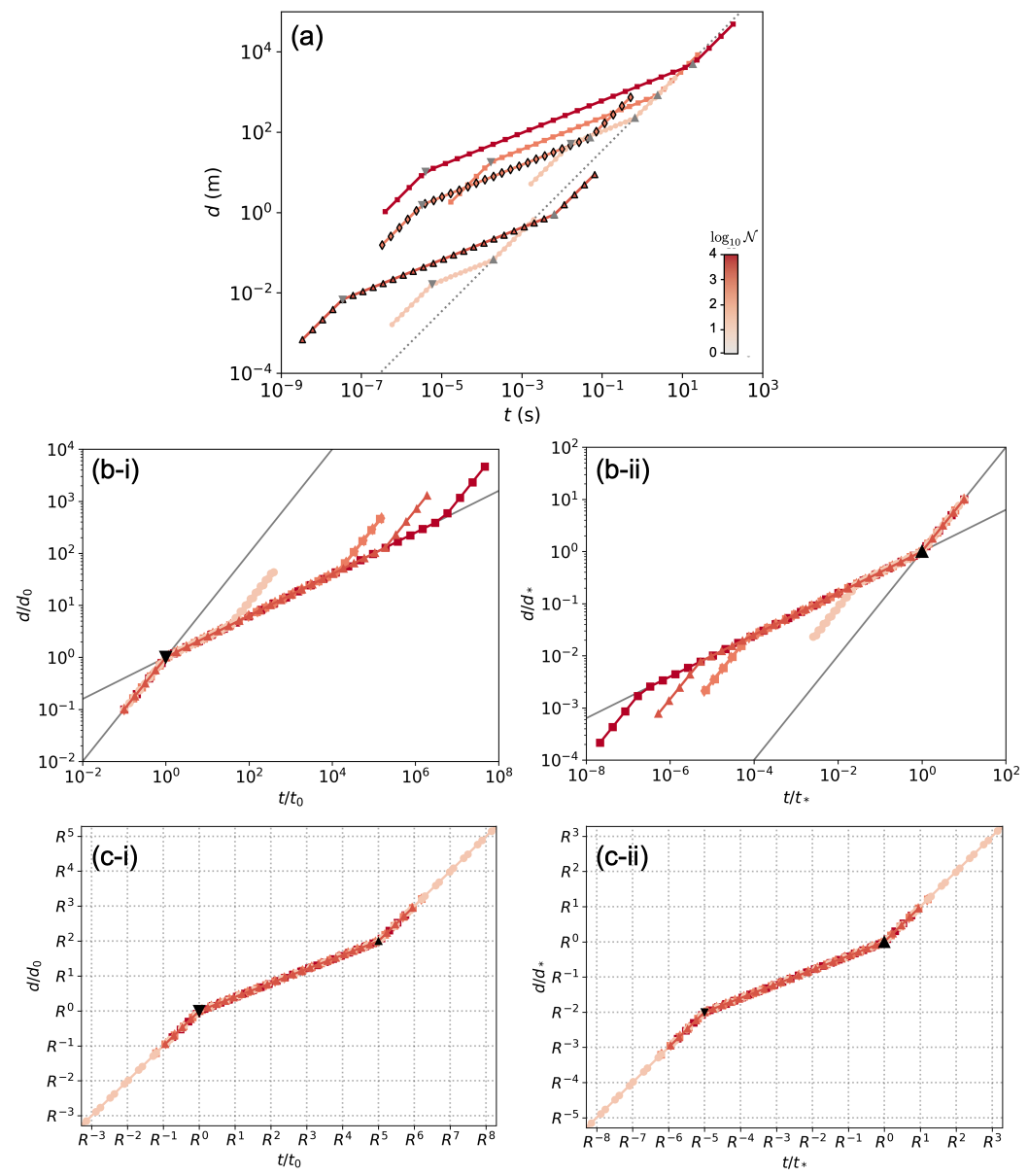}
\caption{Idealized blast radii of explosions, based on the values of $c_0$, $K$ and $c_s$ for a selection of the data sets (Trinity, Dominic Housatonic, Porzel 1957, Kingery 1962, Kleine 2010, Grun 1991). (a) The data are represented in standard units (second, meter). The dotted grey line is the speed of sound in the air at sea level, in normal atmospheric conditions. The diamonds highlighted in black correspond to the Wigwam underwater nuclear explosion (Porzel 1957). The triangles highlighted in black correspond to a laser-induced explosion in rarefied xenon (Grun 1991). The grey triangles correspond to the crossovers of the dynamics, i.e. to the points of coordinates ($t_0$, $d_0$; $\blacktriangledown$), and ($t_*$, $d_*$; $\blacktriangle$). (b) The data are represented with objective units based on ($t_0$, $d_0$) (b-i), or ($t_*$, $d_*$) (b-ii). In these representations, data sets with similar values of initial Mach number ($\mathcal{N}\equiv c_0/c_s$) fully overlap. Data sets corresponding to different initial Mach numbers only overlap for two out of the three regimes, those intersecting at the logarithmic origin of the plot, i.e. the point of coordinates (1,1). (c) The data are represented with objective units, ($t_0$, $d_0$) (c-i), or ($t_*$, $d_*$) (c-ii) , and an objective radix based on the initial Mach number, $R\equiv \mathcal{N}^\frac{1}{3}$. Note that any other power of $\mathcal{N}$ would produce the same overlap, the exponent $\frac{1}{3}$ is chosen such that the two crossovers occur at integer coordinates in $\log_R$. 
\label{Sfig6}}
\end{figure*}

\begin{figure*}[h!]
\centering
\includegraphics[width=17cm,clip]{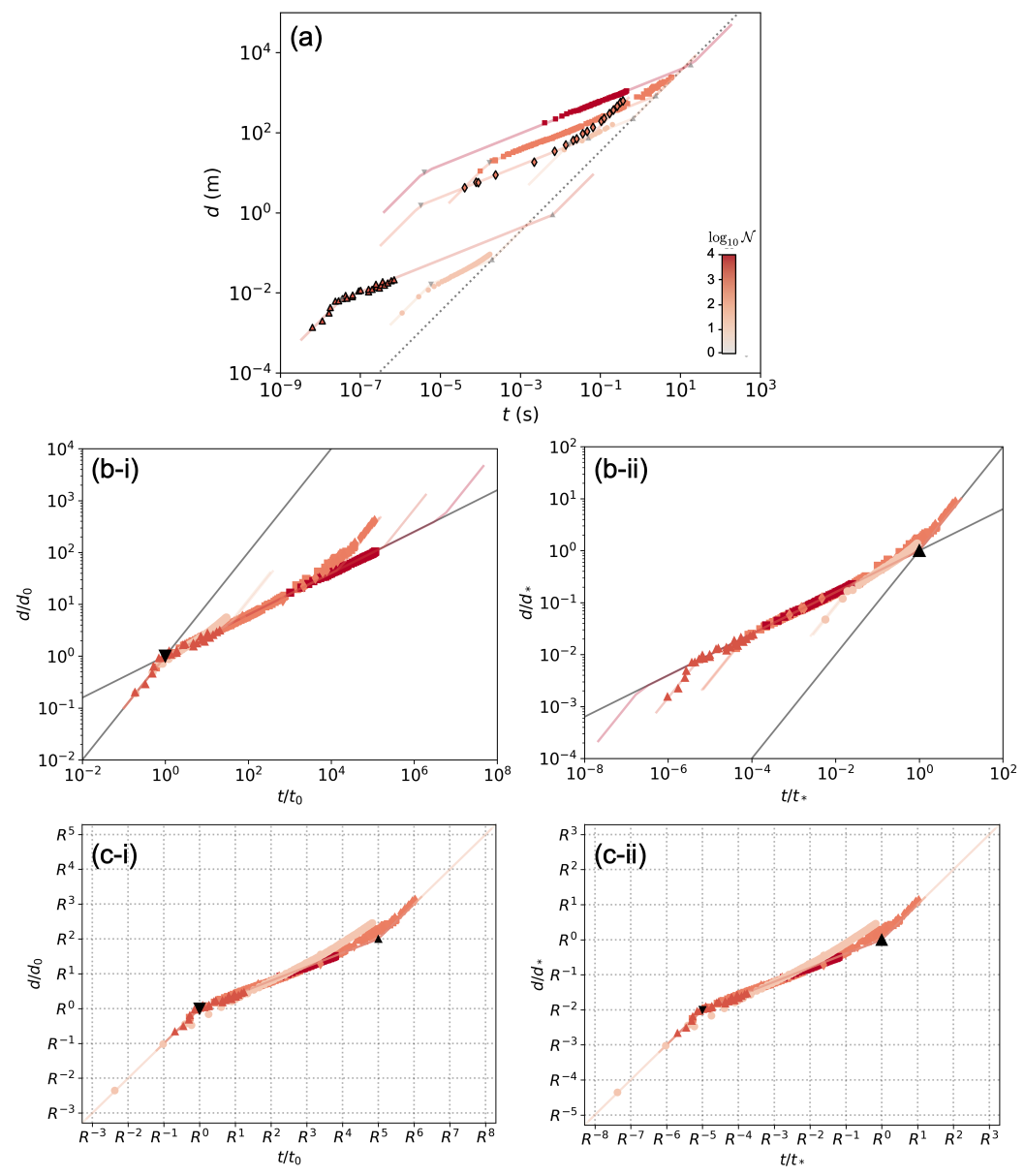}
\caption{Measured blast radii of explosions for a selection of the data sets (Trinity, Dominic Housatonic, Porzel 1957, Kingery 1962, Kleine 2010, Grun 1991). All sub-plots mirror those of Fig.~\ref{Sfig6} (see legend of this figure for details). 
\label{Sfig7}}
\end{figure*}

\begin{figure*}[h!]
\centering
\includegraphics[width=17cm,clip]{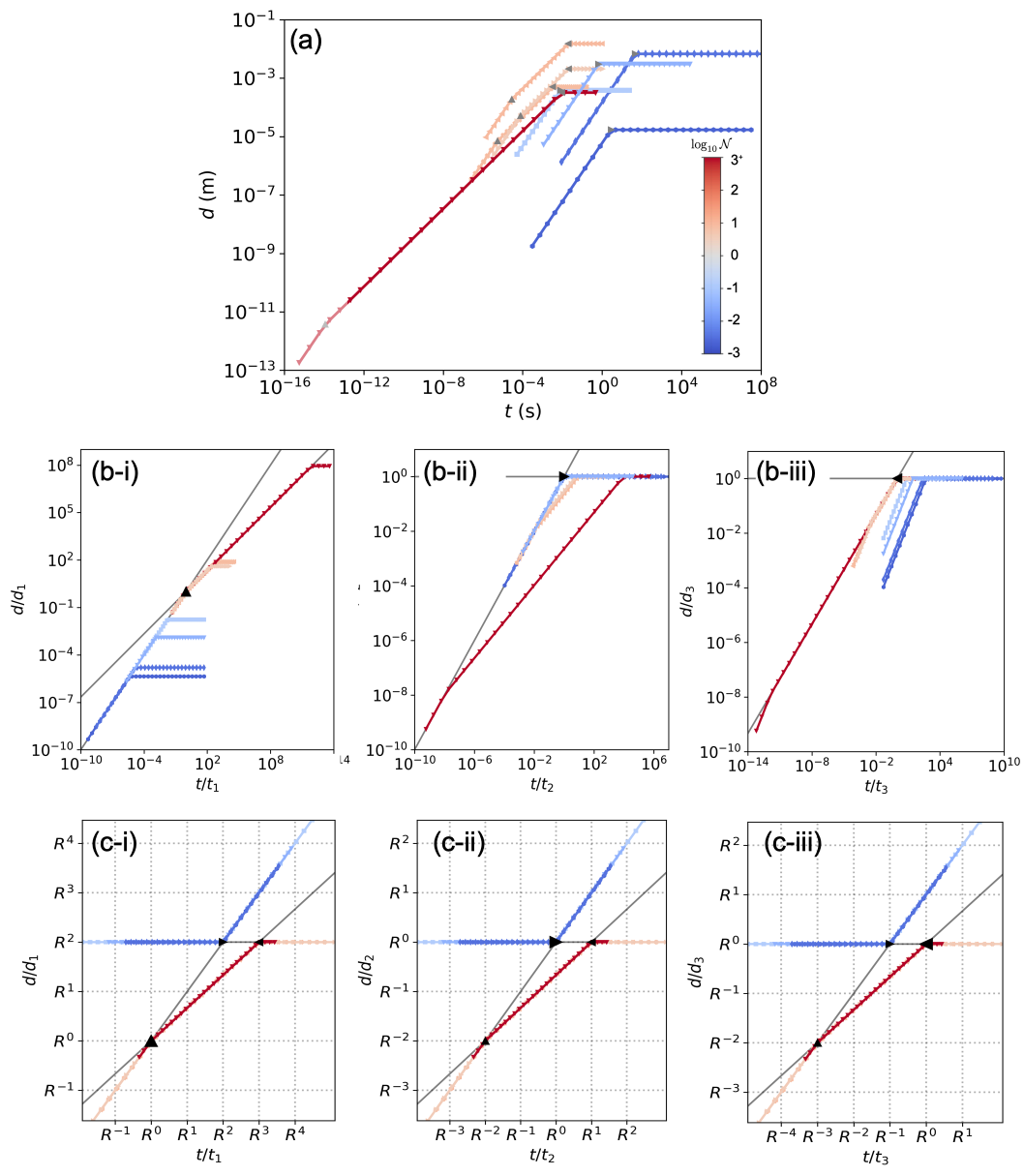}
\caption{Idealized radii for spreading, coalescence and pinching dynamics, based on the values of $c_v$, $K_i$ and $D$ for a selection of the data sets (Eddi2013 Fig5a 105deg, Yao2005 100000cS 0p5cm, Aarts2008 Fig9 bubb17, Rahman2019 Fig6 6p65, McKinley2000 Fig4, Burton2004 Fig5, Bolanos2009 Fig7 O9, Goldstein2010 Fig5). (a) The grey triangles correspond to the crossovers of the dynamics, i.e. to the points of coordinates ($t_1$, $d_1$; $\blacktriangle$), ($t_2$, $d_2$; \rotatebox[origin=c]{-90}{$\blacktriangle$}), and ($t_3$, $d_3$; \rotatebox[origin=c]{90}{$\blacktriangle$}). (b) The data are represented with objective units based on ($t_1$, $d_1$) (b-i), ($t_2$, $d_2$) (b-ii), or ($t_3$, $d_3$) (b-iii). In these representations, data sets with similar values of the inverse Ohnesorge number ($\mathcal{N}\equiv c_v (\delta D)^\frac{1}{2}/K_i^\frac{3}{2}$) fully overlap. Data sets corresponding to different values of $\mathcal{N}$ only overlap for two out of the three regimes, those intersecting at the logarithmic origin of the plot, i.e. the point of coordinates (1,1). (c) The data are represented with objective units, ($t_1$, $d_1$) (c-i), ($t_2$, $d_2$) (c-ii), or ($t_3$, $d_3$) (c-iii) , and an objective radix based on $R\equiv \mathcal{N}$. 
\label{Sfig8}}
\end{figure*}

\begin{figure*}[h!]
\centering
\includegraphics[width=17cm,clip]{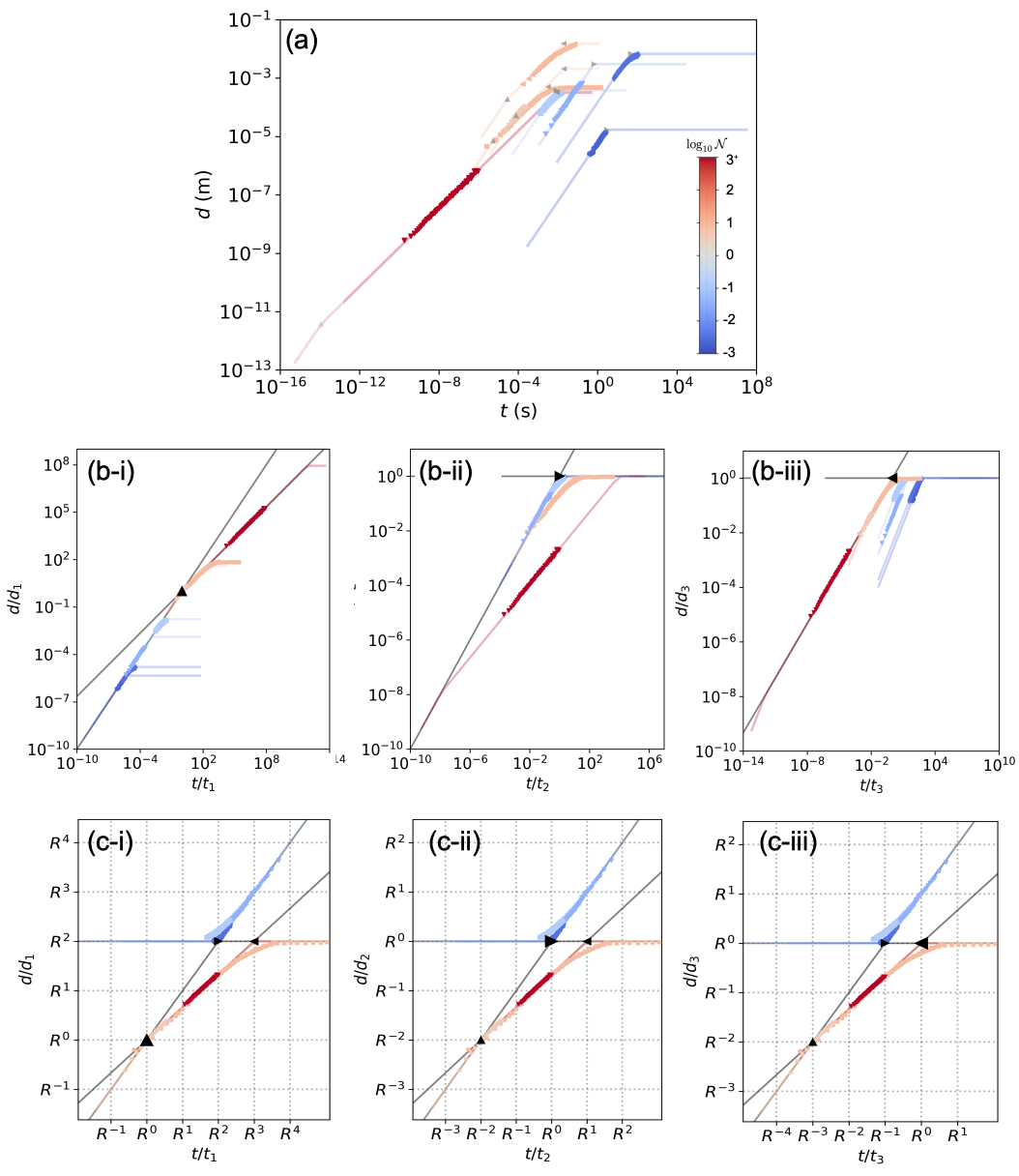}
\caption{Measured radii for a selection of spreading, coalescence and pinching dynamics (Eddi2013 Fig5a 105deg, Yao2005 100000cS 0p5cm, Aarts2008 Fig9 bubb17, Rahman2019 Fig6 6p65, McKinley2000 Fig4, Burton2004 Fig5, Bolanos2009 Fig7 O9, Goldstein2010 Fig5). All sub-plots mirror those of Fig.~\ref{Sfig8} (see legend of this figure for details). 
\label{Sfig9}}
\end{figure*}

\clearpage

\section{From subjective to objective representations}
In this section we summarize how to progressively overlap data sets by choosing objective units and an objective base. We use the two examples presented in the article: explosions, and capillary dynamics of spreading/coalescing/pinching droplets and bubbles. 

\subsection{Explosions\label{Nplotsexplo}}
In order to ovoid crowded figures we will only consider the following data sets: Trinity (i.e. Taylor 1950, Mack 1946 and Bainbridge 1976), Dominic Housatonic, Porzel 1957 (Wigwam test), Kingery 1962 (100 tons of TNT), Kleine 2010 (10 mg of silver azide), Grun 1991 (laser-induced). For these data sets, the kinematic parameters $K$, $c_0$, and $c_s$ can be obtained directly from the data, or they can be estimated from the mechanical parameters with good confidence. 

\subsubsection{Idealized trajectories}
Before we represent the actual data, we may consider idealized artificial data sets constructed in the following way: 
\begin{align}
\text{if } t<t_0~:~~d=c_0 t\\
\text{if } t_0<t<t_*~:~~d=K t^\frac{2}{5}\\
\text{if } t>t_*~:~~d=c_s t
\end{align}
The results of such piece-wise function for the selected data sets are shown in Fig.~\ref{Sfig6}a. Note that the final regime depends solely on the speed of sound in the medium surrounding the explosion, it does not depend on the energy nor the ejected mass. The dotted grey line in Fig.~\ref{Sfig6}a represents the speed of sound in the air at sea level with normal atmospheric conditions, $c_s\simeq 343$~m/s. Most of the explosions represented in the figure were conducted in the air and so they share this asymptotic behavior. The Dominic Housatonic test (upper curve) was conducted at an altitude of 3700~m, hence a slightly lower value for the speed of sound ($c_s\simeq 263$~m/s). Two data sets show more pronounced differences in final speed, they are highlighted in black. The diamonds correspond to the Wigwam nuclear test (Porzel 1957). This test was conducted underwater, where the speed of sound is significantly higher ($c_s\simeq 1450$~m/s). Conversely the highlighted diamonds show an asymptotic speed lower than the speed of sound in the air. The data correspond to a laser-induced explosion in rarefied xenon (Grun 1991). The speed of sound in such medium was significantly lower than in the air ($c_s\simeq 136$~m/s). 

In Fig.~\ref{Sfig6}a, for each curve the locations of the two crossovers, ($t_0$,$d_0$) and ($t_*$,$d_*$), are respectively marked by downward ($\blacktriangledown$) and upward ($\blacktriangle$) pointing triangles. In standard (but subjective) units these turning points of the dynamics seem to occur at different locations. However, if the units are based on these special coordinates, the triangles and the associated regimes can be overlapped. For instance, in Fig.~\ref{Sfig6}b-i, all early crossovers ($t_0$,$d_0$;$\blacktriangledown$) are overlapped by plotting $d/d_0$ vs $t/t_0$. In Fig.~\ref{Sfig6}b-ii, all late crossovers ($t_*$,$d_*$;$\blacktriangle$) are overlapped by plotting $d/d_*$ vs $t/t_*$.

Once the units have been objectively selected, as in Fig.~\ref{Sfig6}b-i or Fig.~\ref{Sfig6}b-ii, data sets with the same value of initial Mach number $\mathcal{N}\equiv c_0/c_s$ fully overlap. For instance although the explosions of 100~tons of TNT (Kingery 1962) and 10~mg of silver azide (Kleine 2010) appear quite different in subjective units, the data overlap in Fig.~\ref{Sfig6}b-i or Fig.~\ref{Sfig6}b-ii, since the values of the initial Mach number are very similar ($\mathcal{N}\simeq 9\simeq 8$). 

In Fig.~\ref{Sfig6}b-i or Fig.~\ref{Sfig6}b-ii, data with the same value of $\mathcal{N}$ overlap, whereas data sets with different values of $\mathcal{N}$ only partially overlap. By construction, in Fig.~\ref{Sfig6}b-i,  data points following the two regimes intersecting at the point of coordinates (1,1) overlap. So all points following $d\simeq c_0 t$ and $d\simeq K t^\frac{2}{5}$ overlap. In Fig.~\ref{Sfig6}b-ii all points following $d\simeq K t^\frac{2}{5}$ and  $d\simeq c_s t$ overlap. No matter which system of objective units we choose, there is always one out of the three regimes that does not overlap. As we saw in the article a complete overlap can be achieved by rescaling the number base as well as the units. 

In a logarithmic plot, translations of the curves is achieved by multiplying or dividing the axes. Such translations are achieved by the rescaling of units going from Fig.~\ref{Sfig6}a to Fig.~\ref{Sfig6}b-i and ii (c.f. section~\ref{gifs} for animated versions). Such translations only results in a partial overlap of the data. In Fig.~\ref{Sfig6}b-i and ii, the curves with different colors, i.e. with different values of $\mathcal{N}$ do not have he same `size', i.e. the number of decades separating the two turning points shrinks as the value of  $\mathcal{N}$ decreases towards 1. Rescaling the base for each data sets corresponds to shrinking and dilating the different curves such that they completely overlap. Such complete overlap is reached in Fig.~\ref{Sfig6}c-i and ii. The only difference between the two sub-panels is the choice of origin. Once the units and the base are defined objectively, all curves such that $\mathcal{N}>1$ overlap. We will discuss the case $\mathcal{N}<1$ in the next section. 

\subsubsection{Actual data}
Fig.~\ref{Sfig7} gives the same representations as in Fig.~\ref{Sfig6}, this time with the actual rather than idealized data sets of the six selected explosions. Comparing Fig.~\ref{Sfig6} and Fig.~\ref{Sfig7}, a few differences can be noticed: 
\begin{itemize}
\item \textbf{Range of the data}: The main difference between the idealized and actual data is also the most trivial. The idealized data were constructed so as to extend from $t=0.1 t_0$ to $t=10 t_*$, in order to systematically capture the three regimes of the explosions. In contrast, the actual data were restricted by the experimental limitations in each situation. For instance the experiment on the laser-induced explosion (Grun 1991, triangles highlighted in black) could not captured the late dynamics (the chamber in which the explosion was produced was not large enough). Another example is the Wigwam underwater nuclear test, where in this case the initial regime could not be captured, the time and spacial resolution was not good enough.  
\item \textbf{Sampling of the data}: The second difference between the idealized and actual data is also quite trivial. The idealized data are sampled logarithmically in a uniform way. There are 30 points on each curves spaced evenly over the range of the data. In contrast, the actual data can be more unevenly sampled. In particular, if data are acquired at a linear rate their sampling will be poor at short times once represented on a logarithmic plot. 
\item \textbf{Noise}: Whereas the idealized data perfectly follow the piece-wise function, the actual data may deviate from it. These deviations are of two types: systematic or random. We will discuss the systematic deviations next, they occur near the crossovers. The random deviations occur away from the crossovers, when the idealized data perfectly follow a power law. The actual data usually display some amount of noise. For instance, in Fig.~2d of the main text, the laser-induced explosion by~\citet{Grun1991} (dark red triangles) do not exactly lineup on Taylor's regime (the horizontal line of ordinate 1). We chose not to represent the errorbars associated with experimental measurements on the plots to preserve their readability, but the variations of the data away from the idealized values are usually within experimental uncertainty. 
\item \textbf{Systematic deviations at the crossovers}: We believe that this last difference is the most significant. For instance, for the Trinity explosion, in Fig.~3a of the main text, it is quite clear that the radius of the explosion starts to depart from Taylor's regime before $t=t_*$. The data points do not actually pass by the black dot of coordinates $(t_*,d_*)$, instead the trajectory of the explosion front smoothly transitions from Taylor's regime to the regime at constant sound speed $c_s$. As we shall see in section~\ref{crossdev}, our radical scaling approach offers a new outlook on this well-known effect~\cite{Dewey1964}. 
\end{itemize} 

\subsection{Droplets and bubbles\label{Nplotsdrop}}
In the case of explosions we only considered detonations, such that $c_0>c_s$, i.e. $\mathcal{N}>1$. We could not find reliable data on deflagrations ($\mathcal{N}<1$) that were not influenced by perturbing effects like friction, gravity, etc. (beyond the mechanical quantities $m$, $E$, $\Sigma$ and $\rho$, which characterize the detonations we analyzed). To discuss the qualitative difference between a radix larger or smaller than one, we used the example of spreading, coalescing and pinching droplets and bubbles, for which examples abound for both $\mathcal{N}<1$ and $\mathcal{N}>1$, where now $\mathcal{N}\equiv c_v D^\frac{1}{2}/K_i^\frac{3}{2}$ (the inverse Ohnesorge number). In order to ovoid crowded figures we will only consider the following data sets: Eddi2013 Fig5a 105deg, Yao2005 100000cS 0p5cm, Aarts2008 Fig9 bubb17, Rahman2019 Fig6 6p65, McKinley2000 Fig4, Burton2004 Fig5, Bolanos2009 Fig7 O9, Goldstein2010 Fig5. For these data sets, the kinematic parameters $K_i$, $c_v$, and $D$ can be obtained directly from the data, or they can be estimated from the mechanical parameters with good confidence. 

\subsubsection{Idealized trajectories}
Before we represent the actual data, we again consider idealized artificial data sets. Now that we are dealing with both $\mathcal{N}<1$ and $\mathcal{N}>1$, these two cases must be reflected in the definitions of the piece-wise functions. 

If $\mathcal{N}>1$, then $t_1<t_2<t_3$~\cite{Fardin2022}. In that case the piece-wise function is as follows:  
\begin{align}
\text{if } t<t_1~:~~d=c_v t\\
\text{if } t_1<t<t_3~:~~d=K_i t^\frac{2}{3}\\
\text{if } t>t_3~:~~d=D
\end{align}

If $\mathcal{N}<1$, then $t_3<t_2<t_1$~\cite{Fardin2022}. In that case the piece-wise function is as follows:  
\begin{align}
\text{if } t<t_2~:~~d=c_v t\\
\text{if } t>t_2~:~~d=D
\end{align}

The results of such piece-wise function for the selected data sets are shown in Fig.~\ref{Sfig8}a. For each curve the locations of the two crossovers, ($t_1$,$d_1$) and ($t_2$,$d_2$), and ($t_3$,$d_3$), are respectively marked by triangles pointing upward ($\blacktriangle$), right (\rotatebox[origin=c]{-90}{$\blacktriangle$}), and left (\rotatebox[origin=c]{90}{$\blacktriangle$}). Note that the curve corresponding to the data set `Burton2004 Fig5' (dark red) needs to be continued down to very small scales in order to reach the crossover with the viscous regime, due to the high surface tension and density of the fluid used in these experiments (mercury). This leads to a predicted crossover in the atomic range, which is probably unrealistic, since other mechanisms beyond viscosity may take over the dynamics in this arena. We ignore these effects for didactic purposes. 

In standard (but subjective) units the turning points of the dynamics seem to occur at different locations. However, if the units are based on these special coordinates, the triangles and the associated regimes can be overlapped. For instance, in Fig.~\ref{Sfig8}b-i, all early crossovers ($t_1$,$d_1$;$\blacktriangle$) are overlapped by plotting $d/d_1$ vs $t/t_1$. In Fig.~\ref{Sfig8}b-ii, all visco-capillary crossovers ($t_2$,$d_2$; \rotatebox[origin=c]{-90}{$\blacktriangle$}) are overlapped by plotting $d/d_2$ vs $t/t_2$. In Fig.~\ref{Sfig8}b-iii, all inertio-capillary crossovers ($t_3$,$d_3$; \rotatebox[origin=c]{90}{$\blacktriangle$}) are overlapped by plotting $d/d_3$ vs $t/t_3$. (c.f. section~\ref{gifs} for animated versions.)

As was the case for explosions, once the units have been objectively selected, data sets with the same value of $\mathcal{N}\equiv c_v (\delta D)^\frac{1}{2}/K_i^\frac{3}{2}$ fully overlap (Note that the inclusion of the numerical correction $\delta$ is necessary for a full overlap). Data sets with different values of $\mathcal{N}$ only partially overlap. No matter which system of objective units we choose, there is always one out of the three regimes that does not overlap. A complete overlap is achieved by rescaling the number base as well as the units, as shown in Fig.~\ref{Sfig8}c. Once the units and the base are defined objectively, all curves such that $\mathcal{N}>1$ (red shades) overlap, and all curves such that $\mathcal{N}<1$ (blue shades) overlap on an alternate master curve. As mentioned in the main text, in Fig.~\ref{Sfig8}c, dynamics with $\mathcal{N}<1$ or $\mathcal{N}>1$ proceed in opposite directions, since $\mathcal{N}^{n+1}>\mathcal{N}^{n} \leftrightarrow \mathcal{N}>1$. For curves with a red shade the time increases from left to right, and the radius from bottom to top. For curves with a blue shade the time increases from right to left, and the radius from top to bottom.

\subsubsection{Actual data}
Fig.~\ref{Sfig9} gives the same representations as in Fig.~\ref{Sfig8}, this time with the actual rather than idealized data sets of the height selected spreading, pinching and coalescence experiments. The comparison between Fig.~\ref{Sfig9} and Fig.~\ref{Sfig8} reveals the same differences noticed in the case of explosions. The range and sampling of the actual and idealized data are different. The actual data main show some noise. More importantly: the data seem to systematically deviate from the piece-wise master curves when approaching crossovers. The actual trajectories smooth out the kinks at the turning points (c.f. section~\ref{crossdev}).

\newpage

\section{Deviations at the crossovers\label{crossdev}}
\begin{figure*}
\centering
\includegraphics[width=17cm,clip]{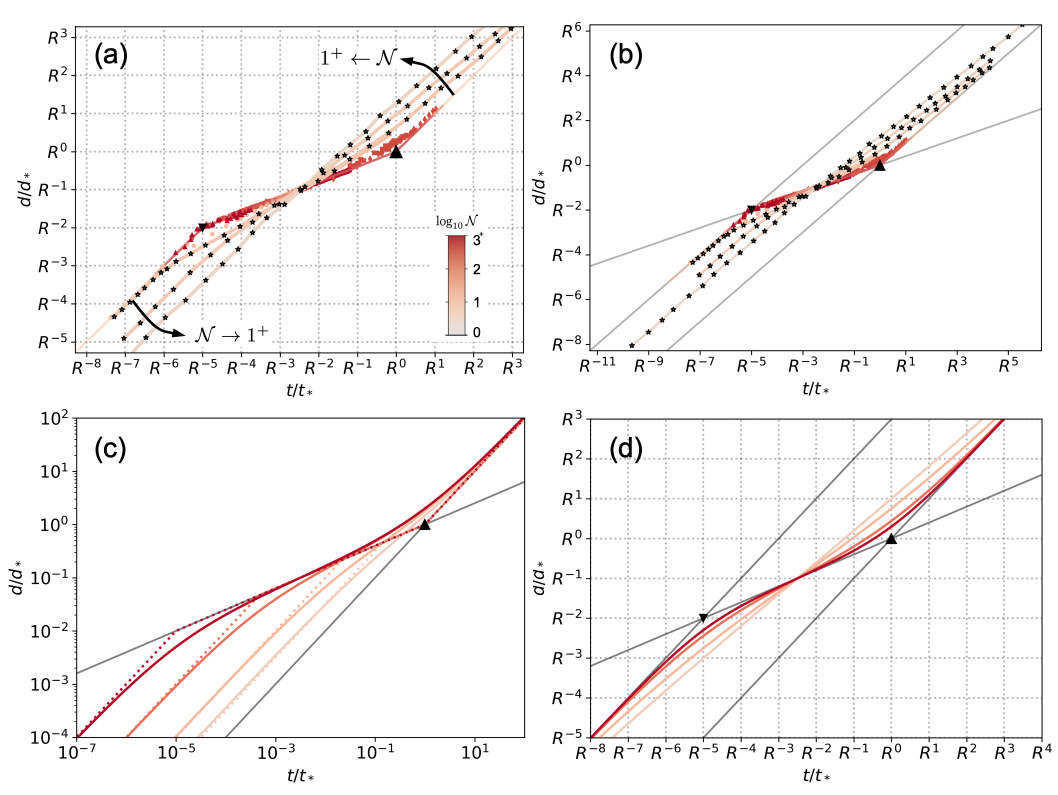}
\caption{Illustration of the smooth transitions in the vicinity of crossovers. (a) A representative set of data (Trinity, Dominic Housatonic, Porzel 1957, Kingery 1962, Kleine 2010, Grun 1991) is plotted in objective units ($t_*$,$d_*$), with an objective base ($R\equiv \mathcal{N}^\frac{1}{3}$). Also included are three vapor cloud explosions with values of $\mathcal{N}$ increasingly close to 1 (stars)~\cite{Tang1999}. As $\mathcal{N}\rightarrow 1^+$, the data show an increasingly smooth transition between the regimes. (b) The same data are plotted on a larger scale. (c) Four theoretical explosions with $\mathcal{N}=10^3$, $10^2$, $10^1$ and $10^\frac{1}{2}$ are represented in objective units ($t_*$,$d_*$). The doted lines follow the piece-wise power laws, $d/d_*=\mathcal{N} (t/t_*)$, then $d/d_*= (t/t_*)^\frac{2}{5}$, and finally $d/d_*=t/t_*$. The continuous lines provide a smooth interpolation(see text for details). (d) The theoretical curves from panel-c are represented with an objective base ($R\equiv \mathcal{N}^\frac{1}{3}$). 
\label{Sfig10}}
\end{figure*}
When comparing idealized and actual data (Fig.~\ref{Sfig6}-\ref{Sfig9}), the most significant difference concerns the neighborhood of the crossovers. Whereas the idealized data are constructed with sharp transitions between the consecutive power laws, the actual data usually show smoother transitions. Such behavior deserves further investigation, but it can already be qualitatively explained. 

Let us consider detonations ($\mathcal{N}>1$) as an example. In Fig.~3 of the main text, we included three examples of vapor cloud explosions~\cite{Tang1999} with small values of initial Mach numbers ($\mathcal{N}\simeq 2-5$). As the value of $\mathcal{N}$ is reduced the width of Taylor's regime progressively shrinks. In the limit $\mathcal{N}=1$, $c_0=c_s$, the initial and late regimes overlap and Taylor's regime disappears. When we represented explosions with an objective base (Fig.~5 of the main text) we did not include these vapor cloud explosions. They are here shown in Fig.~\ref{Sfig10}a and b. In contrast to the explosions with larger values of the initial Mach number $\mathcal{N}$, these three examples seem to be stretched, the smooth transitions between regimes taking over most of the dynamics. This behavior can be reproduced by introducing smoothed out versions of the piece-wise idealized curves, as shown in Fig.~\ref{Sfig10}c, using the Hopkinson-Crantz units ($t_*$, $d_*$). When the base is the conventional 10, the main visual effect of a smaller value of $\mathcal{N}$ is the shrinking range of Taylor's regime. Once the same theoretical curves are represented with an objective base ($R\equiv \mathcal{N}^\frac{1}{3}$), as shown in Fig.~\ref{Sfig10}d, the transitions become overwhelming as $\mathcal{N}\rightarrow 1^+$. 

We believe that this behavior should be quite insensible to the precise type of smoothing function, but for the sake of completeness we here give details about the function we used in Fig.~\ref{Sfig10}c and d. In the Hopkinson-Crantz units ($t_*$,$d_*$), the idealized master curve with sharp transitions is as follows: 
\begin{align}
\text{if } \tilde{t}<\mathcal{N}^{-\frac{5}{3}}~:~~\tilde{d}=\mathcal{N} \tilde{t}\\
\text{if } \mathcal{N}^{-\frac{5}{3}} <\tilde{t}<1~:~~\tilde{d}=\tilde{t}^\frac{2}{5}\\
\text{if } \tilde{t}>1~:~~\tilde{d}=\tilde{t}
\end{align}
\noindent where we have used $\tilde{t}\equiv t/t_*$ and $\tilde{d}\equiv d/d_*$ to simplify the notations. To smooth this piece-wise function we may use additions. For instance, to capture the transition from Taylor's regime to the sound propagation regime, we may introduce the following function (capturing regime 2 and regime 3): 
\begin{equation}
\tilde{d}_{23}  \equiv  \tilde{t}^\frac{2}{5} + \tilde{t}
\end{equation} 
This function is such that $\tilde{d}_{23} \simeq  \tilde{t}^\frac{2}{5}$ if $\tilde{t} \ll 1$, and $\tilde{d}_{23}\simeq  \tilde{t}$ if $\tilde{t} \gg 1$.  

The transition between the initial regime and Taylor's regime cannot be captured by using a simple addition. Indeed, since the exponent of the initial regime is larger than Taylor's exponent (2/5), simply adding, $\mathcal{N} \tilde{t} + \tilde{t}^\frac{2}{5}$, would generate a curve following Taylor's regime for $t<t_0$, and $d=c_0 t$ for $t>t_0$, the opposite of what is observed. Instead we may use the inverse addition (capturing regime 1 and regime 2): 
\begin{equation}
\frac{1}{\tilde{d}_{12}} \equiv \frac{1}{\mathcal{N} \tilde{t}}   + \frac{1}{\tilde{t}^\frac{2}{5}} 
\end{equation} 

We can use $\tilde{d}_{23}$ and $\tilde{d}_{12}$, in two different ways to construct a smoothed out function capturing the three regimes: 
\begin{align}
\tilde{d}_{123} &\equiv \tilde{d}_{12} + \tilde{t} \\ 
\tilde{d}^{\dagger}_{123} &\equiv \Big(\frac{1}{\mathcal{N} \tilde{t}}+\frac{1}{\tilde{d}_{23}} \Big)^{-1}\\ 
\end{align} 
The first function is biased towards the initial regime, whereas the second function is biased towards the final regime. To obtained a balanced result we use $d_s \equiv (\tilde{d}_{123} \tilde{d}^{\dagger}_{123})^\frac{1}{2}$. The curves drawn in Fig.~\ref{Sfig10}c and d follow this smoothing function.  


\section{Animated figures\label{gifs}}
The files `Fig2b.gif', `Fig3b.gif', `Fig3c.gif', `Fig4b.gif', `Fig4c.gif', `Fig4d.gif', `Fig4e.gif', `Fig5a.gif', and `Fig5b.gif' give animated versions of the corresponding figures in the main text, showing how every data set is included in the graph. Separate plots for each data set are given in section~\ref{datacomments}. The images used to create the animated files are given in the zip archives of the same names. Note that in `Fig5b.gif' the data sets `Bolanos2009 Fig8 G2' and `Bolanos2009 Fig9 G6' do not appear, because although they are on the expected scalings, they fall outside the range of the plot. Indeed, these two experiments respectively have $\mathcal{N}\simeq 1/1.6$ and $\mathcal{N}\simeq 1.1$, hence their objective bases are very close to 1, which leads to the sort of stretching discussed in the case of explosions in Fig~\ref{Sfig10}. 

The files `SI3b.gif' and `SI3c.gif' provide animations showing the transition from standard units (seconds, meters) to objective units, i.e. from Fig.~2b of the main text to Fig.~3b and 3c respectively. For instance, for `SI3b.gif', for each image of the gif, the units $t_n$ and $d_n$ are selected separately for each data set. For the first image of the animation $t_{n=0}=1$~s, and $d_{n=0}=1$~m. For the last image of the animation $t_N=t_*$ and $d_N=d_*$. In between, $t_n$ and $d_n$ are spaced logarithmically between the initial and final values. Basically, $t_n=(t_*/\text{1 s})^{n/N}$, and $d_n=(d_*/\text{1 m})^{n/N}$. 

The file `SI5a.gif' provides an animation showing the transition from a plot with objective units ($t_*$,$d_*$) but a subjective base (10), to a plot with both objective units and base ($R\equiv \mathcal{N}^\frac{1}{3}$), i.e. from Fig.~3b to Fig.~5a of the main text. The radix/base is initially $n=10$ for all data sets and $n=R$ at the end of the animation, where the value of $R\equiv \mathcal{N}^\frac{1}{3}$ varies from one data set to another. In between these bounds the radix is varied linearly. 

Similarly to the animated plots for explosions, the files `SI4c.gif', `SI4d.gif', and `SI4e.gif' give animated transitions from Fig.~4a of the main text to Fig.~4c, 4d, and 4e respectively. The file `SI5b.gif' gives a transition from Fig.~4c to Fig.~5b. Note that in this last case, and in contrast to `SI5a.gif' for explosions, some data sets have $\mathcal{N}<1$, so these data are flipped upside-down once the base goes from a value above 1 to a value below it, since $\mathcal{N}>1 \rightarrow \mathcal{N}^{n+1}>\mathcal{N}^{n}$, but $\mathcal{N}<1 \rightarrow \mathcal{N}^{n+1}<\mathcal{N}^{n}$. 

The animated figures `SelfSim Gif1.gif', `SelfSim Gif2.gif', `SelfSim Gif3.gif', `SelfSim Gif4.gif' illustrate the principle of self-similarity on the example of the Trinity explosion (Taylor 1950). In `SelfSim Gif1.gif' the range of the plot is adjusted to reveal its logarithmic origin, i.e. the point of coordinates 1~s and 1~m, arbitrarily chosen as unit. In `SelfSim Gif2.gif' the location of the origin is changed, i.e. the units are changed, and the corresponding value of the prefactor of the power law are updated accordingly. If the logarithmic origin is located at coordinates $t_i$, $d_i$, the units of the plots are then $t_i$ and $d_i$, and the displayed value of $K$ is in units $d_i.t_i^{-\frac{2}{5}}$. If the units $t_i$ and $d_i$ fall on the power law, then by construction $d_i=Kt_i^\frac{2}{5}$, and so the value of $K$ in these units is $1$, as shown in `SelfSim Gif3.gif'. For a single power law there is an infinity of equally ``good units'', as discussed in the main text. Once an additional power law is considered, as the initial regime $d=c_0 t$, there is only one choice of units that reduces the prefactors of both powers laws to 1, as shown in `SelfSim Gif4.gif', where the value of $c_0$ is given in units $d_i.t_i^{-1}$. (For additional details see: \href{https://youtu.be/wlckZEXOKJo?si=K6O5QtDV5MkujcYr}{Mechanics - Lecture 10}.) \\

\clearpage

\section{Details on experimental data\label{datacomments}}
In this section we provide additional details on the experiments reproduced in the article. The readers are referred to the original publications for the full context. 
\subsection{Explosions\label{dataexplo}}
The initial regime ($d\simeq c_0 t$, dotted line), Taylor's regime ($d\simeq Kt^\frac{2}{5}$, continuous line), and the late regime at sound speed ($d\simeq c_s t$, dashed line) are shown when the range of the data encompasses them. The initial regime appears in grey for data sets with a large uncertainty on the ejected mass $m$. The four dominant mechanical parameters, the ejected mass $m$, the yield/energy $E$, and the density $\rho$ and bulk modulus $\Sigma$ of the ambient medium are given on each plot in standard units (respectively: kg, J, kg/m$^3$, and Pa). 

\begin{table}[h]
  \centering
  \begin{tabular}{ | p{9cm} | p{9cm} | }
    \hline
    \textbf{Trinity (Taylor 1950, Mack 1946, Bainbridge 1976)} & \textbf{O'Connell 1957}  \\
    \begin{minipage}{.5\textwidth}
      \includegraphics[width=\linewidth]{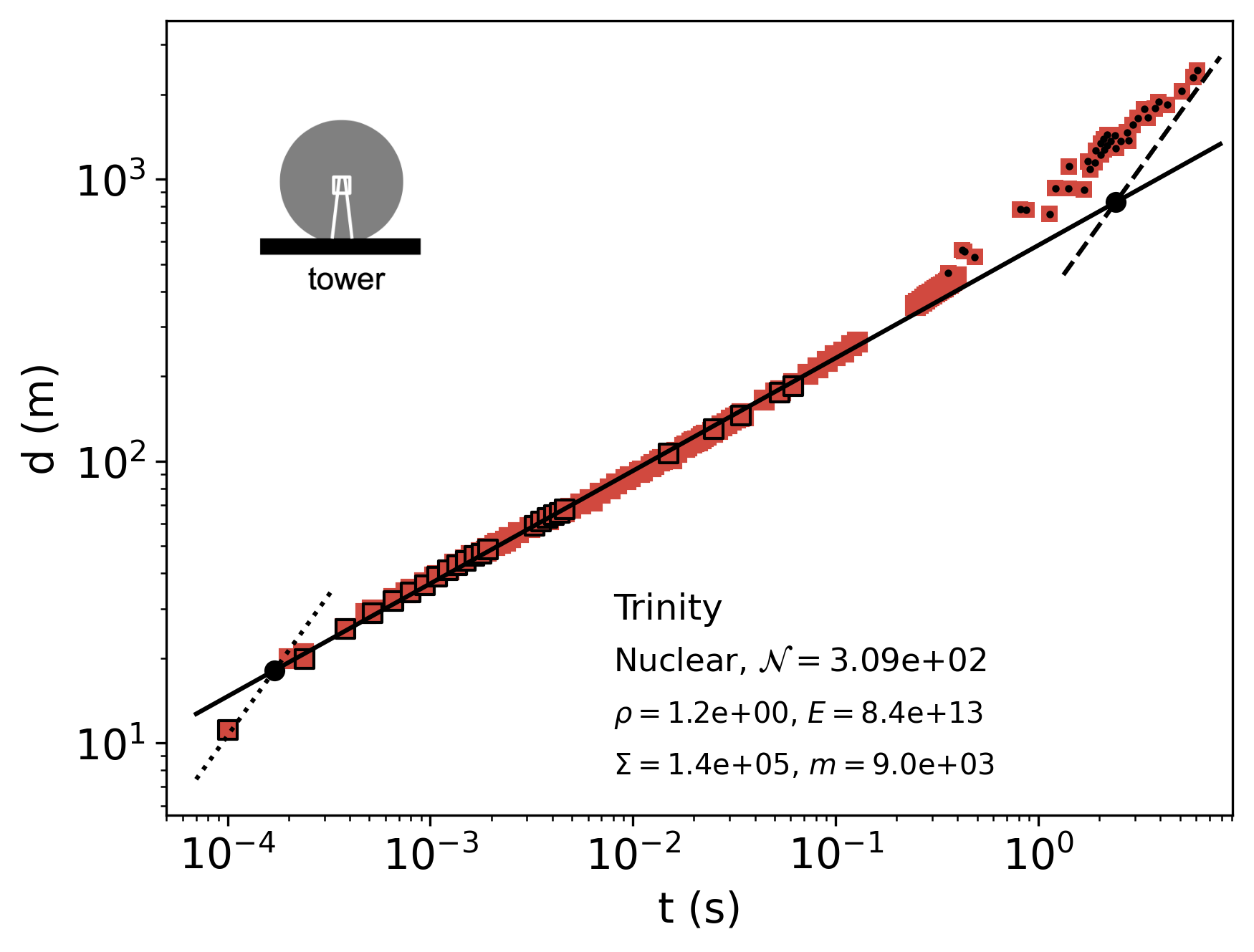}
    \end{minipage}
    &
    \begin{minipage}{.5\textwidth}
      \includegraphics[width=\linewidth]{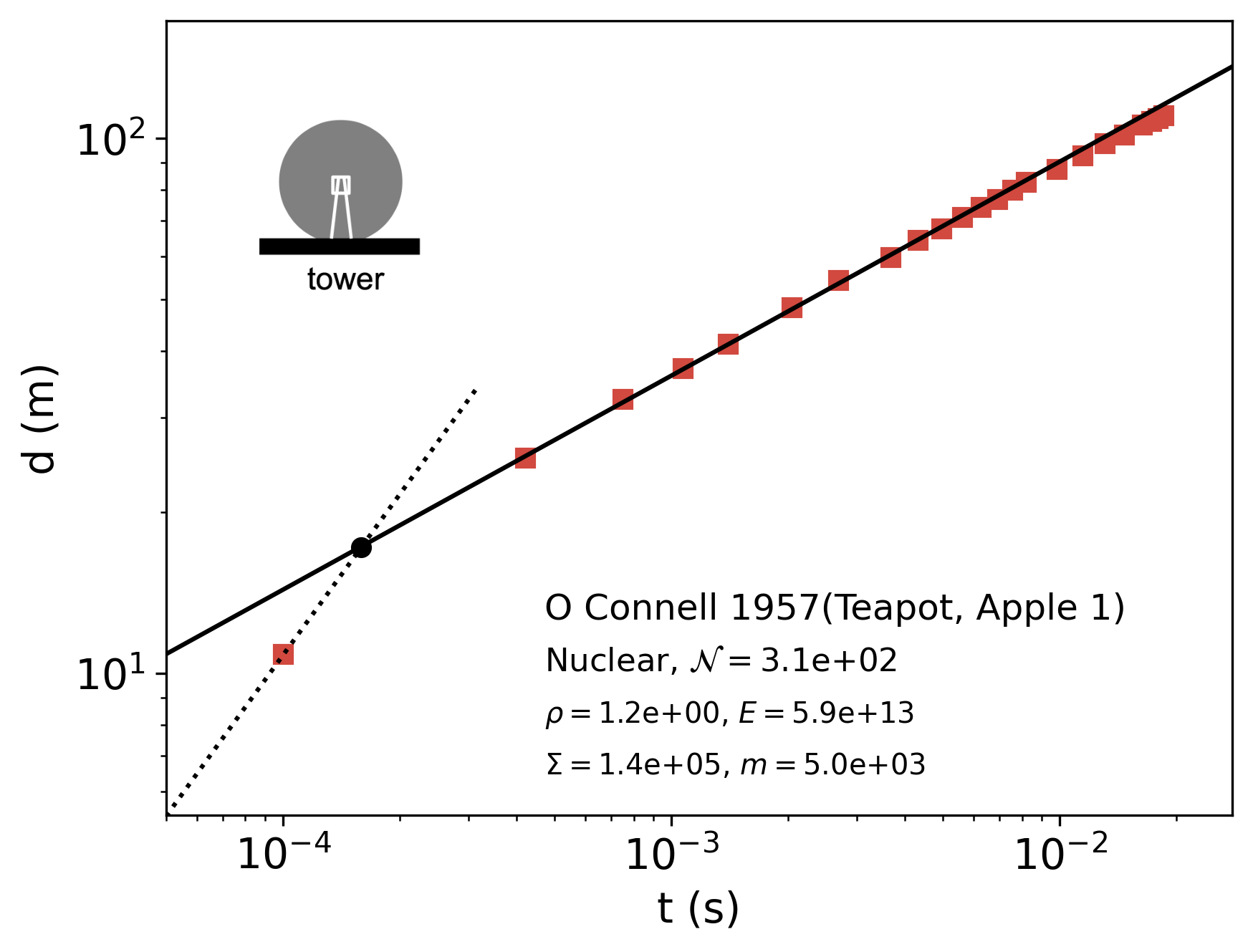}
    \end{minipage}
    \\
\href{https://youtu.be/wki4hg9Om-k?si=xY3e3tuE882KW-O-}{Trinity nuclear test} (July 16, 1945, New Mexico).  ``The Gadget'' (Fat man design) is detonated on top of 30~m tower.     
   & Teapot - Apple 1 test (March 29, 1955, Nevada). The bomb is detonated on top of a 150~m tower.  \\
      \hline \hline
    \textbf{Nguyen 2017} & \textbf{Schmitt 2016 Climax}  \\ 
    \begin{minipage}{.5\textwidth}
      \includegraphics[width=\linewidth]{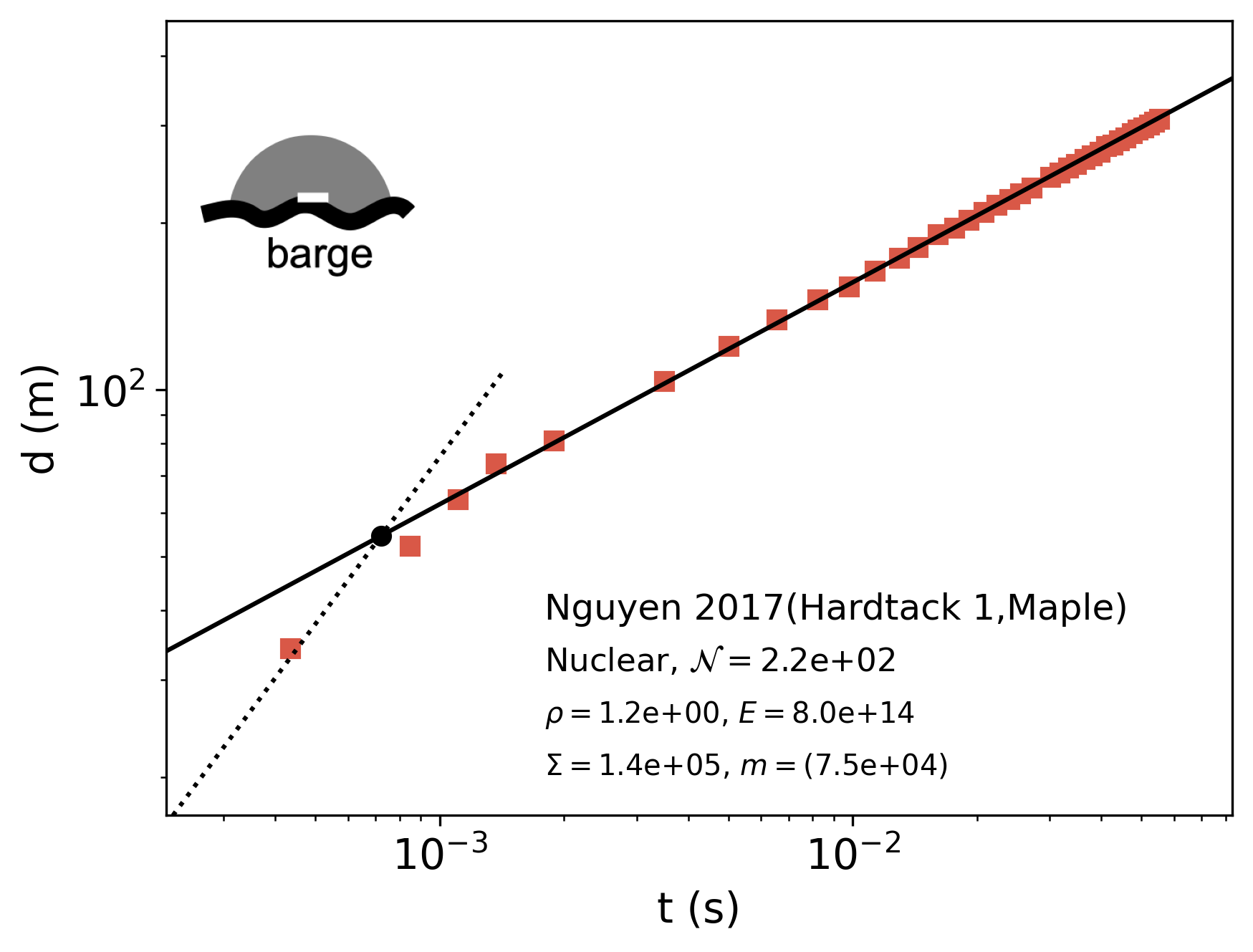}
    \end{minipage}
    &
    \begin{minipage}{.5\textwidth}
      \includegraphics[width=\linewidth]{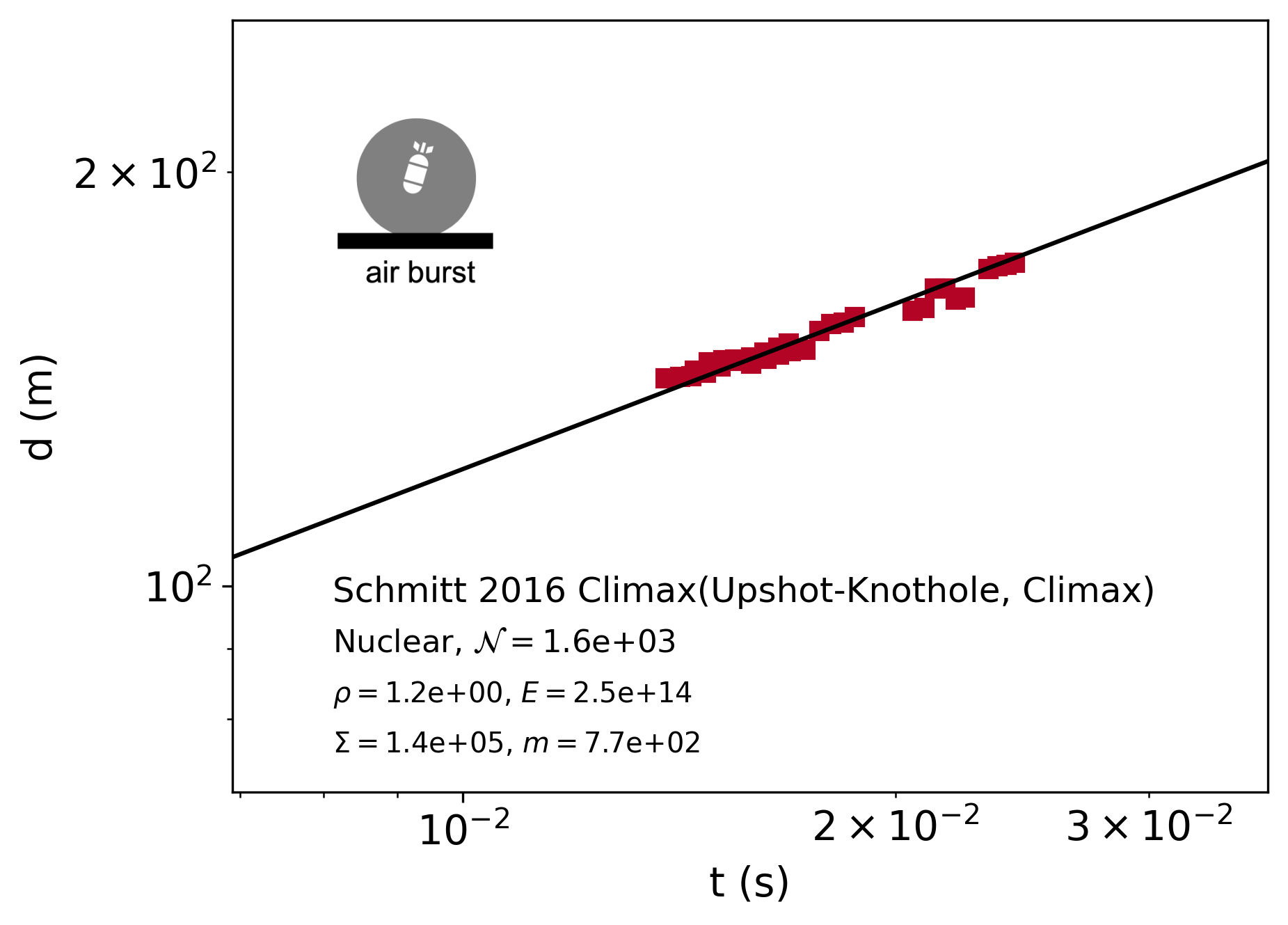}
    \end{minipage}
    \\ 
  \href{https://youtu.be/H5zK2qmo2Pg?si=lsrkrmOw2sQQD0Kb}{Hardtack 1 - Maple nuclear test} (June 11, 1958, Lomilik, Pacific Ocean). Detonated on a barge. & \href{https://youtu.be/fPshfLoxWjM?si=fS4mEzORSHARAV-h}{Upshot Knothole - Climax nuclear test} (June 4, 1953, Nevada). Free air drop of a MK-7 weapon, detonated at 410~m from the ground.  \\ \hline 
  \end{tabular}
\end{table}

\begin{table} 
 \centering 
 \begin{tabular}{ | p{9cm} | p{9cm} | } 
 \hline 
 \textbf{Schmitt 2016 Grable} & \textbf{Dominic Housatonic}  \\ 
 \begin{minipage}{.5\textwidth} 
 \includegraphics[width=\linewidth]{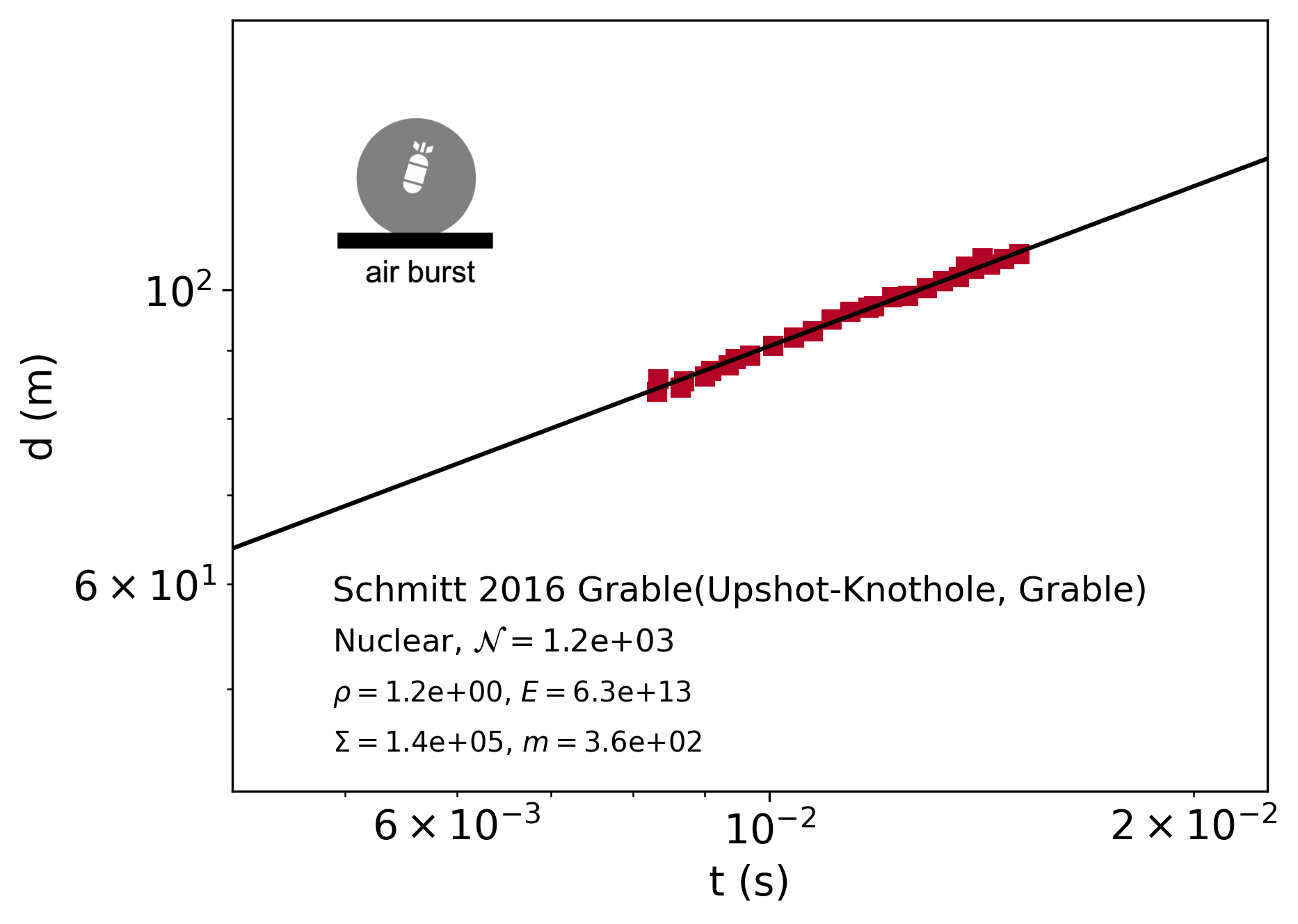}
 \end{minipage}
 & 
 \begin{minipage}{.5\textwidth} 
\includegraphics[width=\linewidth]{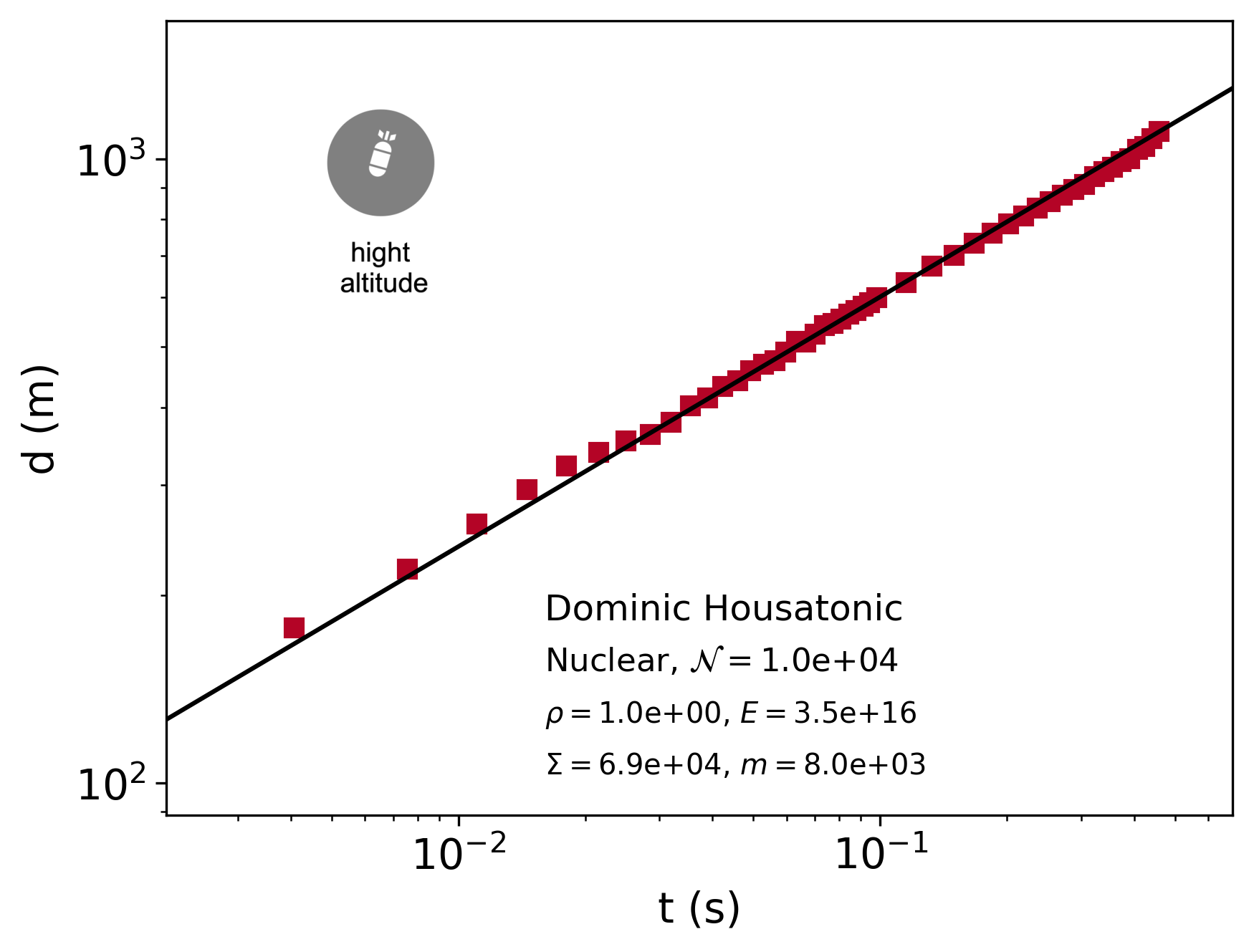} \end{minipage} 
 \\ 
 \href{https://youtu.be/QsB83fAtNQE?si=ur6RgLah6l3UhPwv}{Upshot Knothole - Grable nuclear test} (May 25, 1953, Nevada). W9 AFAP type weapon, launched from an artillery gun, detonated 160~m from the ground.  &  \href{https://youtu.be/IZZ_IsyE_iE?si=Z8kkd72jJm56DBxY}{Dominic - Housatonic nuclear test} (October 30, 1962, Johnston Island, Pacific Ocean). Thermonuclear weapon (Kinglet primary and Ripple II secondary), air dropped, detonated an altitude of 3700~m.\\ \hline \hline 
\textbf{Porzel 1957} & \textbf{Kingery 1962}  \\ 
 \begin{minipage}{.5\textwidth} 
\includegraphics[width=\linewidth]{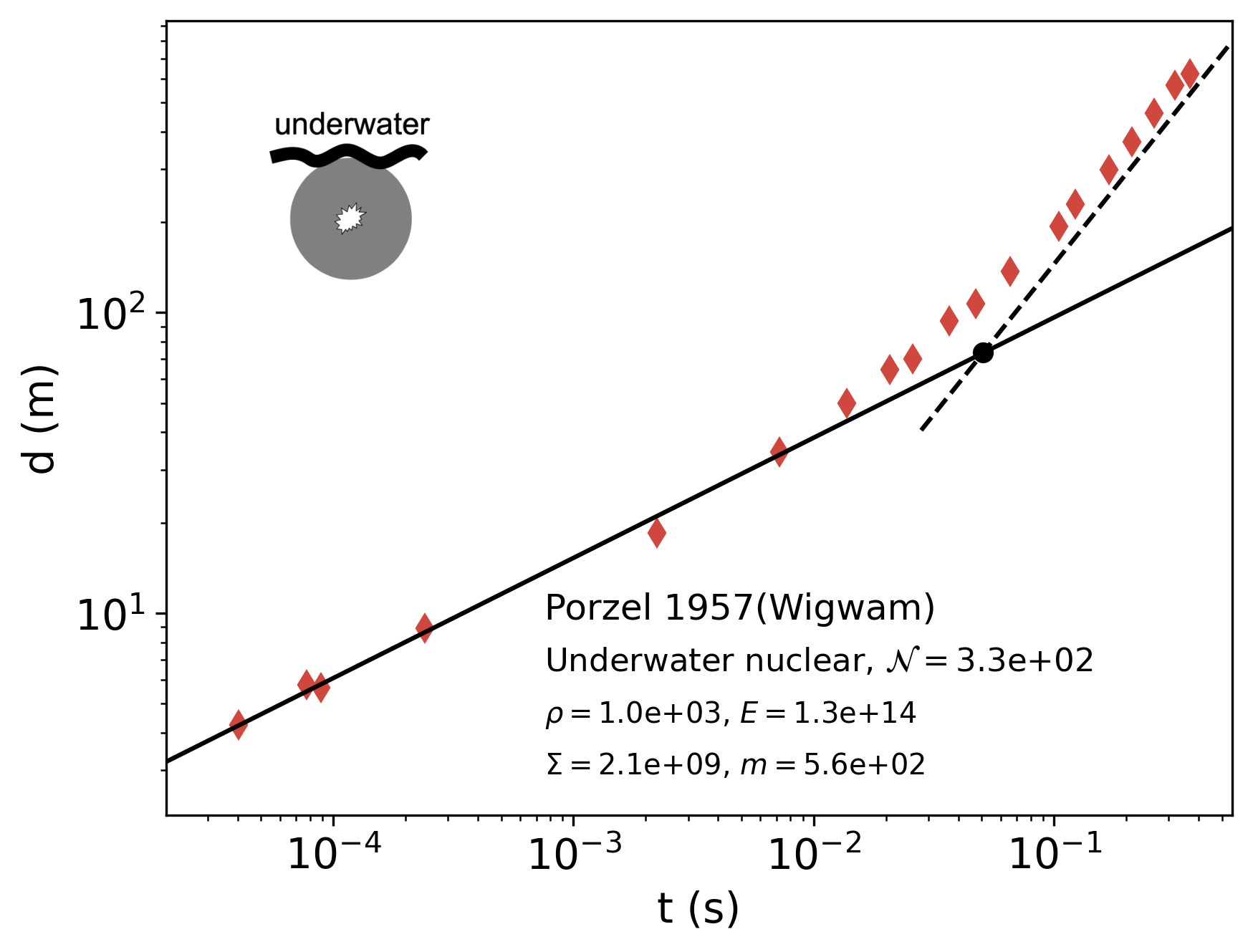} \end{minipage}
 & 
 \begin{minipage}{.5\textwidth} 
\includegraphics[width=\linewidth]{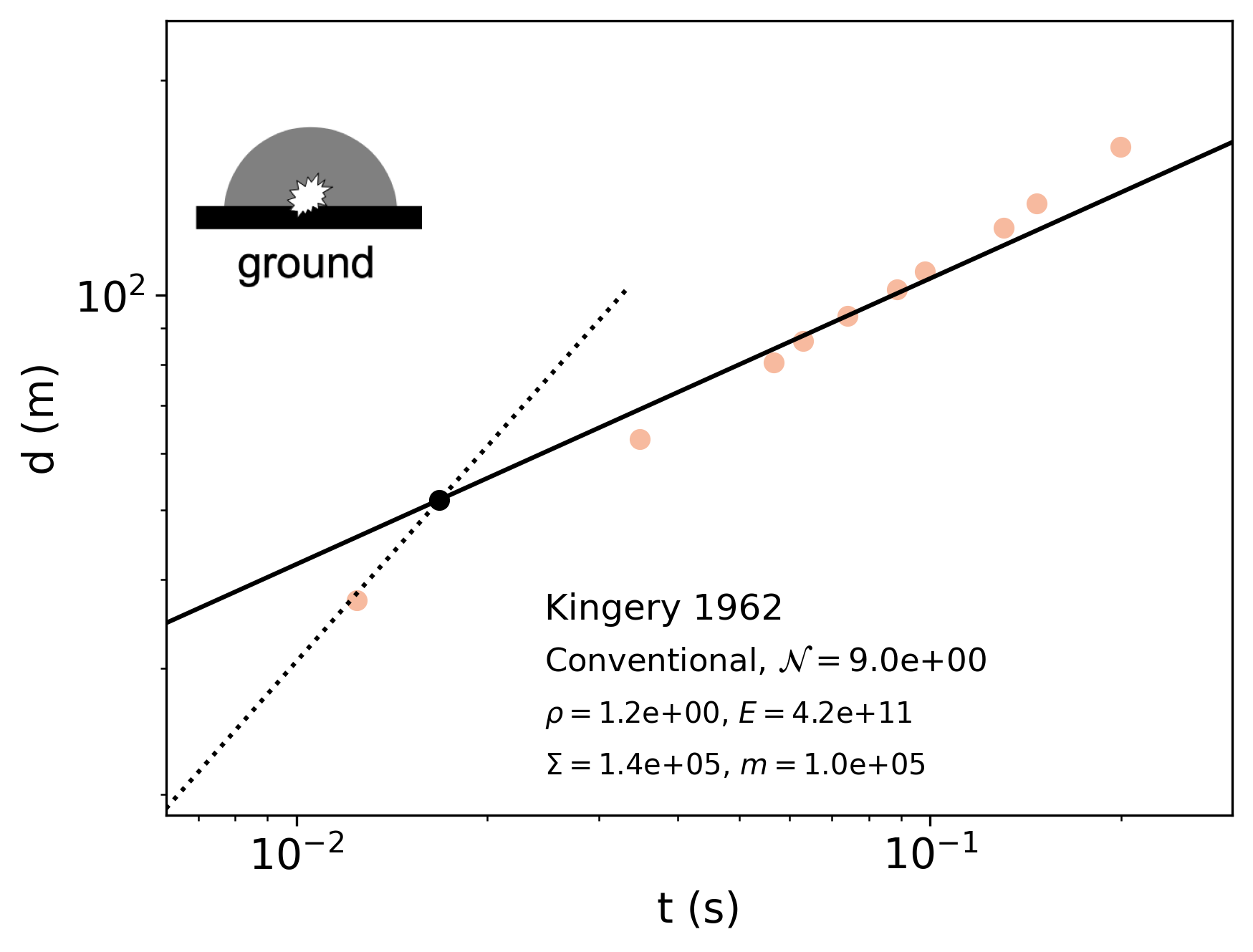} \end{minipage} 
 \\ 
\href{https://youtu.be/ku7R1TSBfjI?si=SRdNecriEgwTiOJq}{Wigwam nuclear test} (May 14, 1955, Pacific Ocean). Mark-90 weapon, detonated underwater at a depth of 300~m.  & 100 tons of TNT detonated on the ground, similar to the \href{https://youtu.be/nN5q8i-kQj0?si=zdJEUA6m8c80jGQC}{trial run of the Trinity test} .  \\ \hline \hline 
\textbf{Aouad 2021} & \textbf{Hargather 2007}  \\ 
 \begin{minipage}{.5\textwidth} 
\includegraphics[width=\linewidth]{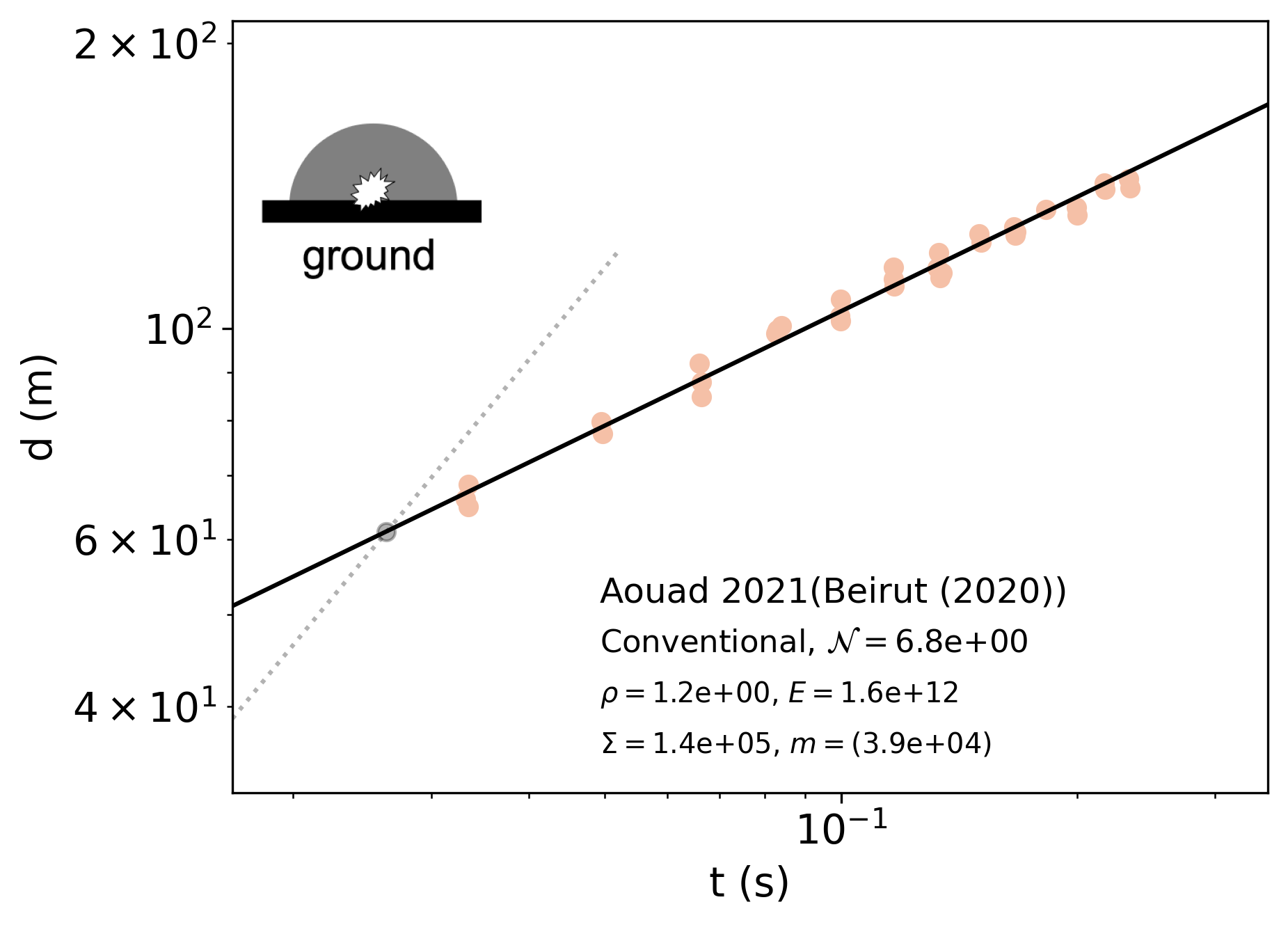} \end{minipage} 
 & 
 \begin{minipage}{.5\textwidth} 
\includegraphics[width=\linewidth]{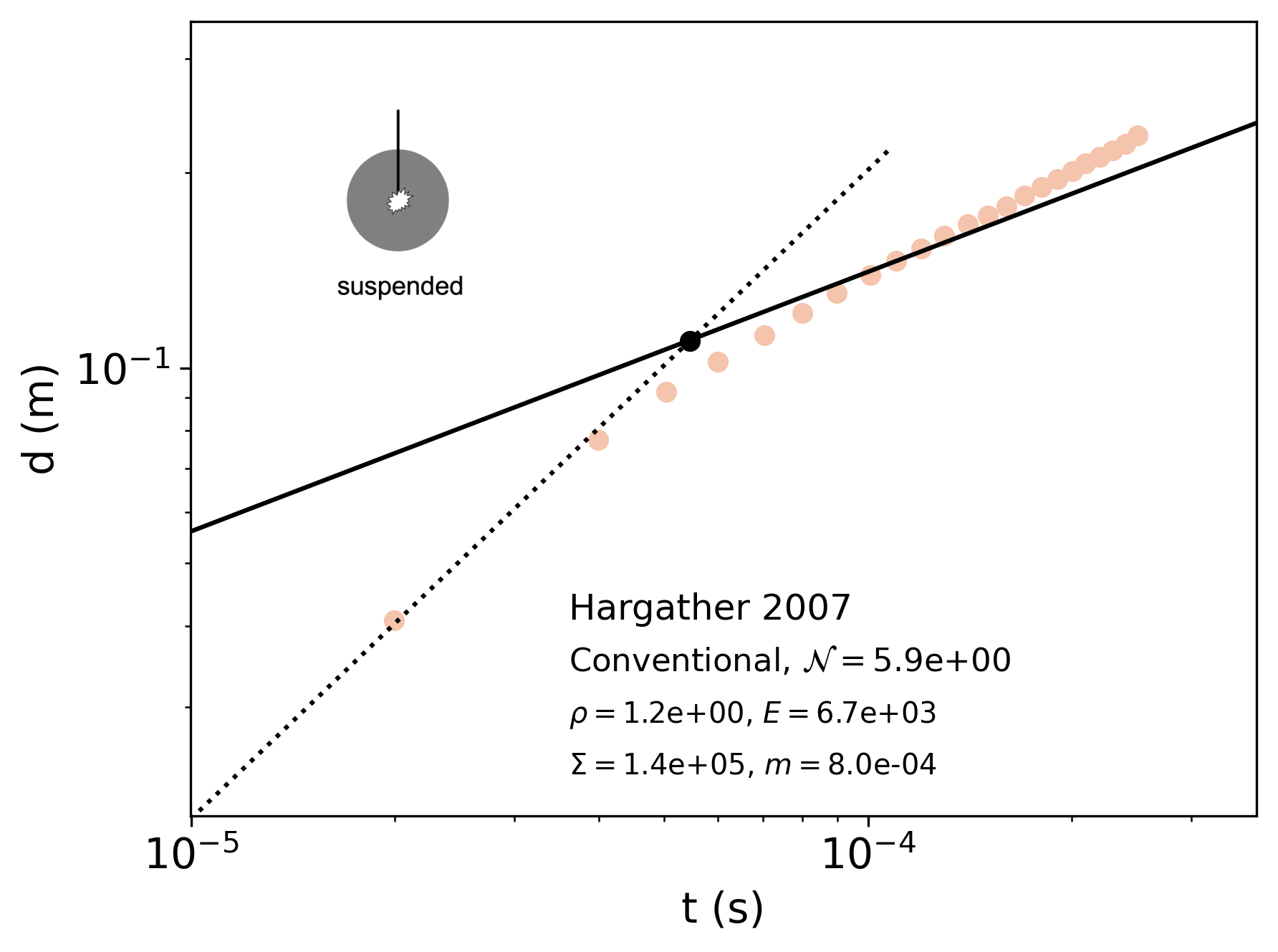}
 \end{minipage} 
 \\ 
2020 Beirut explosion (August 4, 2020), caused by 2.75 kilotons of ammonium nitrate.  & Suspended charge of 0.8~g of Pentaerythritol tetranitrate (PETN).\\ \hline 
\end{tabular} 
 \end{table} 
\begin{table} 
 \centering 
 \begin{tabular}{ | p{9cm} | p{9cm} | } 
 \hline 
 \textbf{Kleine 2010} & \textbf{Grun 1991}  \\ 
 \begin{minipage}{.5\textwidth} 
\includegraphics[width=\linewidth]{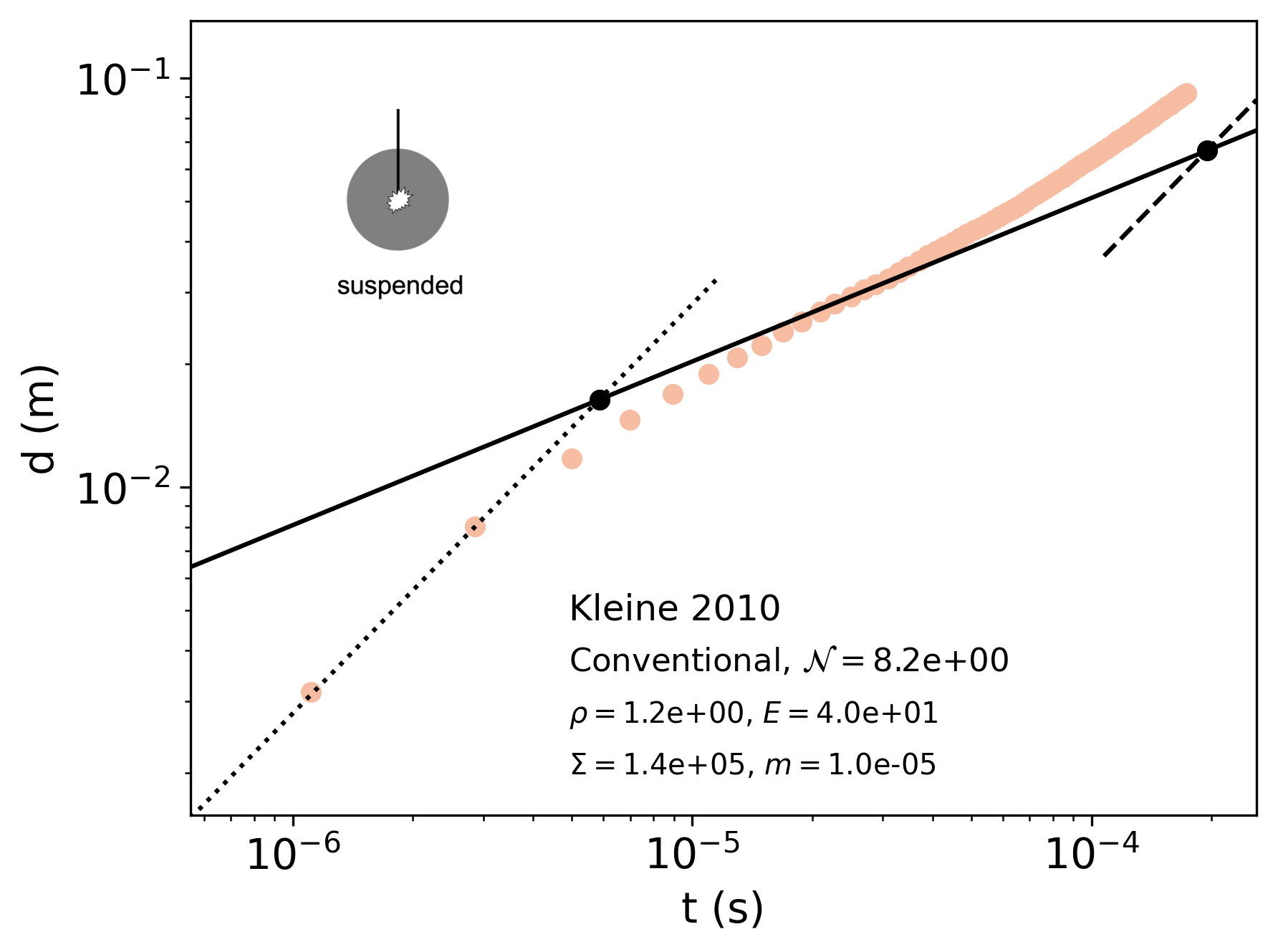}
 \end{minipage}
 & 
 \begin{minipage}{.5\textwidth} 
 \includegraphics[width=\linewidth]{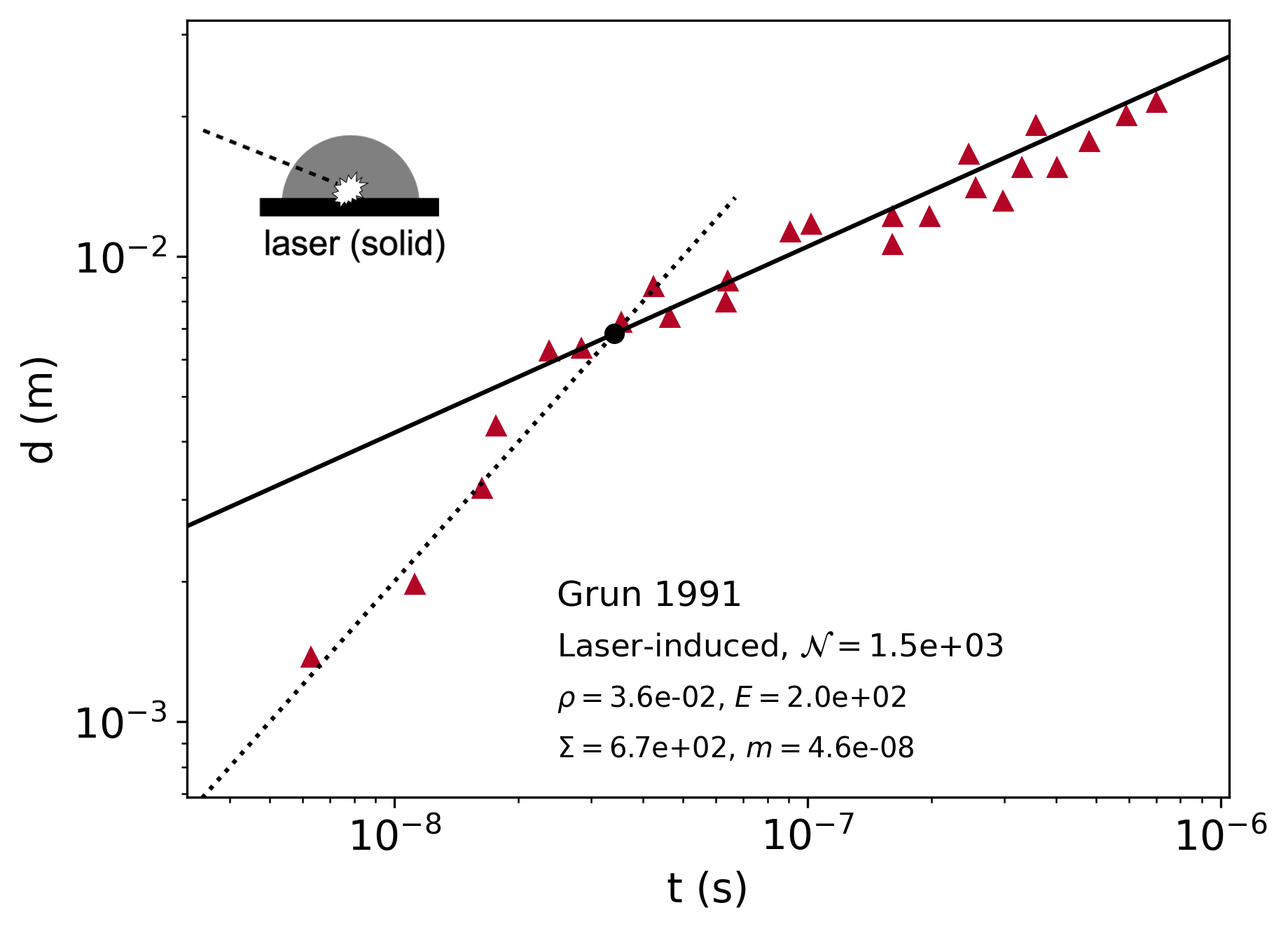}
 \end{minipage} 
 \\ 
Suspended charge of 10~mg of silver azide. & Laser-induced blast on a polystyrene foil, in rarefied xenon. \\ \hline \hline 
\textbf{Porneala 2006} & \textbf{Gatti 1988}  \\ 
 \begin{minipage}{.5\textwidth} 
 \includegraphics[width=\linewidth]{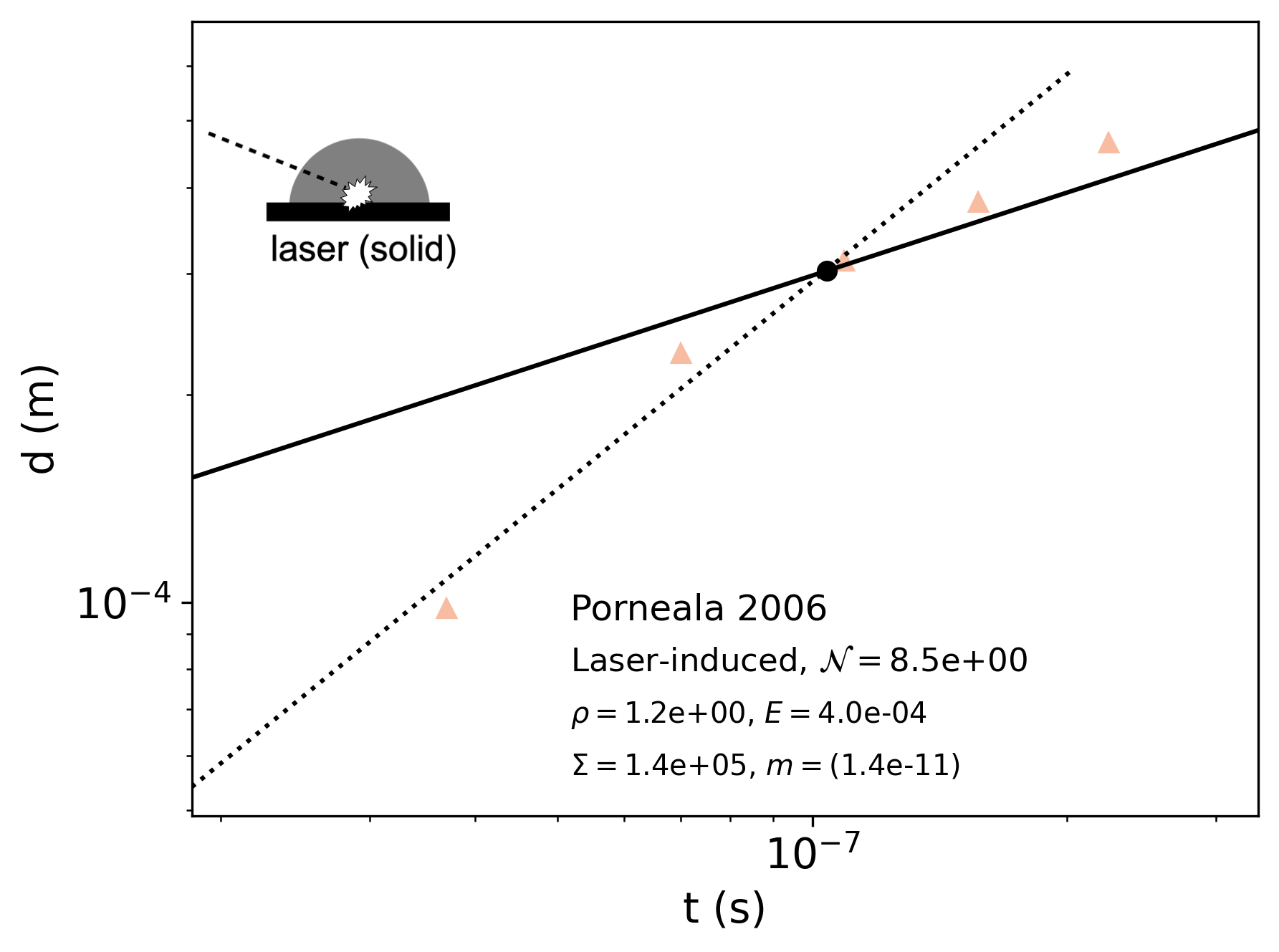}
 \end{minipage}
 & 
 \begin{minipage}{.5\textwidth} 
 \includegraphics[width=\linewidth]{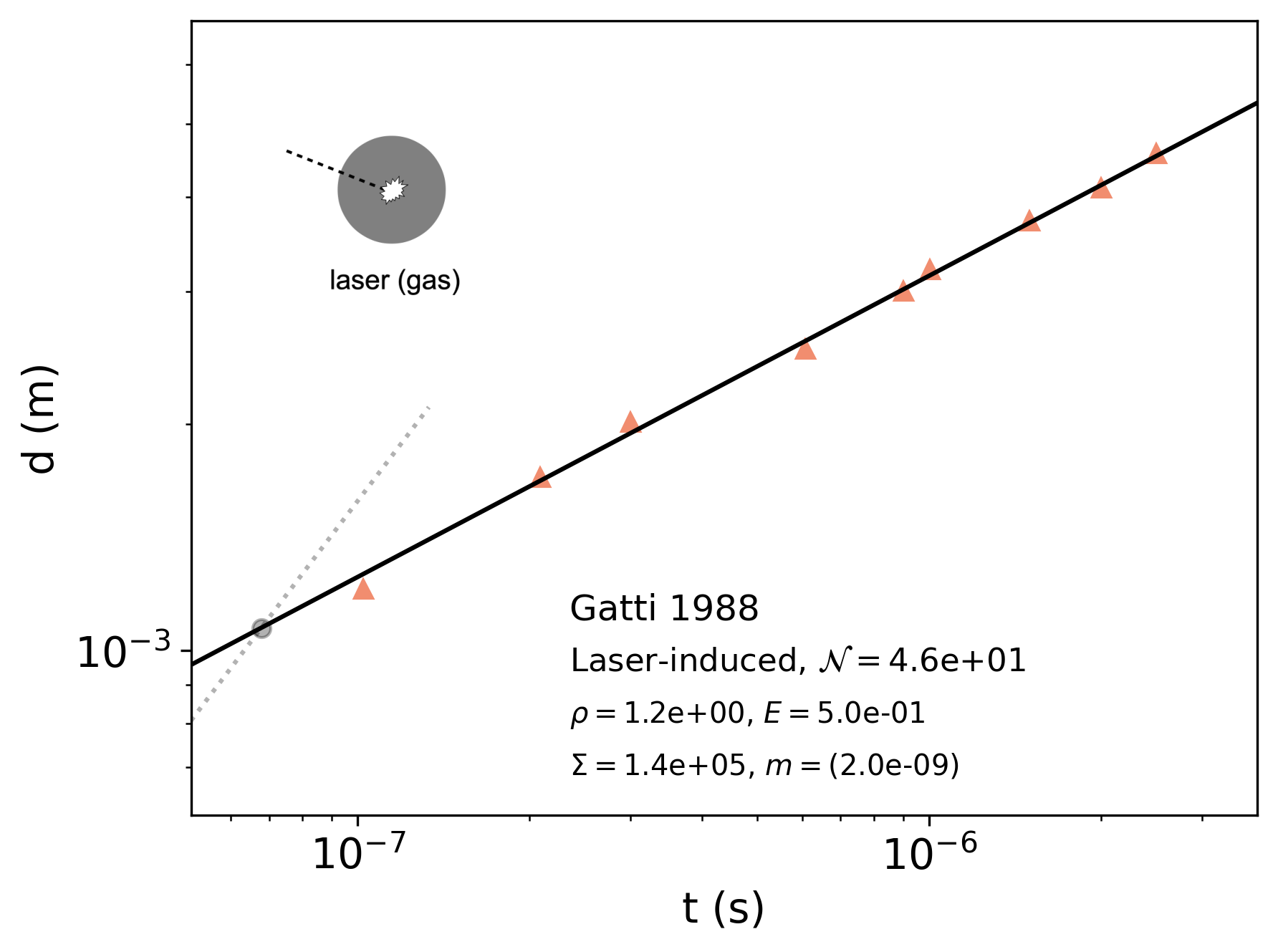}
 \end{minipage} 
 \\ 
Laser-induced blast on an aluminum target, in the air. & Laser-induced blast in the air. \\ \hline 
\end{tabular} 
 \end{table}

\subsection{Droplets and bubbles\label{datadrop}}
The visco-capillary regime ($d\simeq c_v t$, dotted line), the inertio-capillary regime ($d\simeq K_i t^\frac{2}{3}$, continuous line), and the late regime ($d\simeq D$, dashed line) are shown when the range of the data encompasses them. For spreading experiments in completely wetting conditions, Tanner's regime (grey dotted dashed line; Eq.~\ref{Tanner}) is shown for reference. The three dominant mechanical parameters, the viscosity $\eta$, the density $\rho$, and the surface-tension $\Gamma$ are given on each plot in standard units (respectively: Pa.s, kg/m$^3$, and N/m). The value of the droplet or bubble size $D$ is given in meters. 

\newpage

\begin{table} 
 \centering 
 \begin{tabular}{ | p{9cm} | p{9cm} | } 
 \hline 
 \textbf{Eddi2013 Fig4 0p37} & \textbf{Eddi2013 Fig4 0p5}  \\ 
 \begin{minipage}{.5\textwidth} 
  \includegraphics[width=\linewidth]{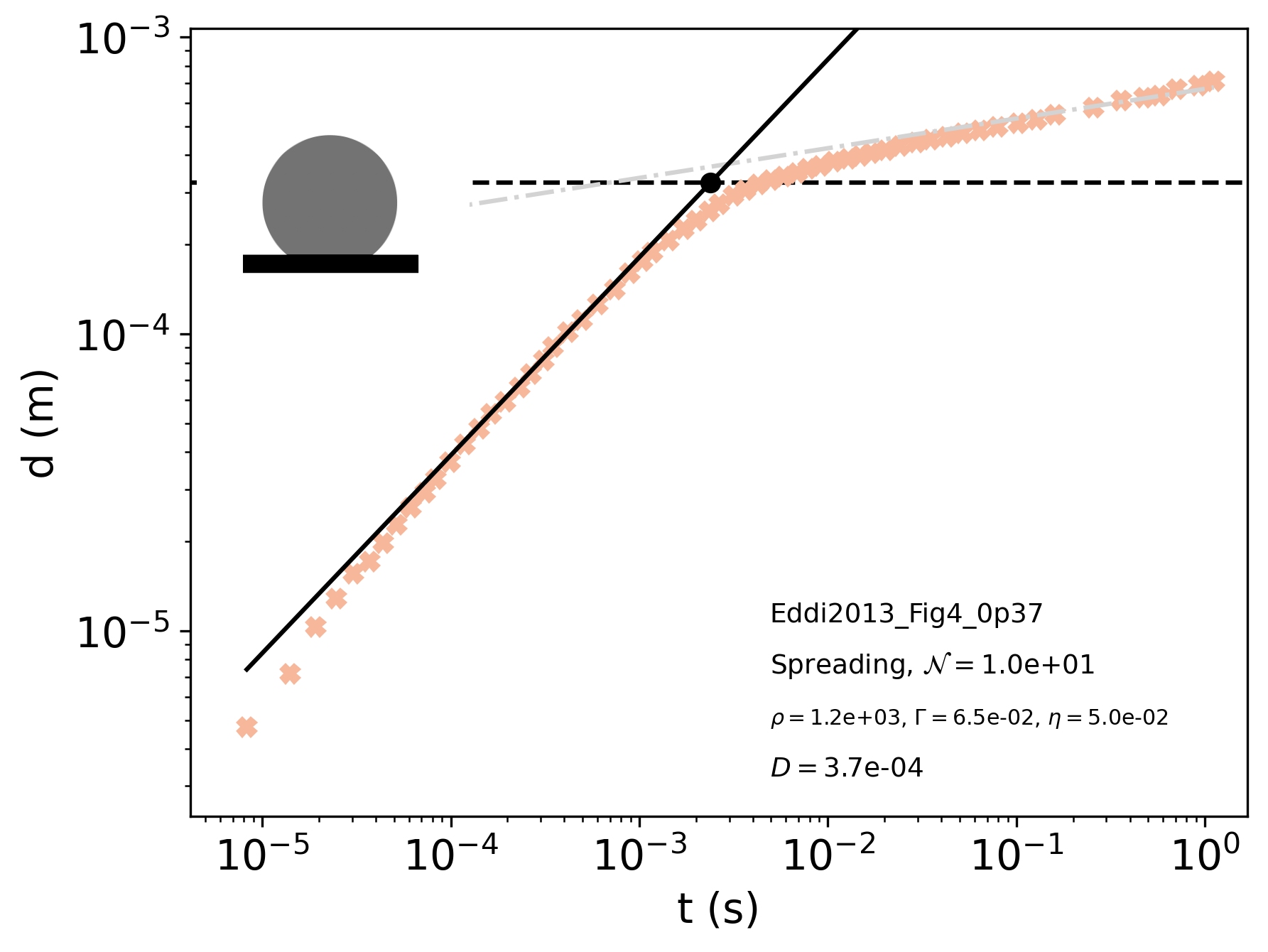}
 \end{minipage}
 & 
 \begin{minipage}{.5\textwidth} 
  \includegraphics[width=\linewidth]{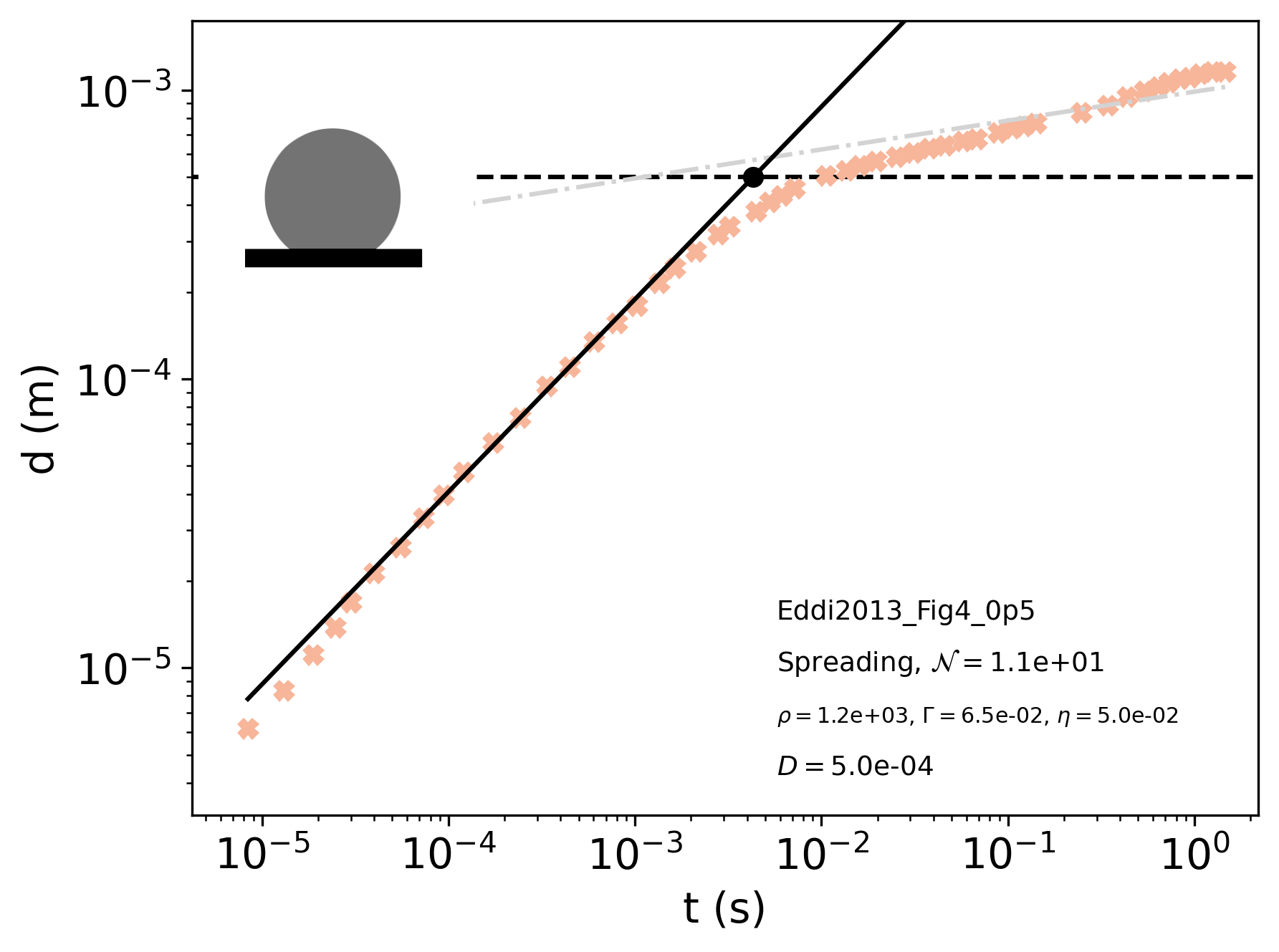} 
 \end{minipage} 
 \\ 
Water-glycerol mixture in ambient air, on hydrophilic glass. & Water-glycerol mixture in ambient air, on hydrophilic glass. \\ \hline \hline 
\textbf{Eddi2013 Fig4 0p63} & \textbf{Eddi2013 Fig5a 105deg}  \\ 
 \begin{minipage}{.5\textwidth} 
   \includegraphics[width=\linewidth]{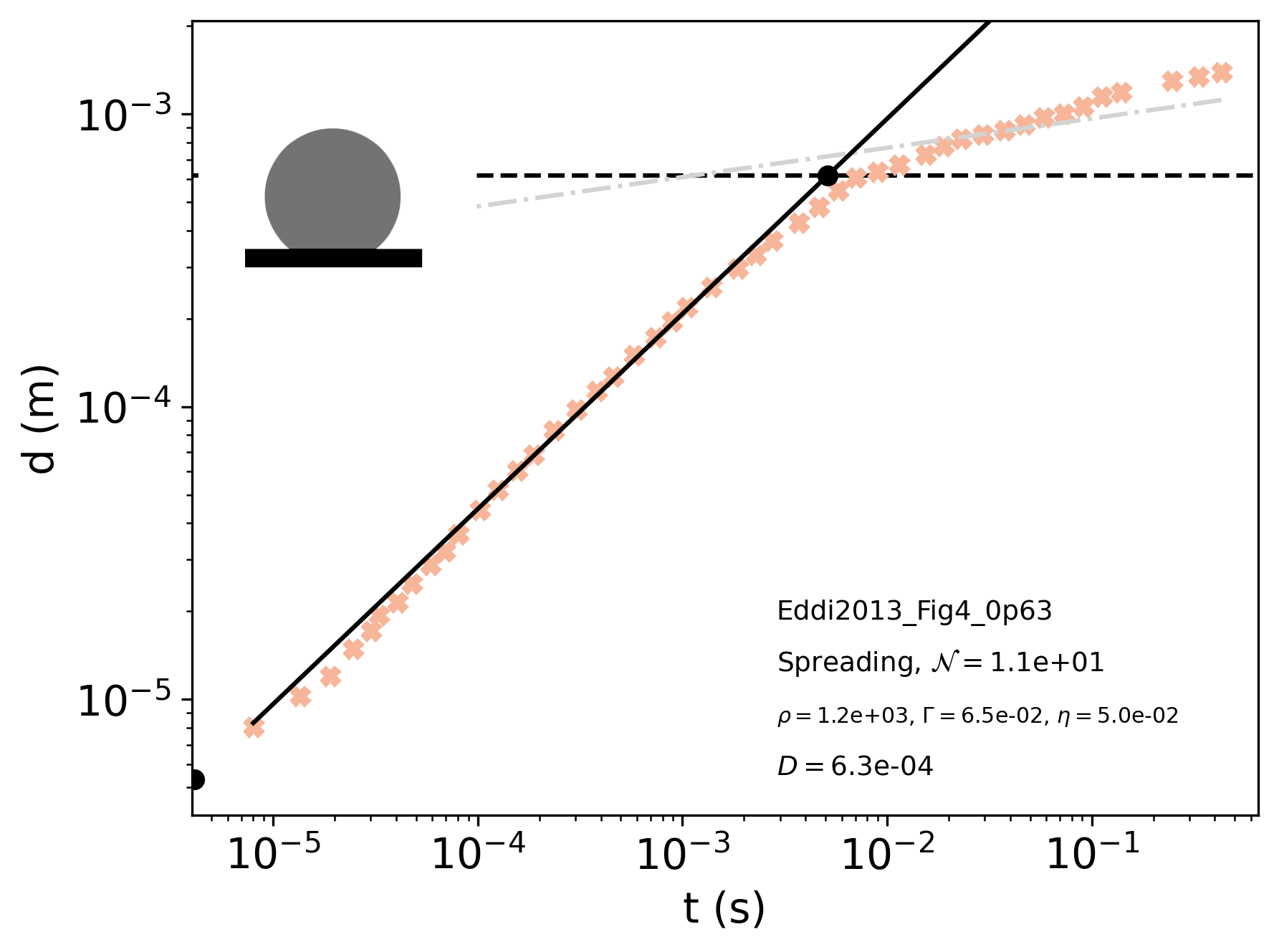} 
 \end{minipage}
 & 
 \begin{minipage}{.5\textwidth} 
   \includegraphics[width=\linewidth]{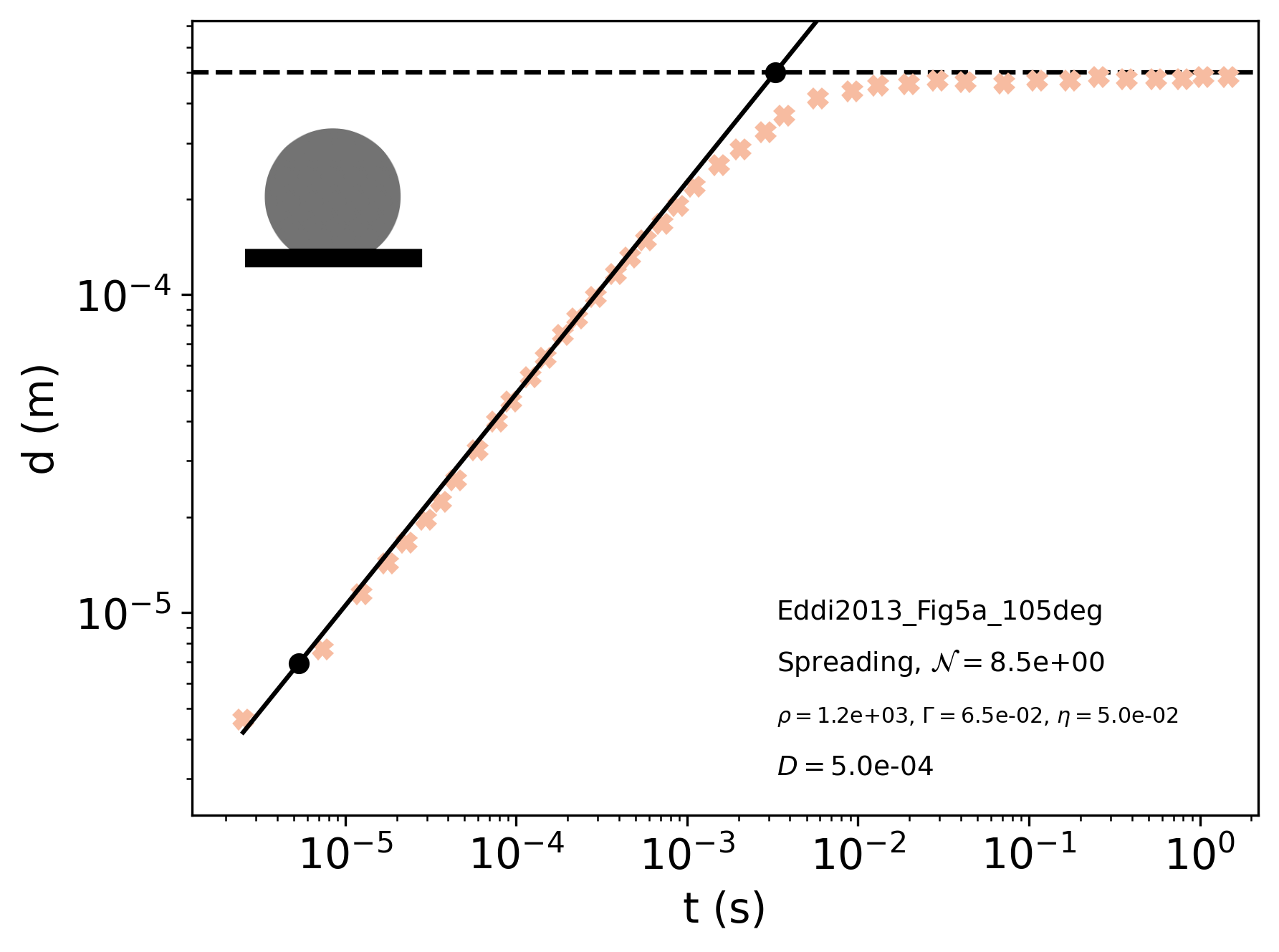} 
 \end{minipage} 
 \\ 
Water-glycerol mixture in ambient air, on hydrophilic glass. & Water-glycerol mixture in ambient air, \newline on hydrophobic fluoropolymer-coated glass.\\ \hline \hline 
\textbf{Eddi2013 Fig6 11} & \textbf{Yao2005 100000cS 0p5cm}  \\ 
 \begin{minipage}{.5\textwidth} 
   \includegraphics[width=\linewidth]{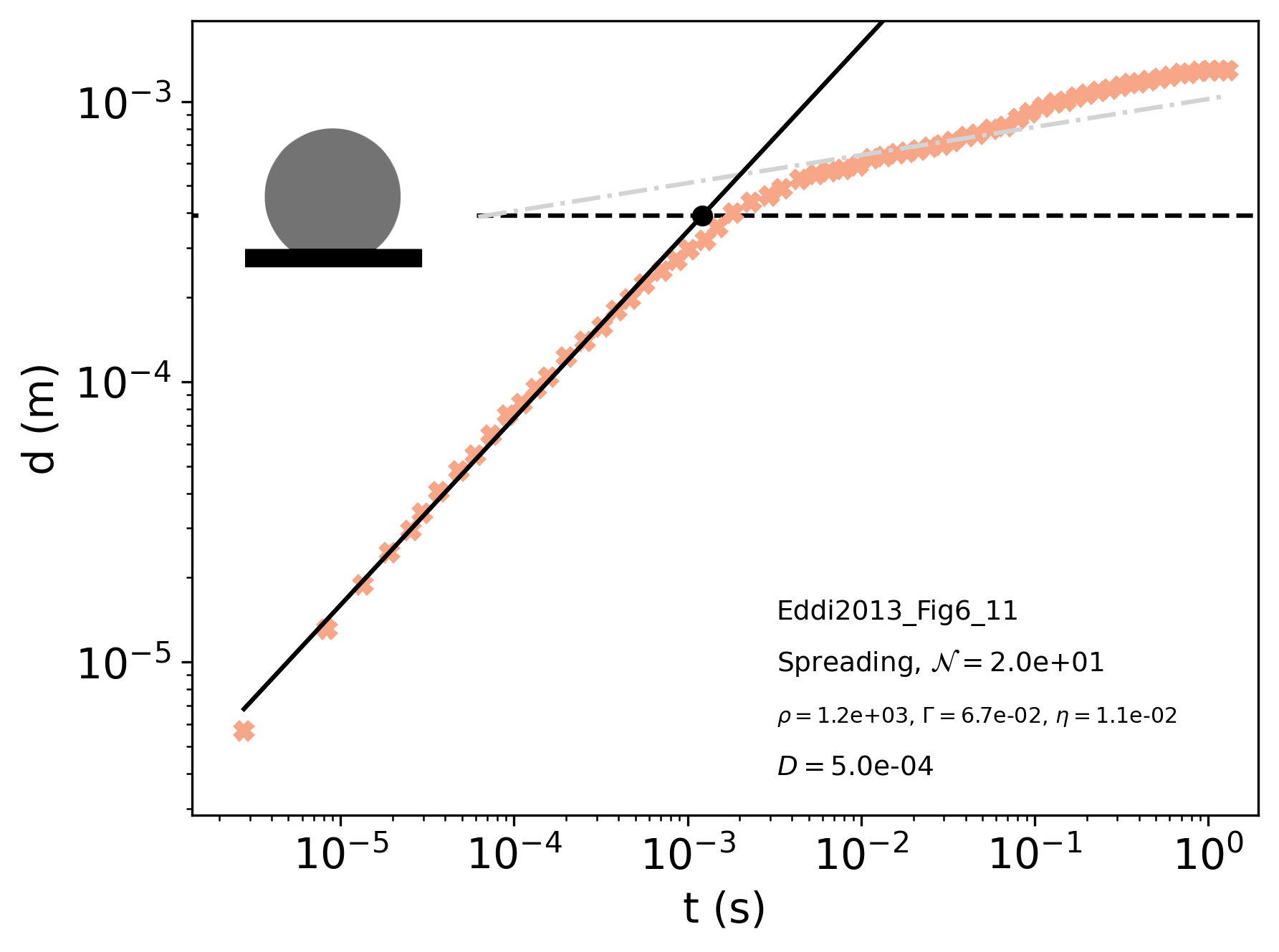} 
 \end{minipage}
 & 
 \begin{minipage}{.5\textwidth} 
   \includegraphics[width=\linewidth]{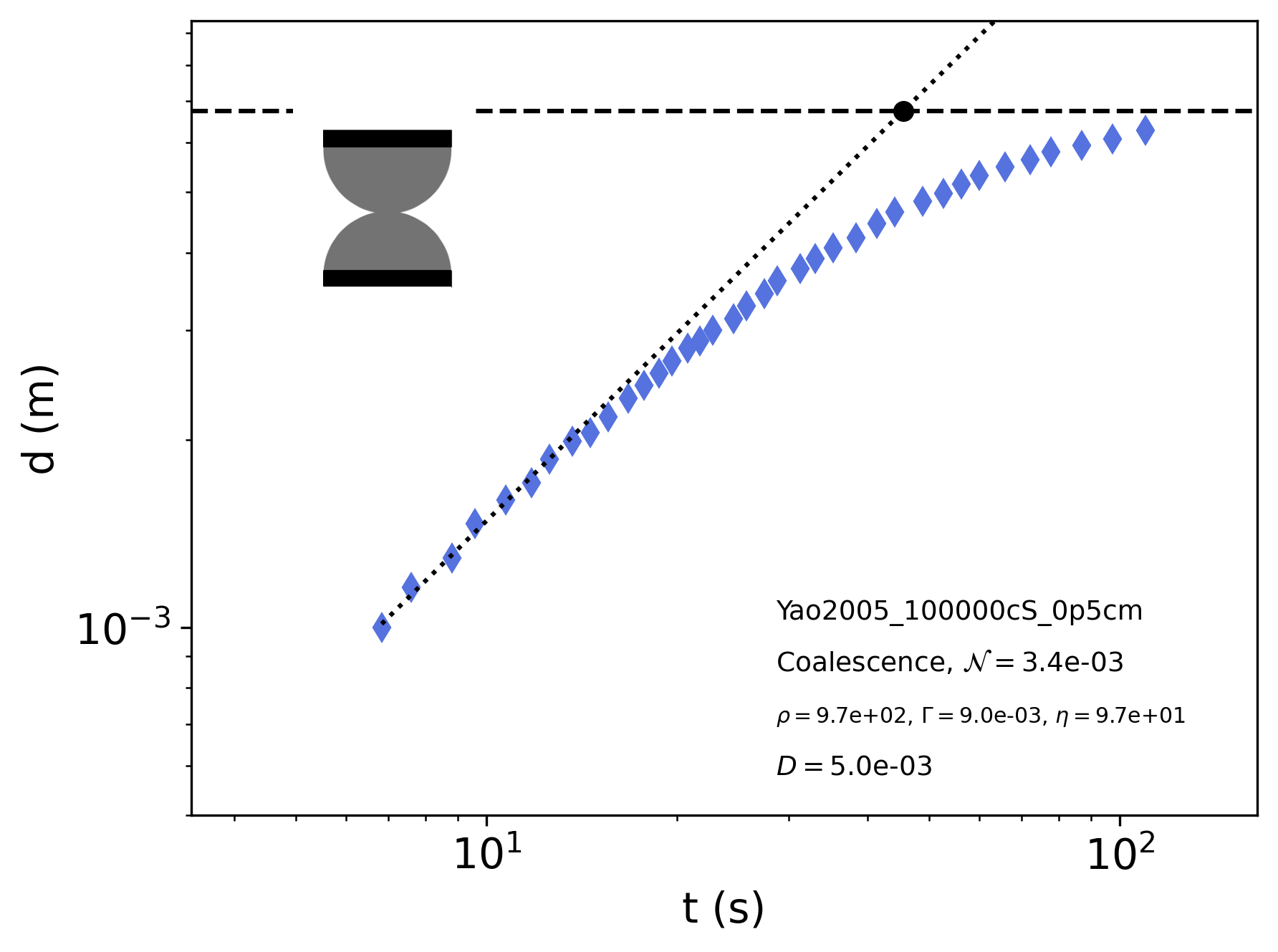} 
 \end{minipage} 
 \\ 
Water-glycerol mixture in ambient air, on hydrophilic glass. & Silicone oil in density-matched water-alcohol mixture.\\ \hline 
\end{tabular} 
 \end{table} 
 
 \begin{table} 
 \centering 
 \begin{tabular}{ | p{9cm} | p{9cm} | } 
 \hline 
 \textbf{Yao2005 10000cS 0p5cm} & \textbf{Yao2005 1000cS 0p5cm}  \\ 
 \begin{minipage}{.5\textwidth} 
  \includegraphics[width=\linewidth]{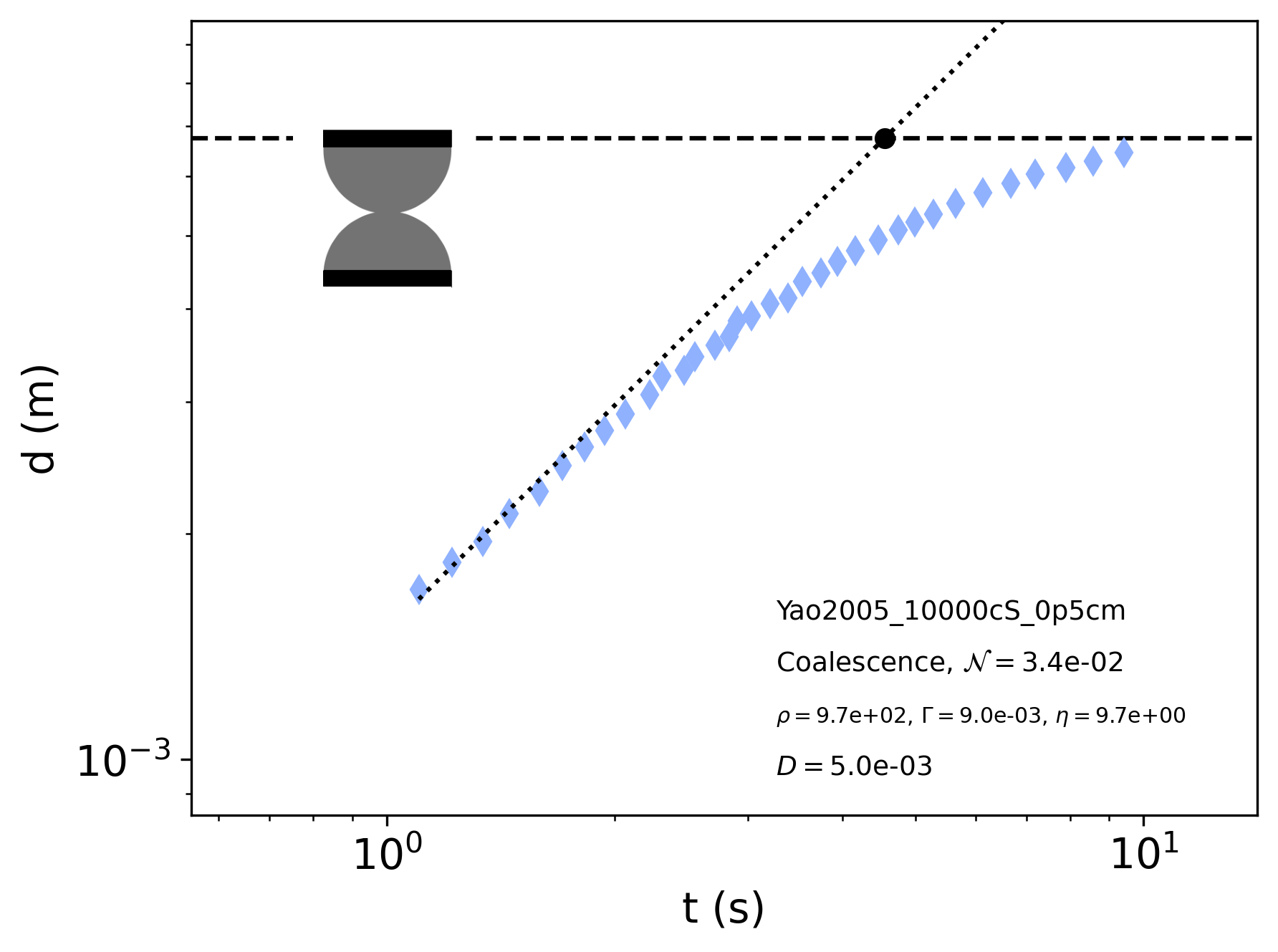}
 \end{minipage}
 & 
 \begin{minipage}{.5\textwidth} 
  \includegraphics[width=\linewidth]{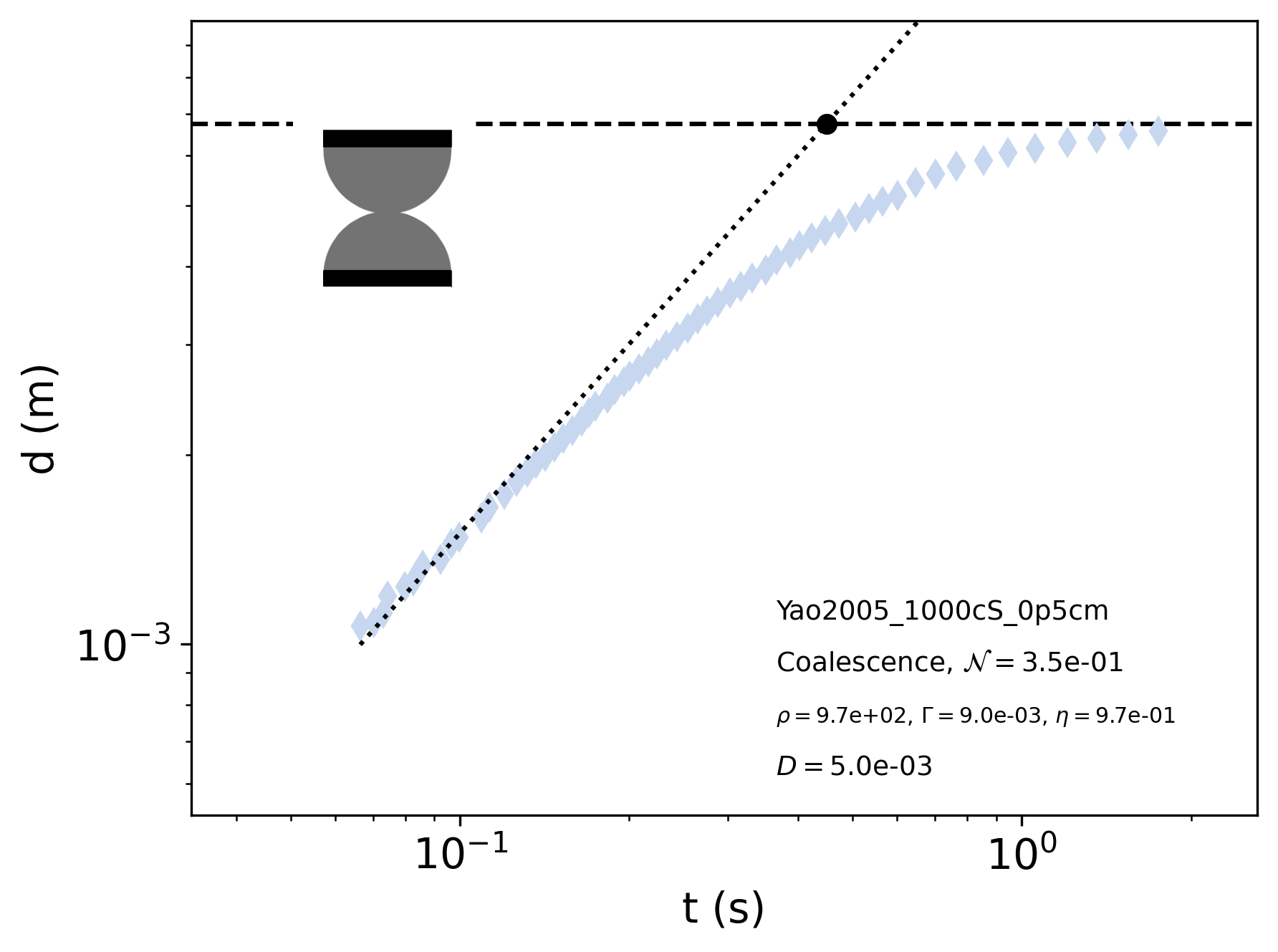} 
 \end{minipage} 
 \\ 
Silicone oil in density-matched water-alcohol mixture. & Silicone oil in density-matched water-alcohol mixture.\\ \hline \hline 
\textbf{Aarts2005 Fig2 1Pas} & \textbf{Aarts2005 Fig2 500mPas}  \\ 
 \begin{minipage}{.5\textwidth} 
   \includegraphics[width=\linewidth]{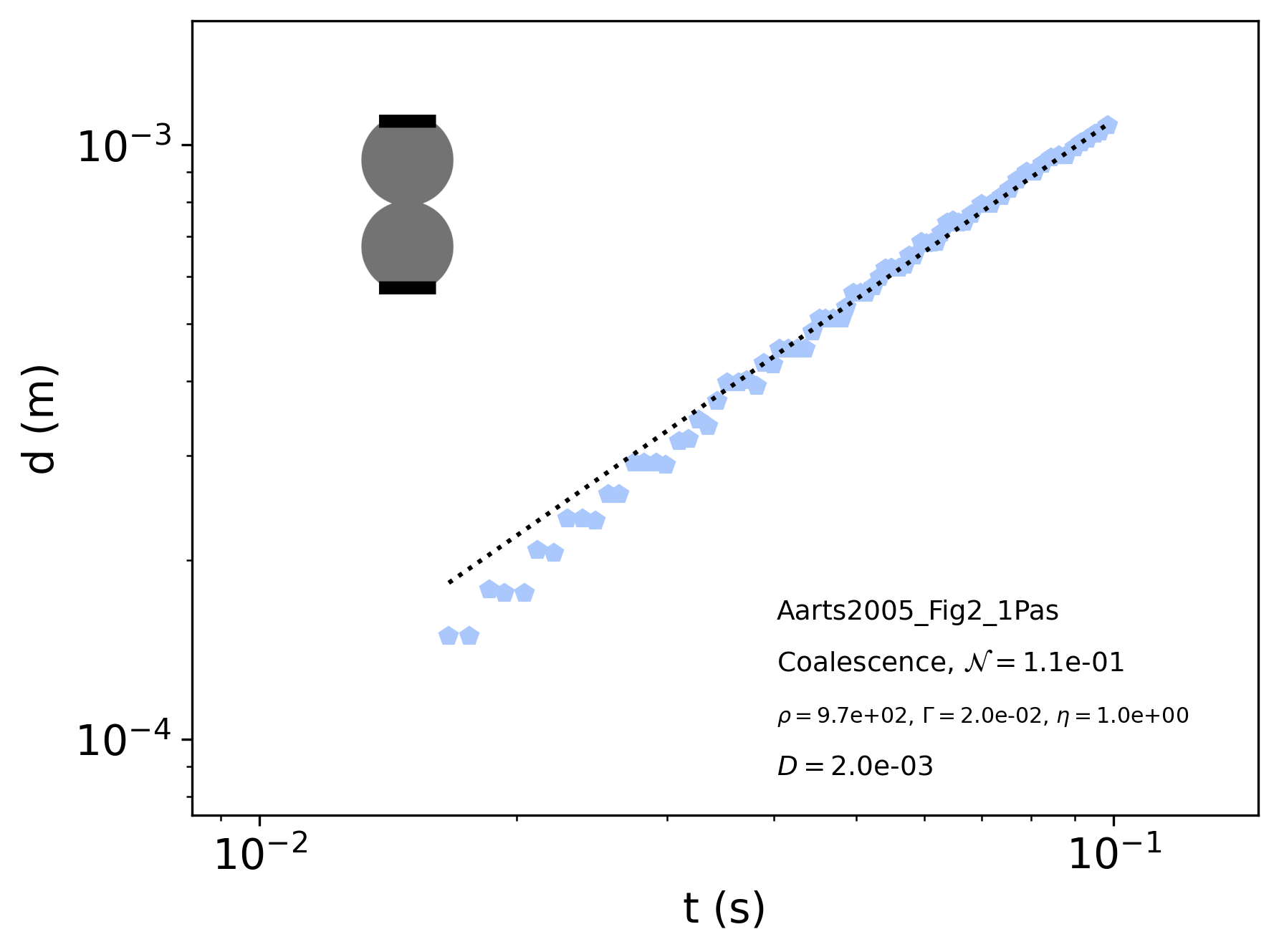} 
 \end{minipage}
 & 
 \begin{minipage}{.5\textwidth} 
   \includegraphics[width=\linewidth]{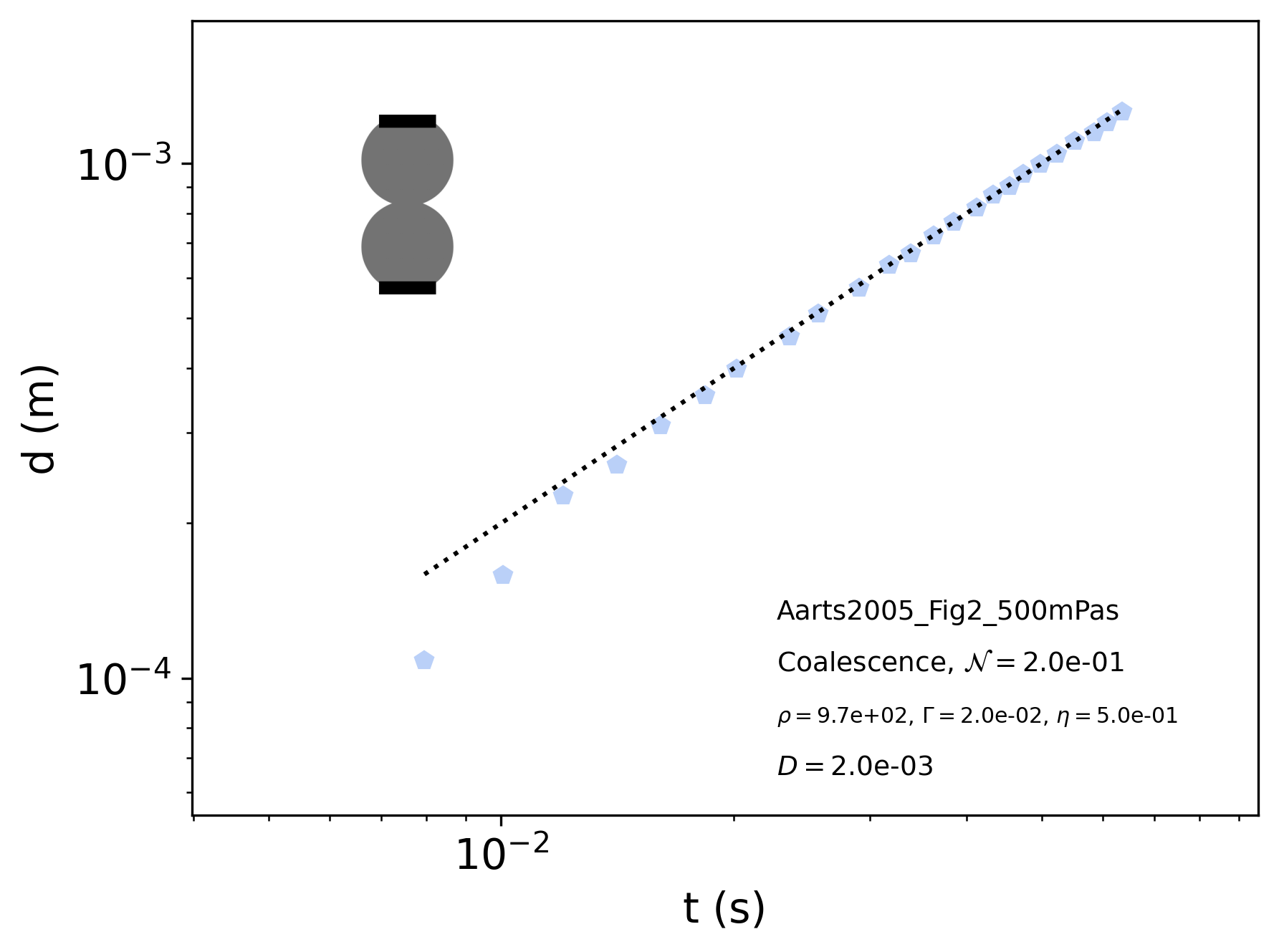} 
 \end{minipage} 
 \\ 
Silicone oil in ambient air. & Silicone oil in ambient air. \\ \hline \hline 
\textbf{Aarts2005 Fig2 300mPas} & \textbf{Aarts2005 Fig3 5mPas}  \\ 
 \begin{minipage}{.5\textwidth} 
   \includegraphics[width=\linewidth]{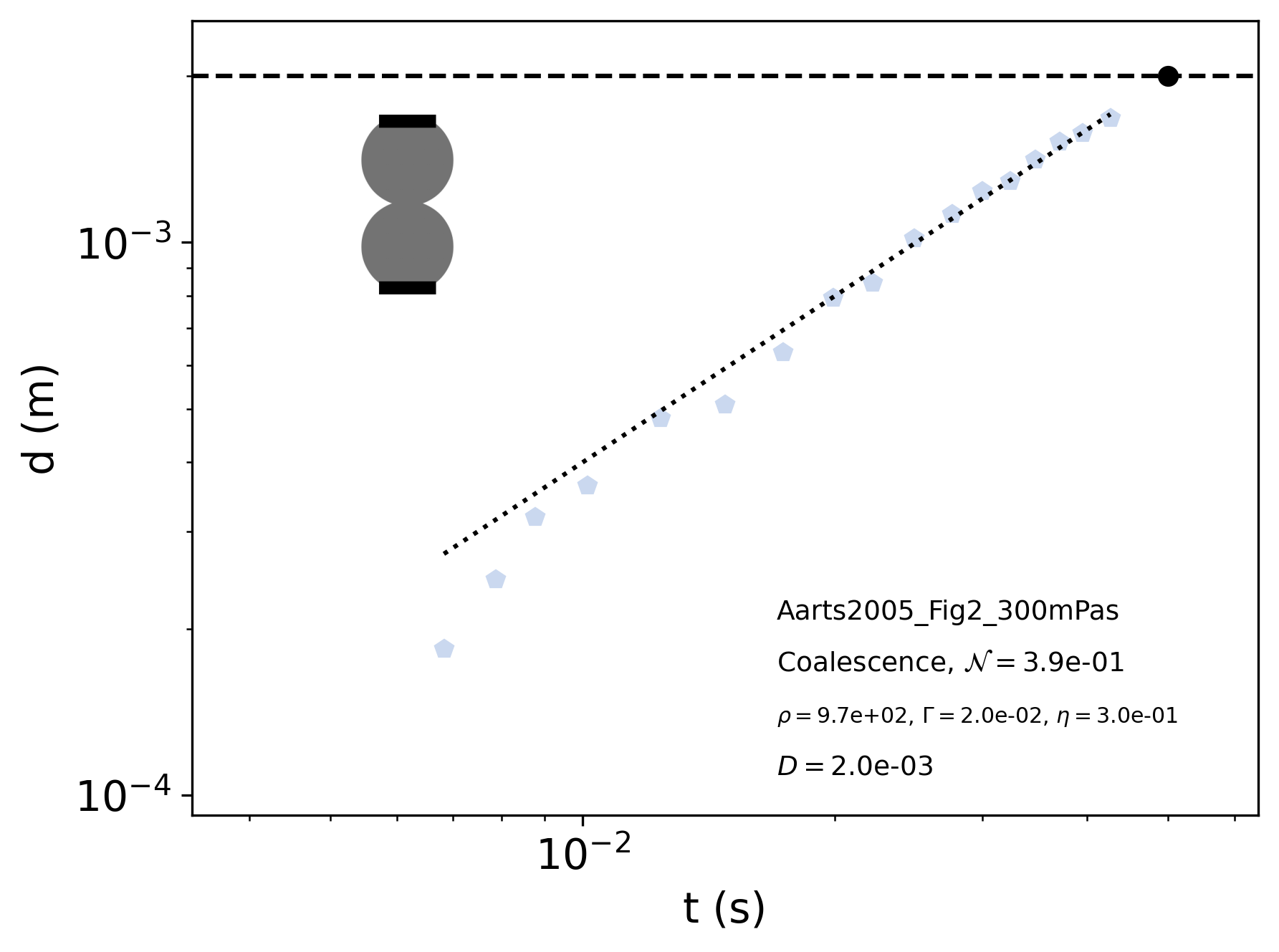} 
 \end{minipage}
 & 
 \begin{minipage}{.5\textwidth} 
   \includegraphics[width=\linewidth]{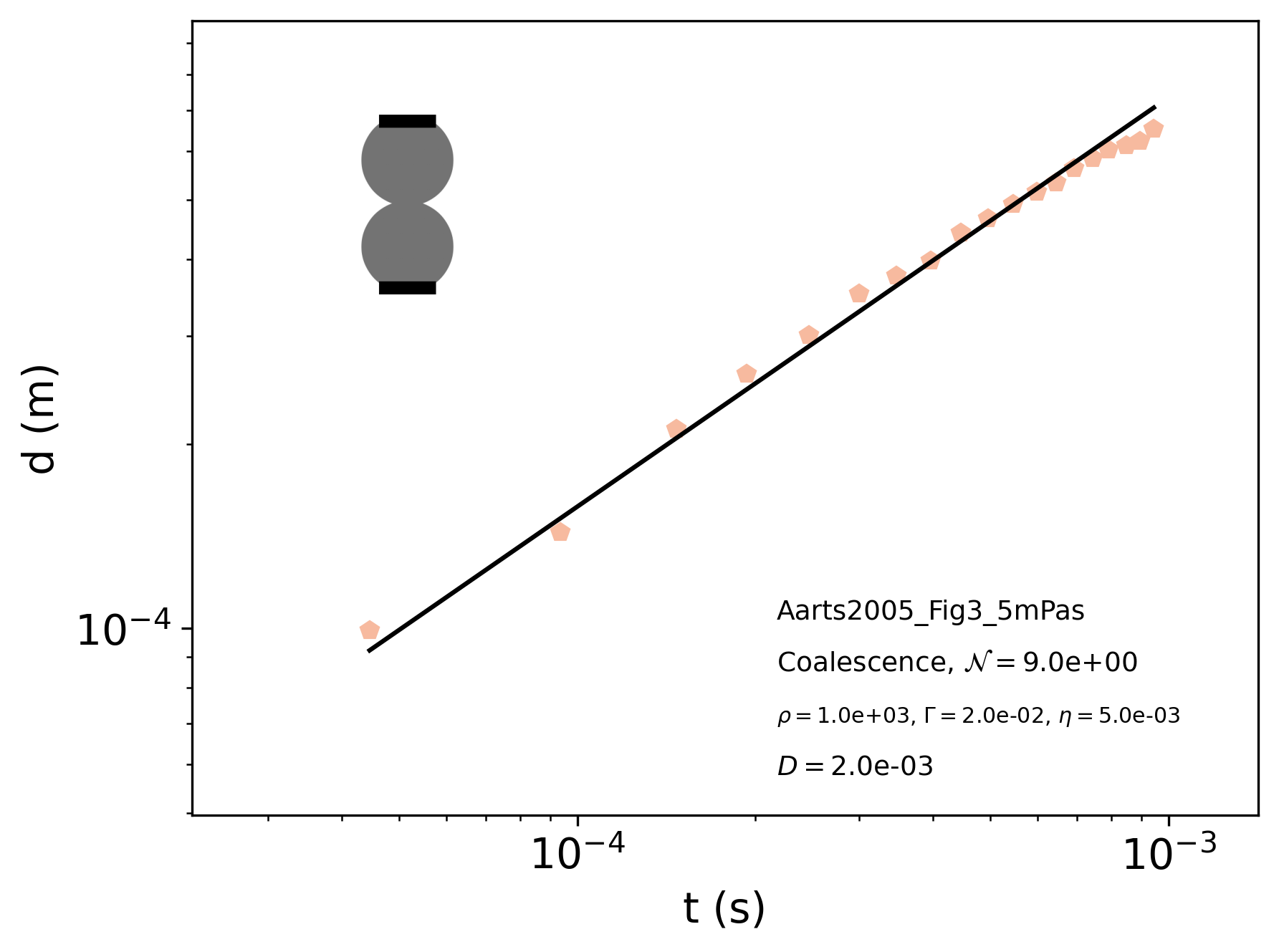} 
 \end{minipage} 
 \\ 
Silicone oil in ambient air. & Silicone oil in ambient air.\\ \hline 
\end{tabular} 
 \end{table}
 
  \begin{table} 
 \centering 
 \begin{tabular}{ | p{9cm} | p{9cm} | } 
 \hline 
 \textbf{Aarts2008 Fig9 bubb17} & \textbf{Aarts2008 Fig9 drop17}  \\ 
 \begin{minipage}{.5\textwidth} 
  \includegraphics[width=\linewidth]{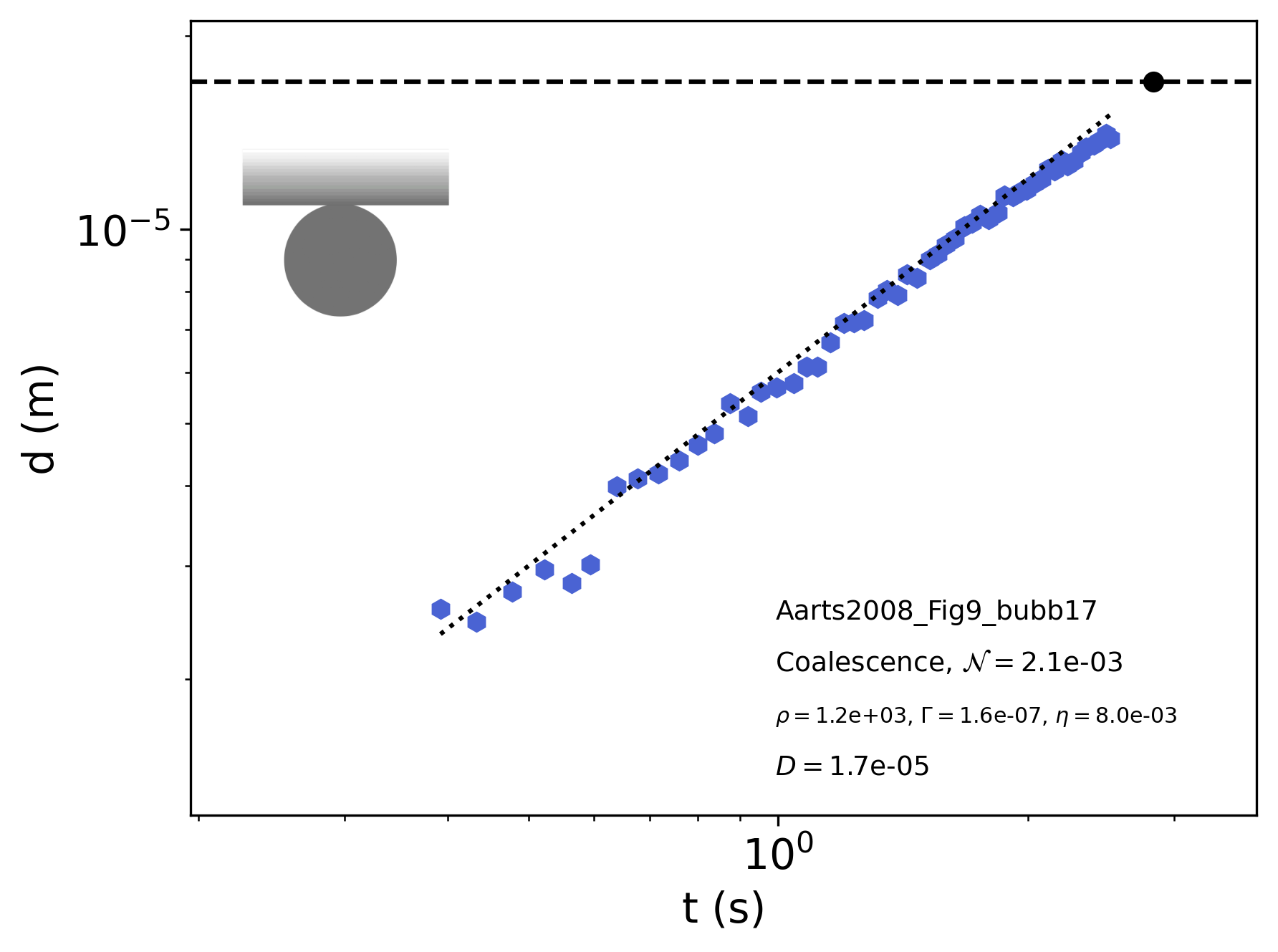}
 \end{minipage}
 & 
 \begin{minipage}{.5\textwidth} 
  \includegraphics[width=\linewidth]{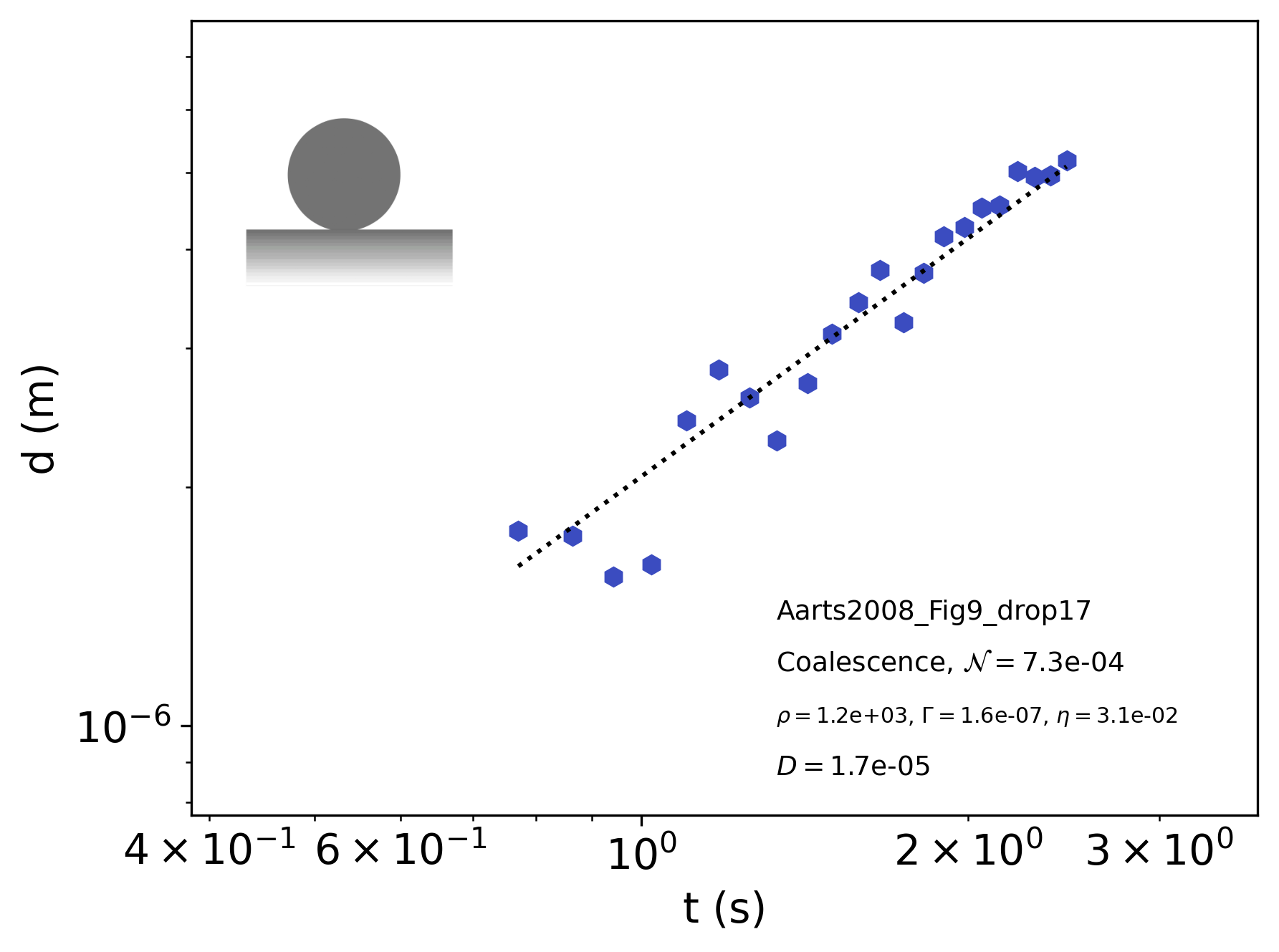} 
 \end{minipage} 
 \\ 
PMMA Poly(styrene) in Decalin, gas phase. \newline The density used is that of the PMMA colloid. & PMMA Poly(styrene) in Decalin, liquid phase. \newline The density used is that of the PMMA colloid. \\ \hline \hline 
\textbf{Paulsen2011 Fig2 230} & \textbf{Rahman2019 Fig6 6p65}  \\ 
 \begin{minipage}{.5\textwidth} 
   \includegraphics[width=\linewidth]{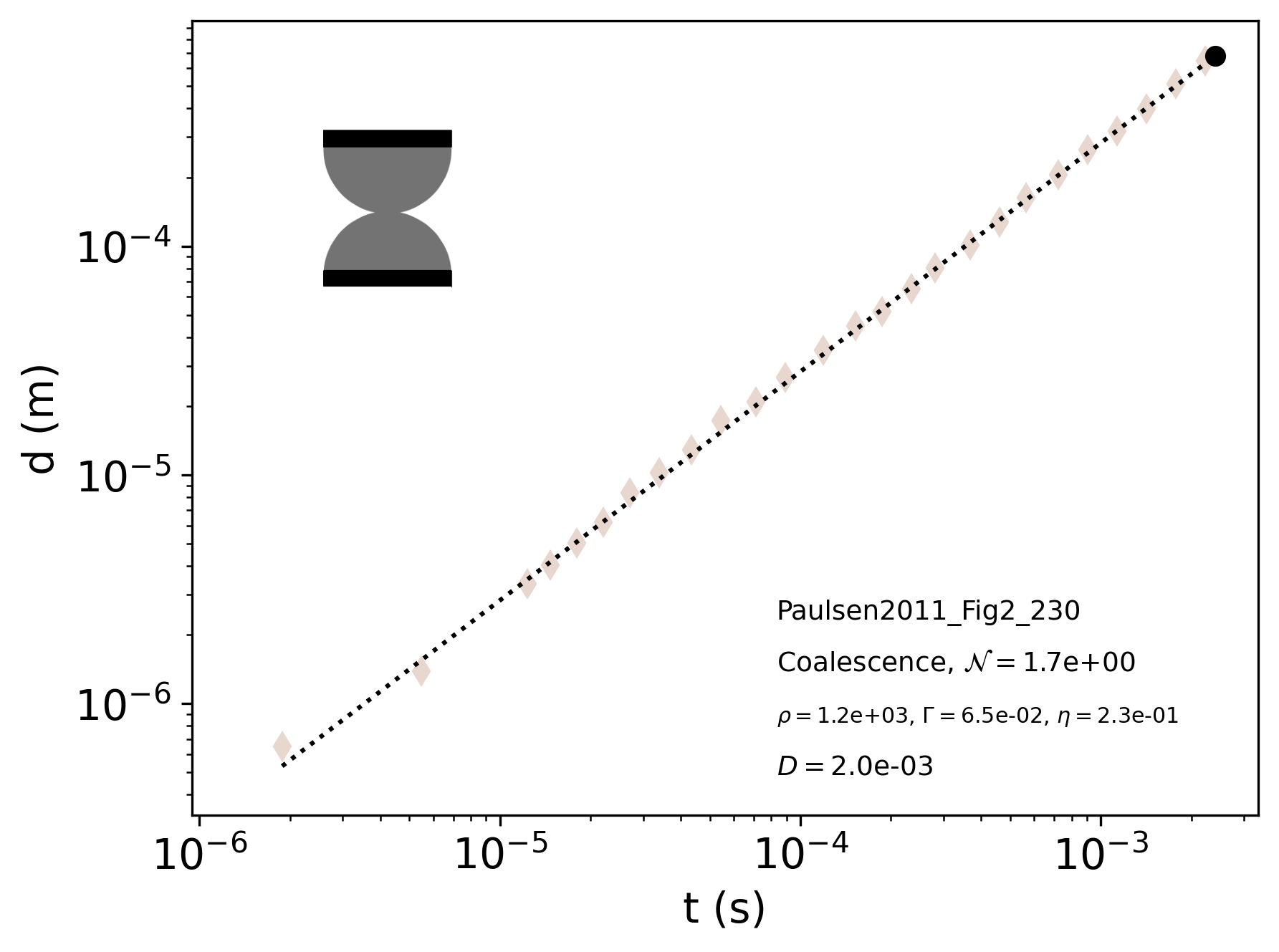} 
 \end{minipage}
 & 
 \begin{minipage}{.5\textwidth} 
   \includegraphics[width=\linewidth]{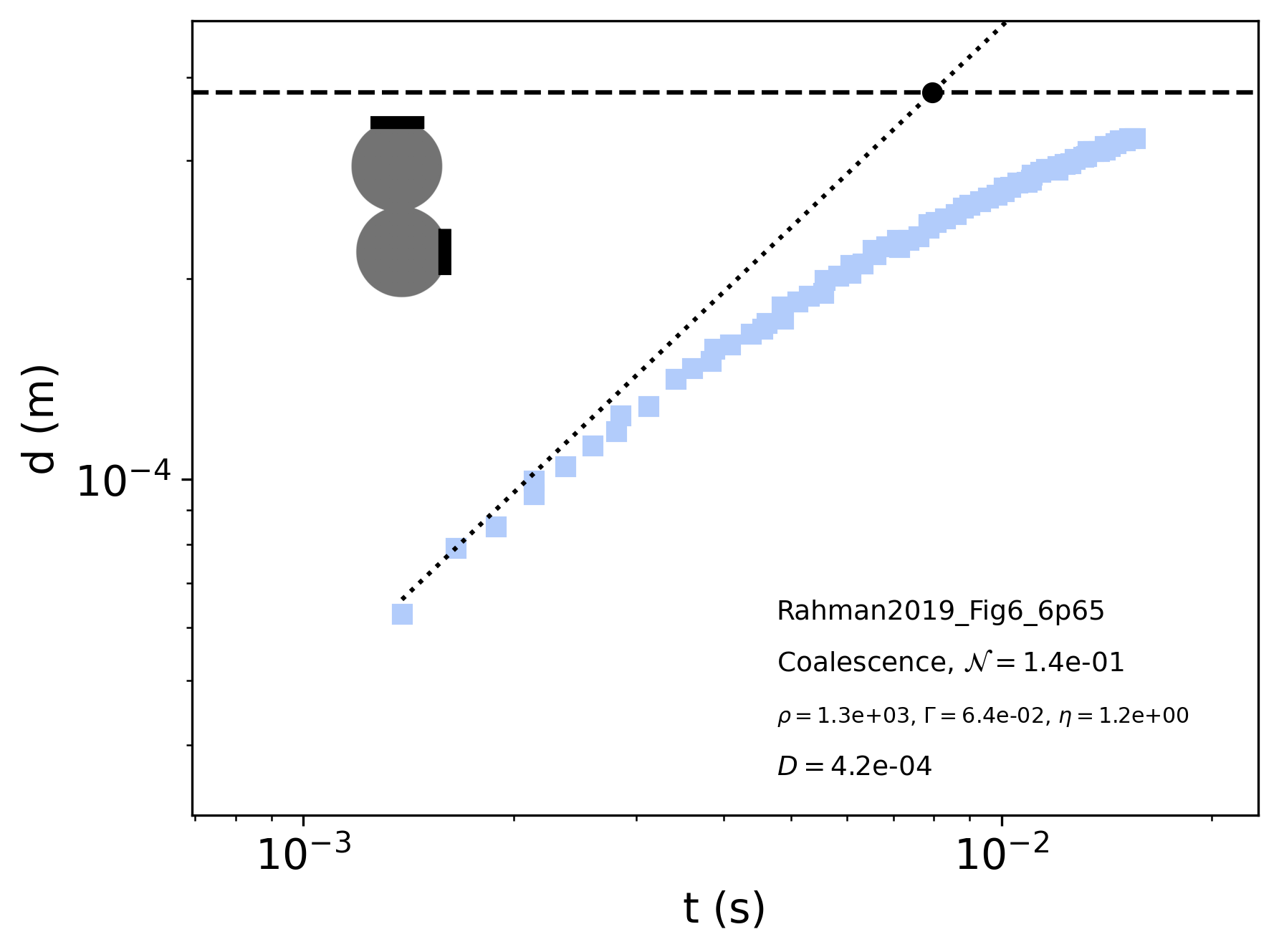} 
 \end{minipage} 
 \\ 
Glycerol-water-NaCl mixture in ambient air.  & Water-glycerol mixture in ambient air. \\ \hline \hline 
\textbf{Eddi2013b Fig3a 81} & \textbf{Chen1997 Fig7}  \\ 
 \begin{minipage}{.5\textwidth} 
   \includegraphics[width=\linewidth]{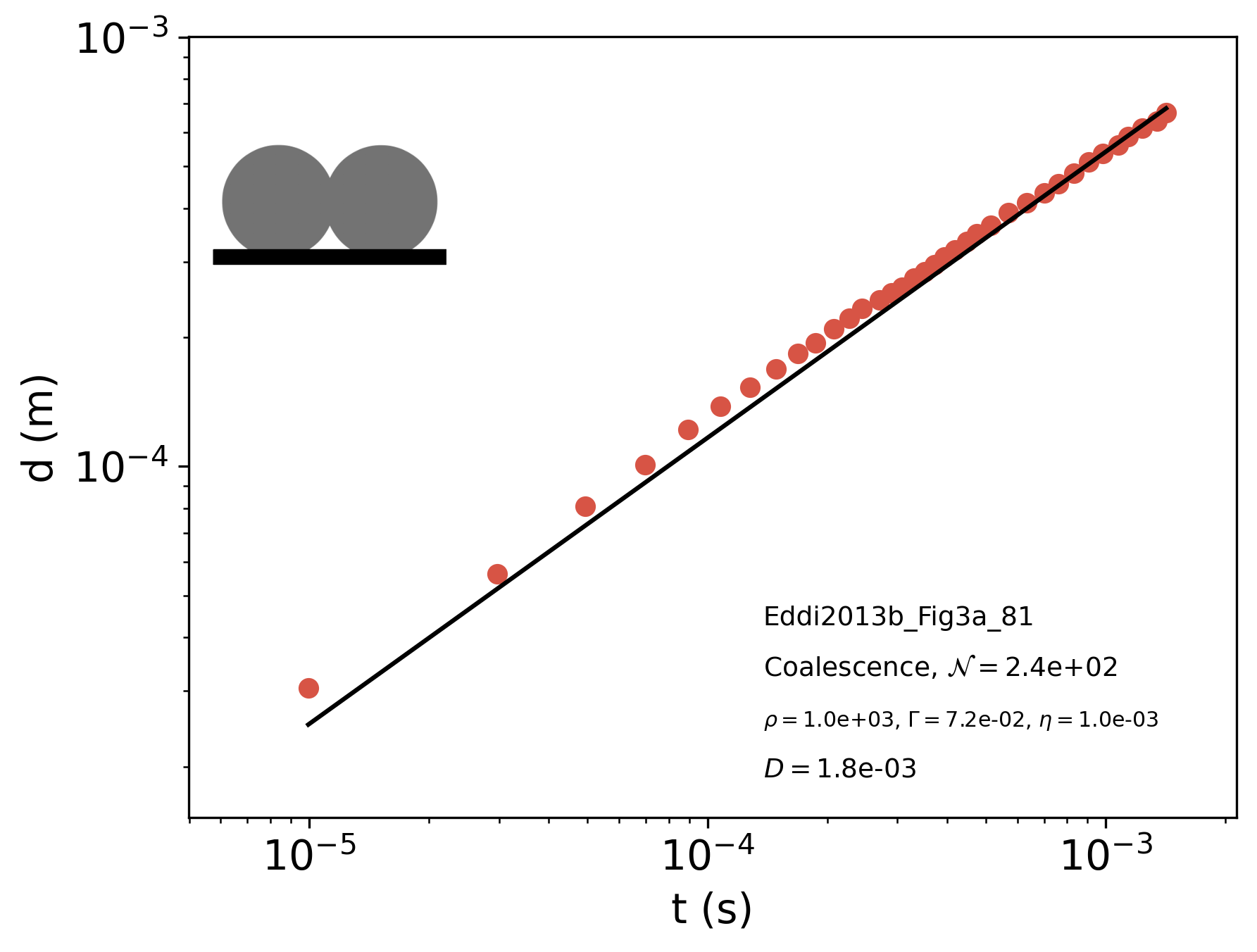} 
 \end{minipage}
 & 
 \begin{minipage}{.5\textwidth} 
   \includegraphics[width=\linewidth]{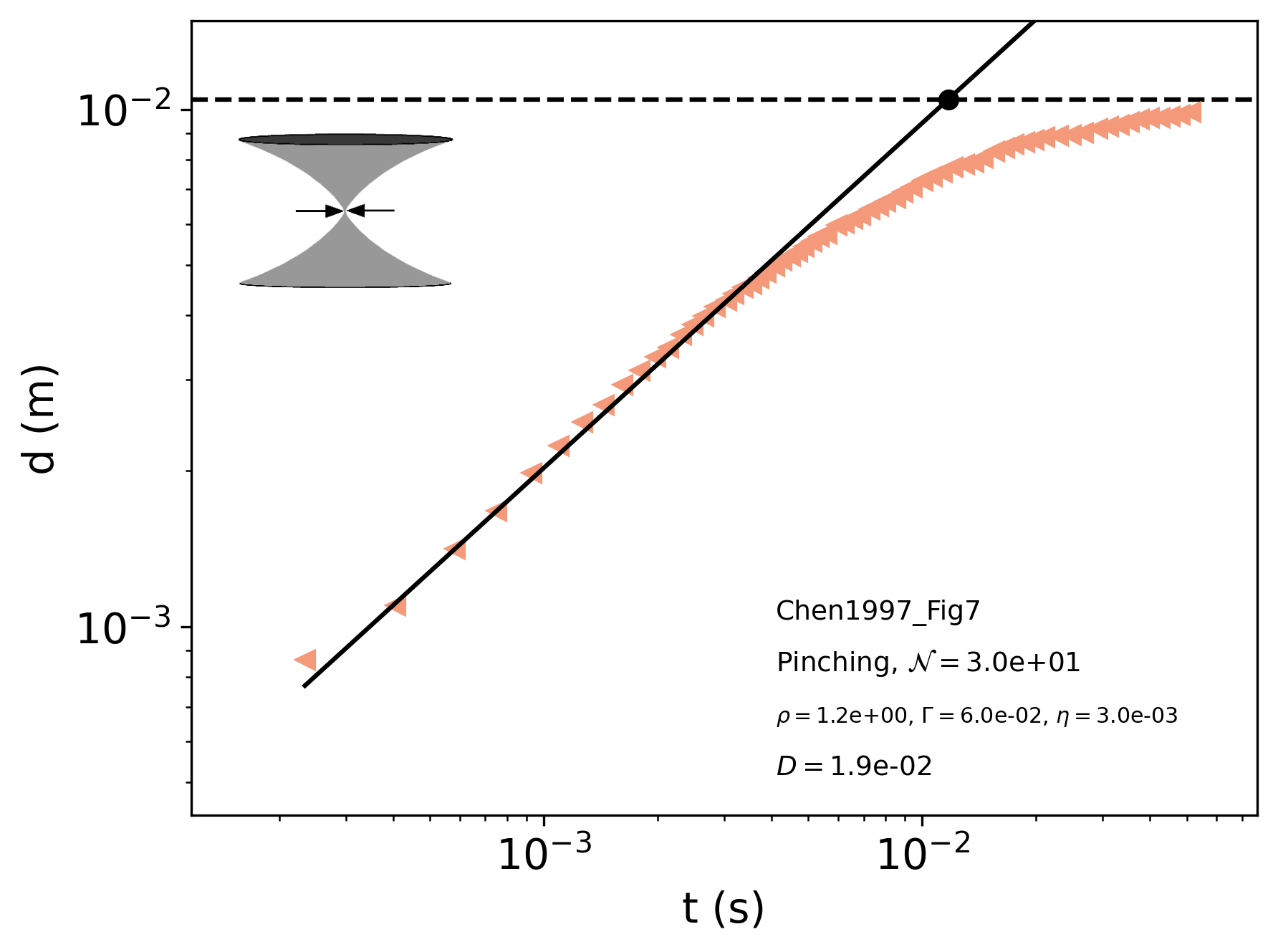} 
 \end{minipage} 
 \\ 
Water in ambient air. & Catenoid soap film in ambient air.\newline Density is that of air, viscosity is that of the soap film.\\ \hline 
\end{tabular} 
 \end{table}
 
   \begin{table} 
 \centering 
 \begin{tabular}{ | p{9cm} | p{9cm} | } 
 \hline 
 \textbf{McKinley2000 Fig4} & \textbf{Chen2002 Fig3}  \\ 
 \begin{minipage}{.5\textwidth} 
  \includegraphics[width=\linewidth]{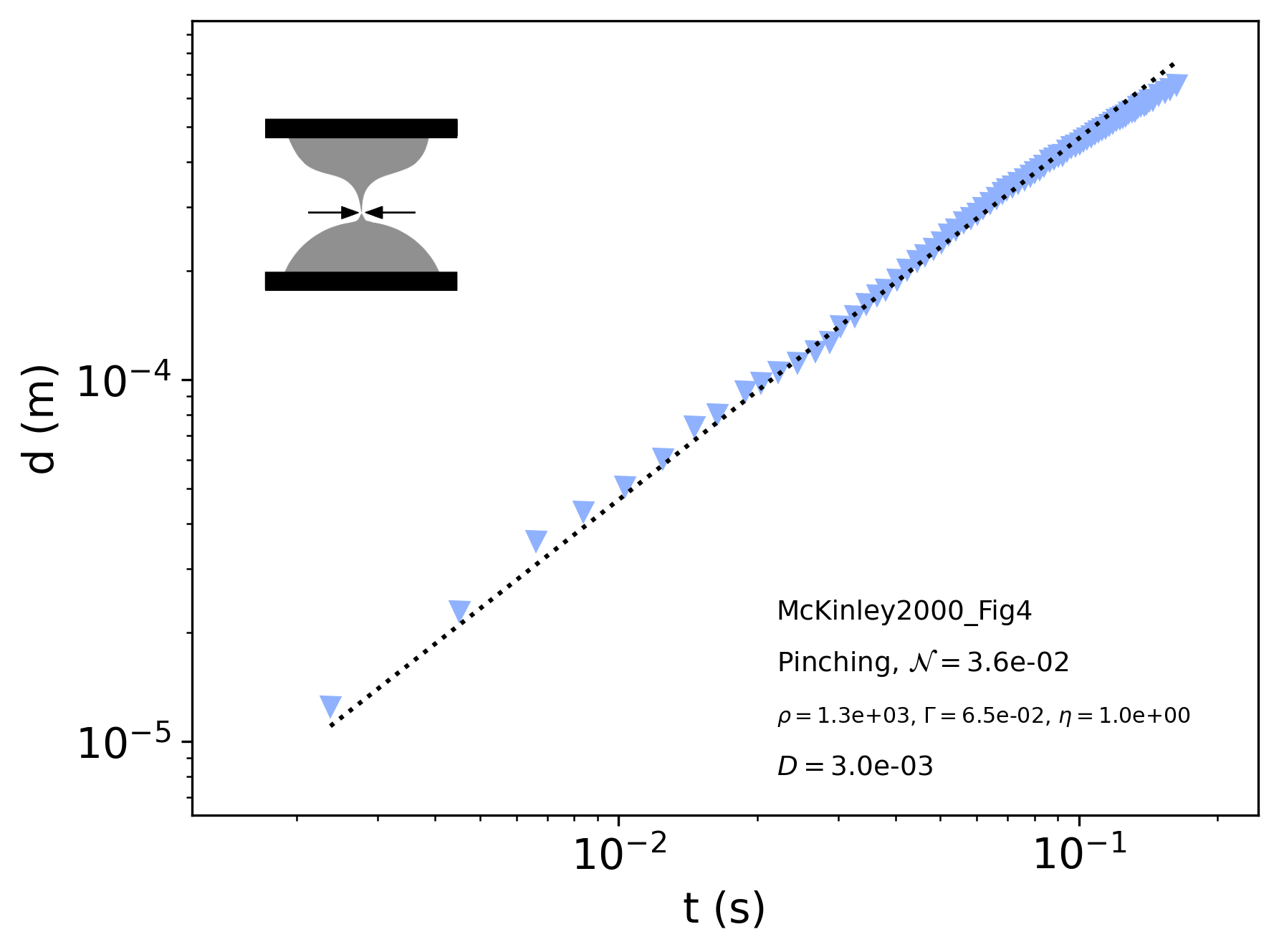}
 \end{minipage}
 & 
 \begin{minipage}{.5\textwidth} 
  \includegraphics[width=\linewidth]{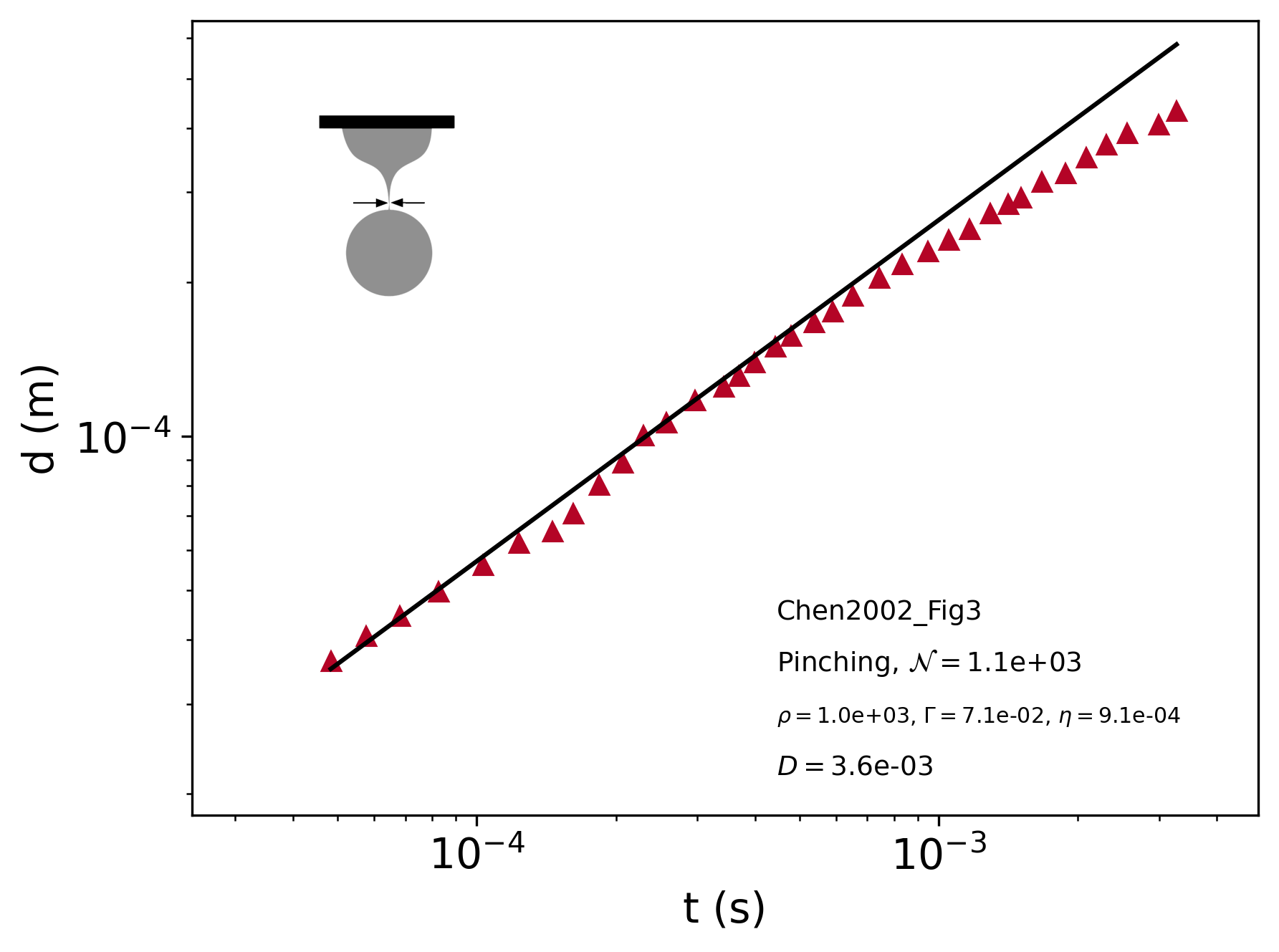} 
 \end{minipage} 
 \\ 
Glycerol in ambient air. & Water in ambient air. \\ \hline \hline 
\textbf{Burton2004 Fig5} & \textbf{Burton2005 1011}  \\ 
 \begin{minipage}{.5\textwidth} 
   \includegraphics[width=\linewidth]{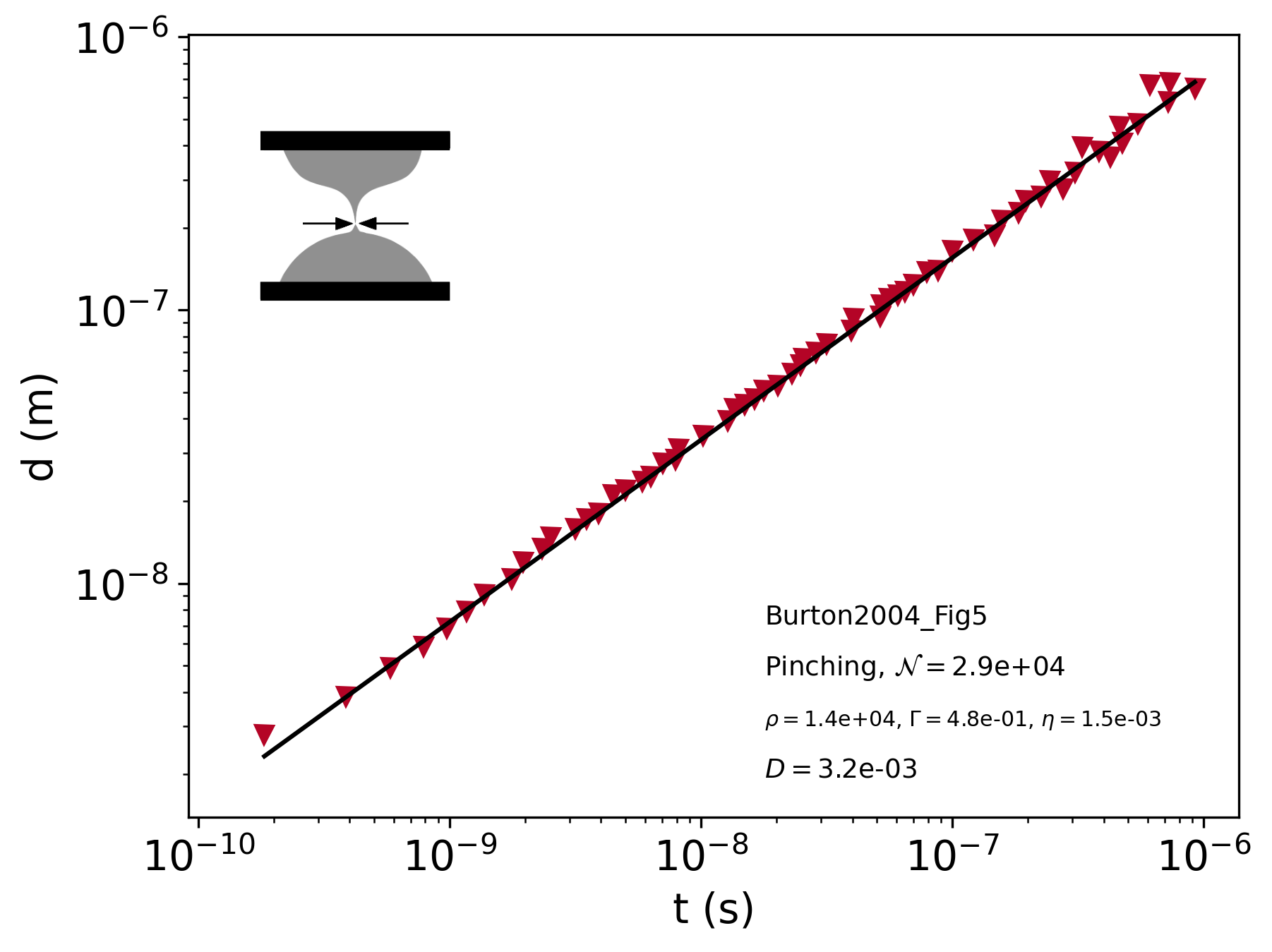} 
 \end{minipage}
 & 
 \begin{minipage}{.5\textwidth} 
   \includegraphics[width=\linewidth]{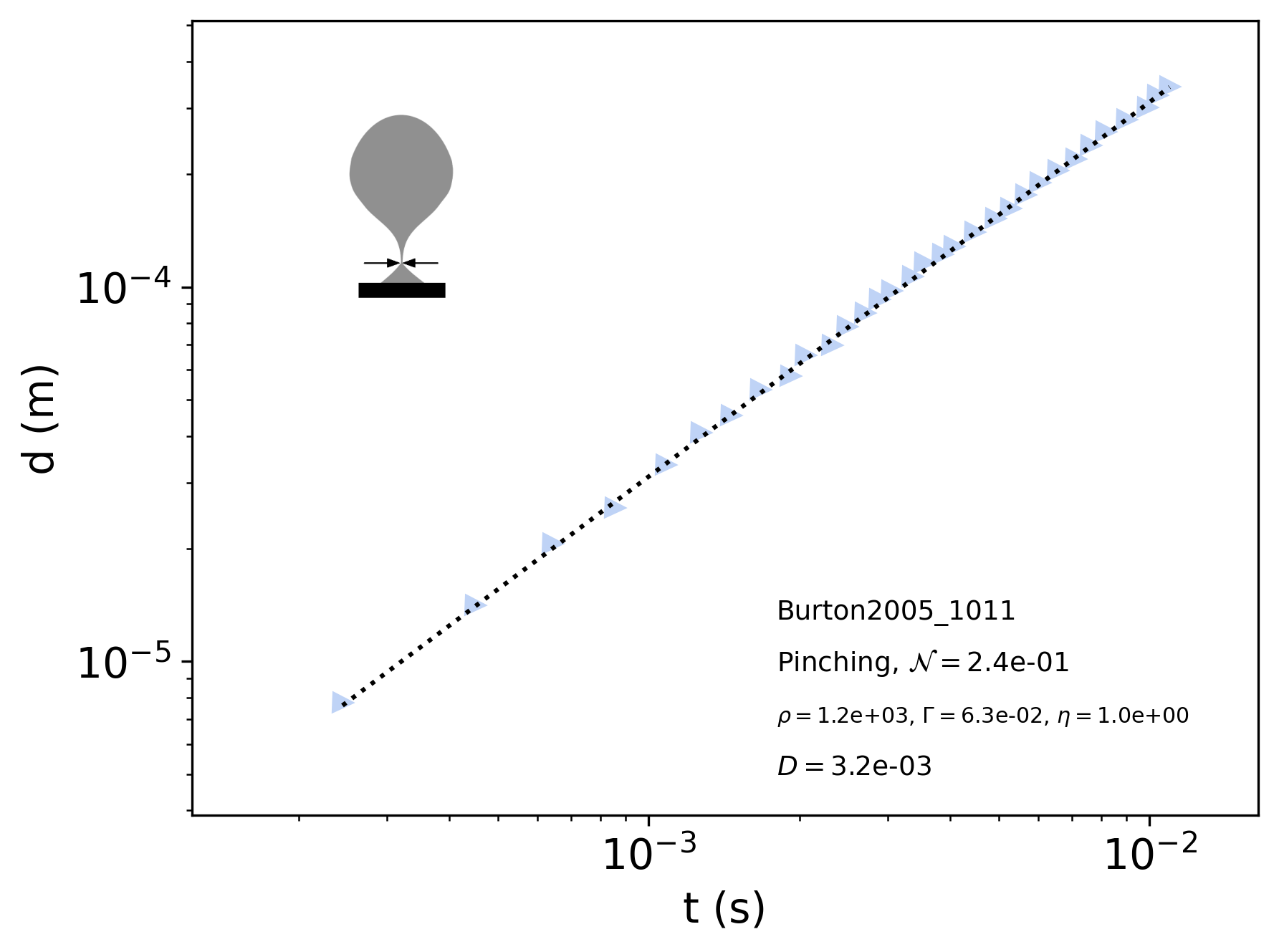} 
 \end{minipage} 
 \\ 
Mercury in ambient air.  & Air bubble in water-glycerol mixture.\newline Density and viscosity are that of the outer fluid. \\ \hline \hline 
\textbf{Bolanos2009 Fig6 O2} & \textbf{Bolanos2009 Fig7 O8}  \\ 
 \begin{minipage}{.5\textwidth} 
   \includegraphics[width=\linewidth]{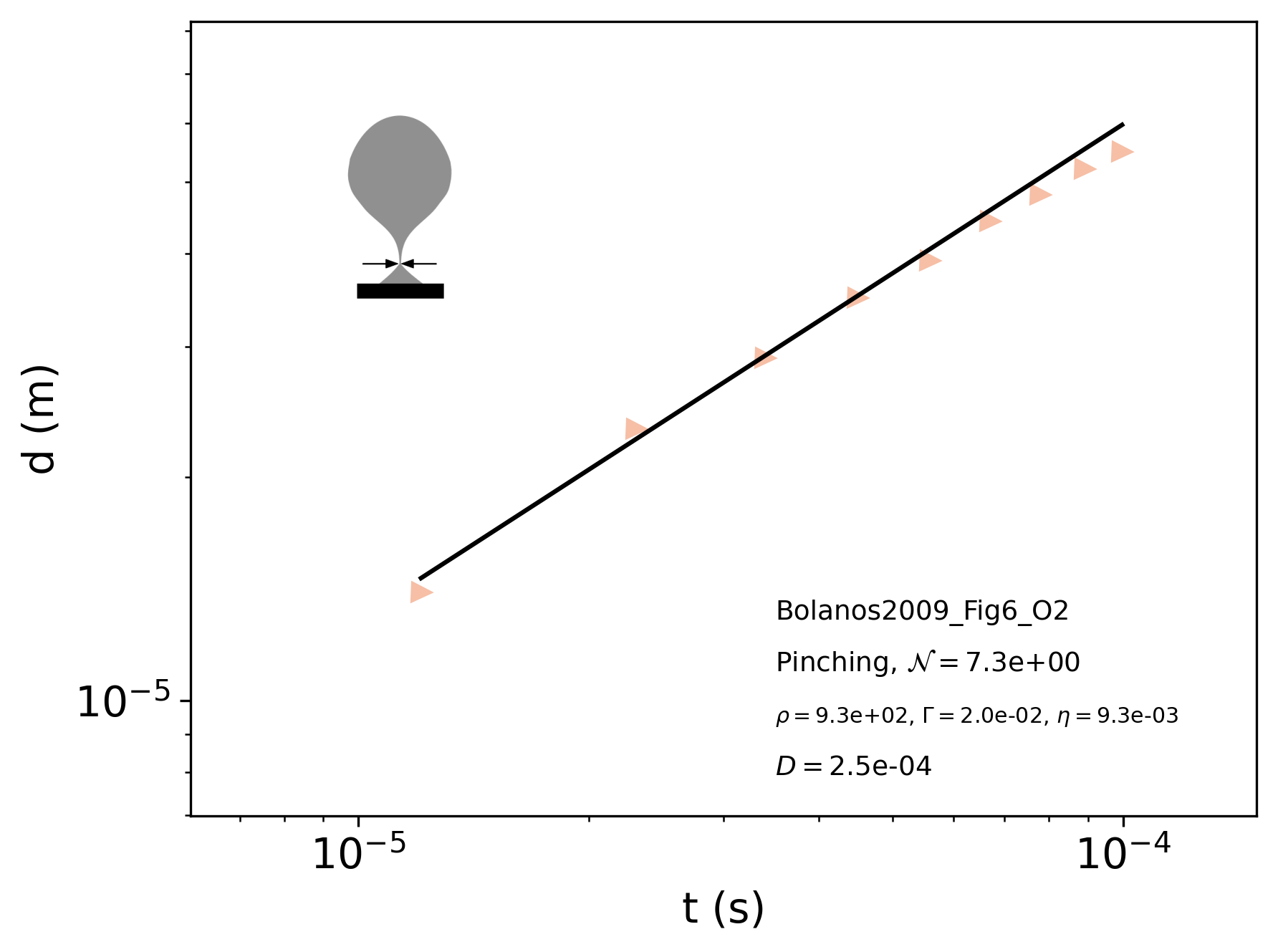} 
 \end{minipage}
 & 
 \begin{minipage}{.5\textwidth} 
   \includegraphics[width=\linewidth]{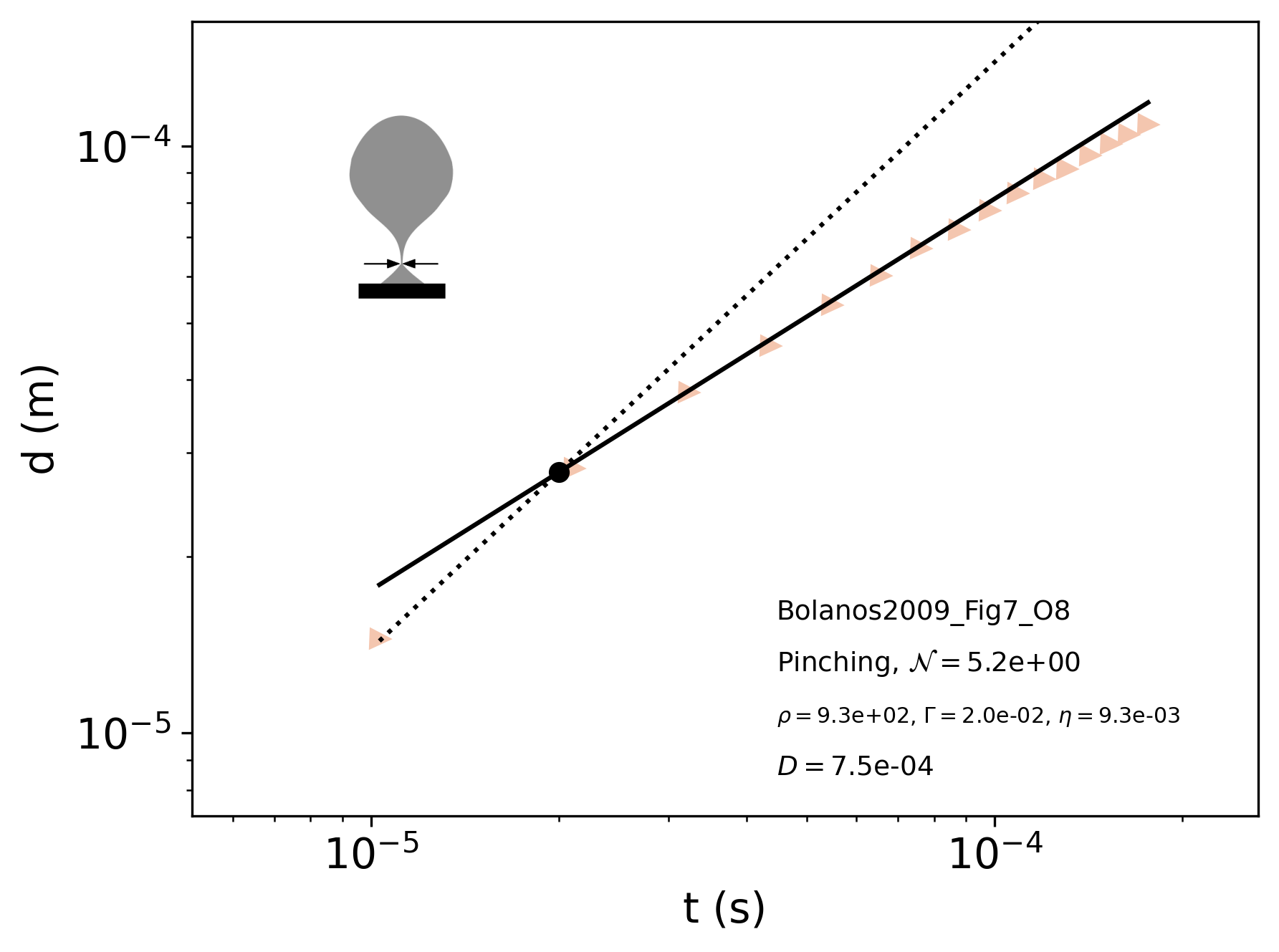} 
 \end{minipage} 
 \\ 
Air bubble in silicone oil. \newline Density and viscosity are that of the outer fluid. & Air bubble in silicone oil. \newline Density and viscosity are that of the outer fluid.\\ \hline 
\end{tabular} 
 \end{table}
 
 \begin{table} 
 \centering 
 \begin{tabular}{ | p{9cm} | p{9cm} | } 
 \hline 
 \textbf{Bolanos2009 Fig7 O9} & \textbf{Bolanos2009 Fig8 G1}  \\ 
 \begin{minipage}{.5\textwidth} 
  \includegraphics[width=\linewidth]{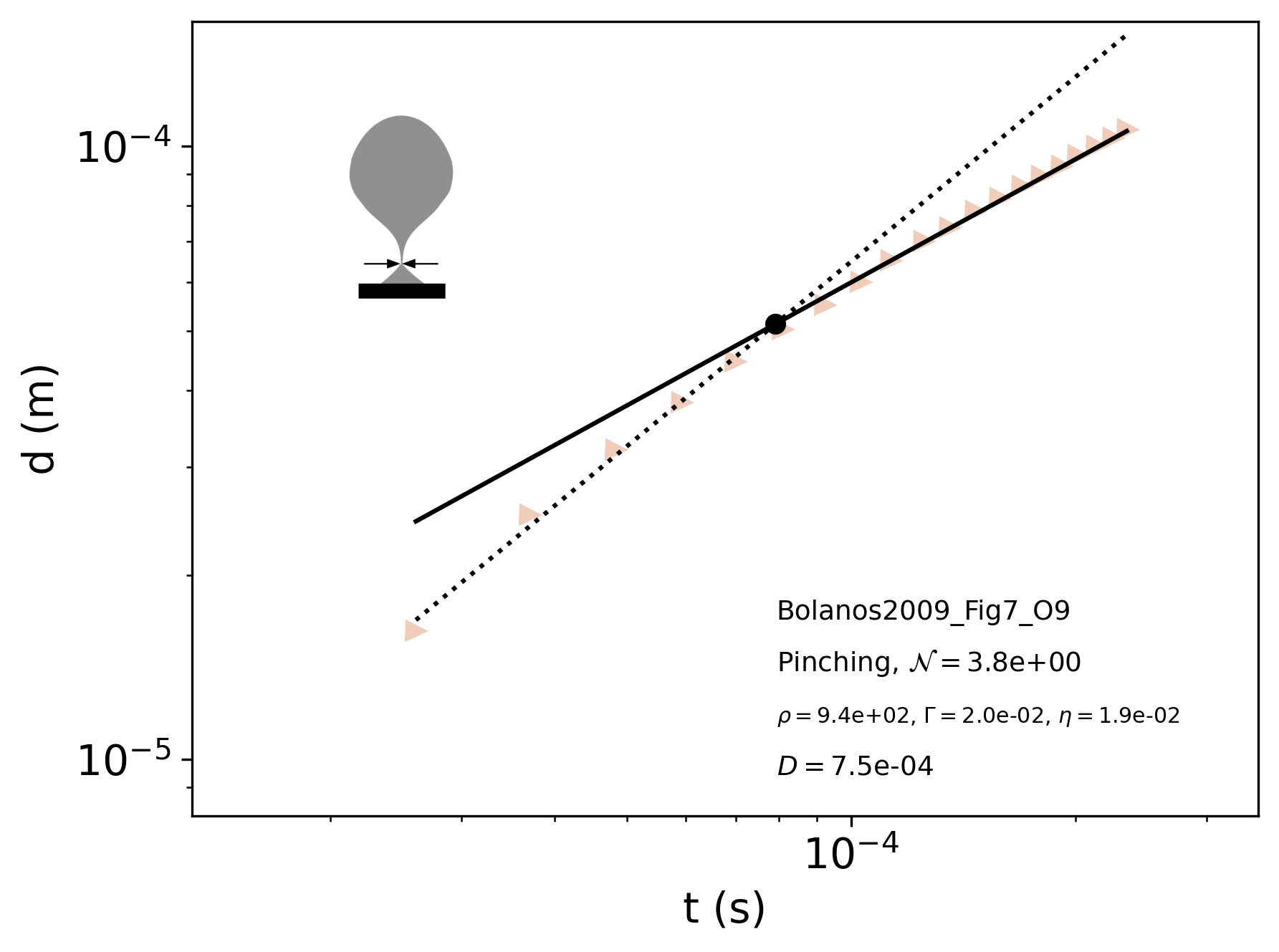}
 \end{minipage}
 & 
 \begin{minipage}{.5\textwidth} 
  \includegraphics[width=\linewidth]{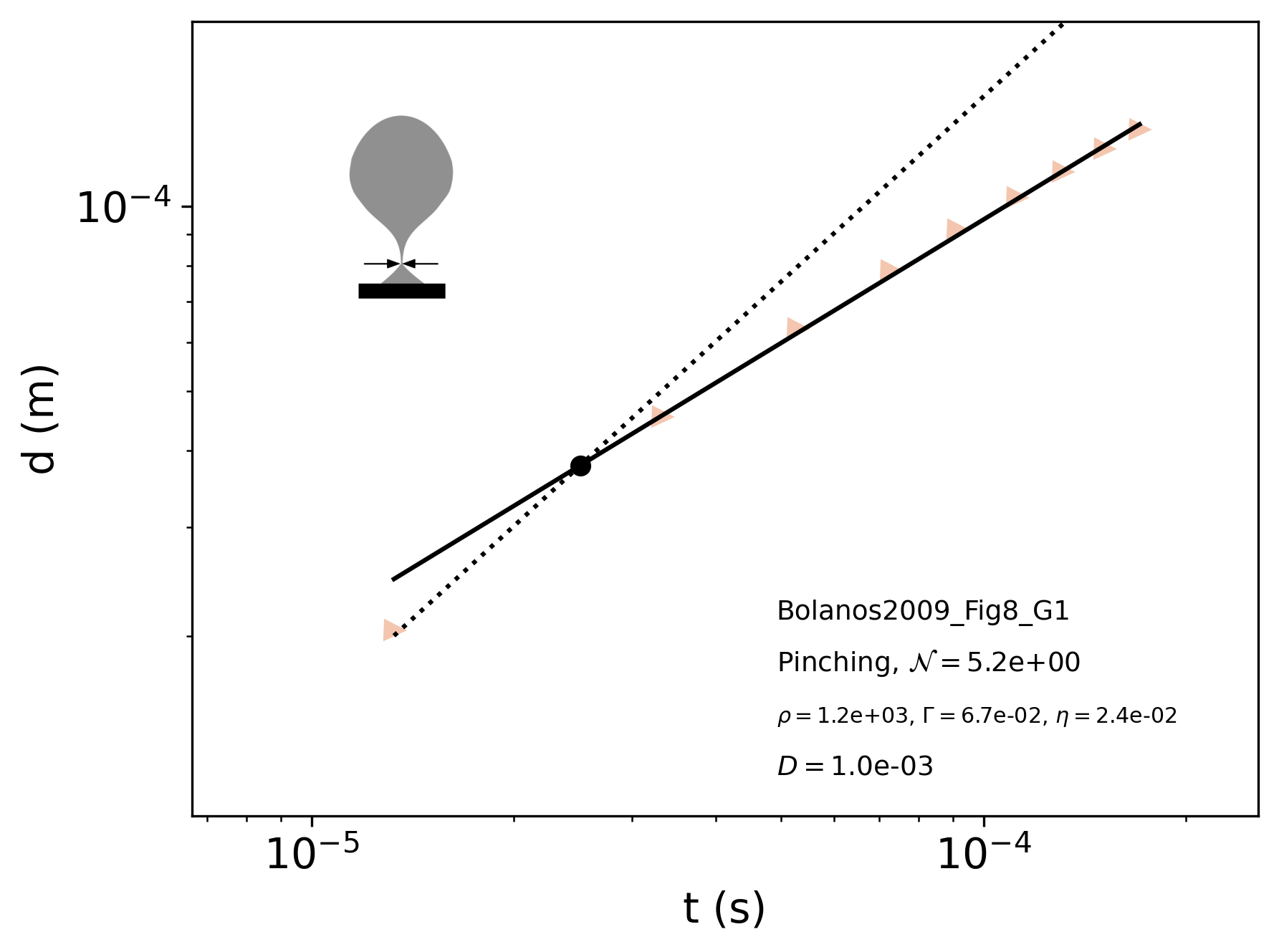} 
 \end{minipage} 
 \\ 
Air bubble in silicone oil. \newline Density and viscosity are that of the outer fluid. & Air bubble in water-glycerol mixture. \newline Density and viscosity are that of the outer fluid.\\ \hline  \hline 
\textbf{Bolanos2009 Fig8 G2} & \textbf{Bolanos2009 Fig8 G4}  \\ 
 \begin{minipage}{.5\textwidth} 
   \includegraphics[width=\linewidth]{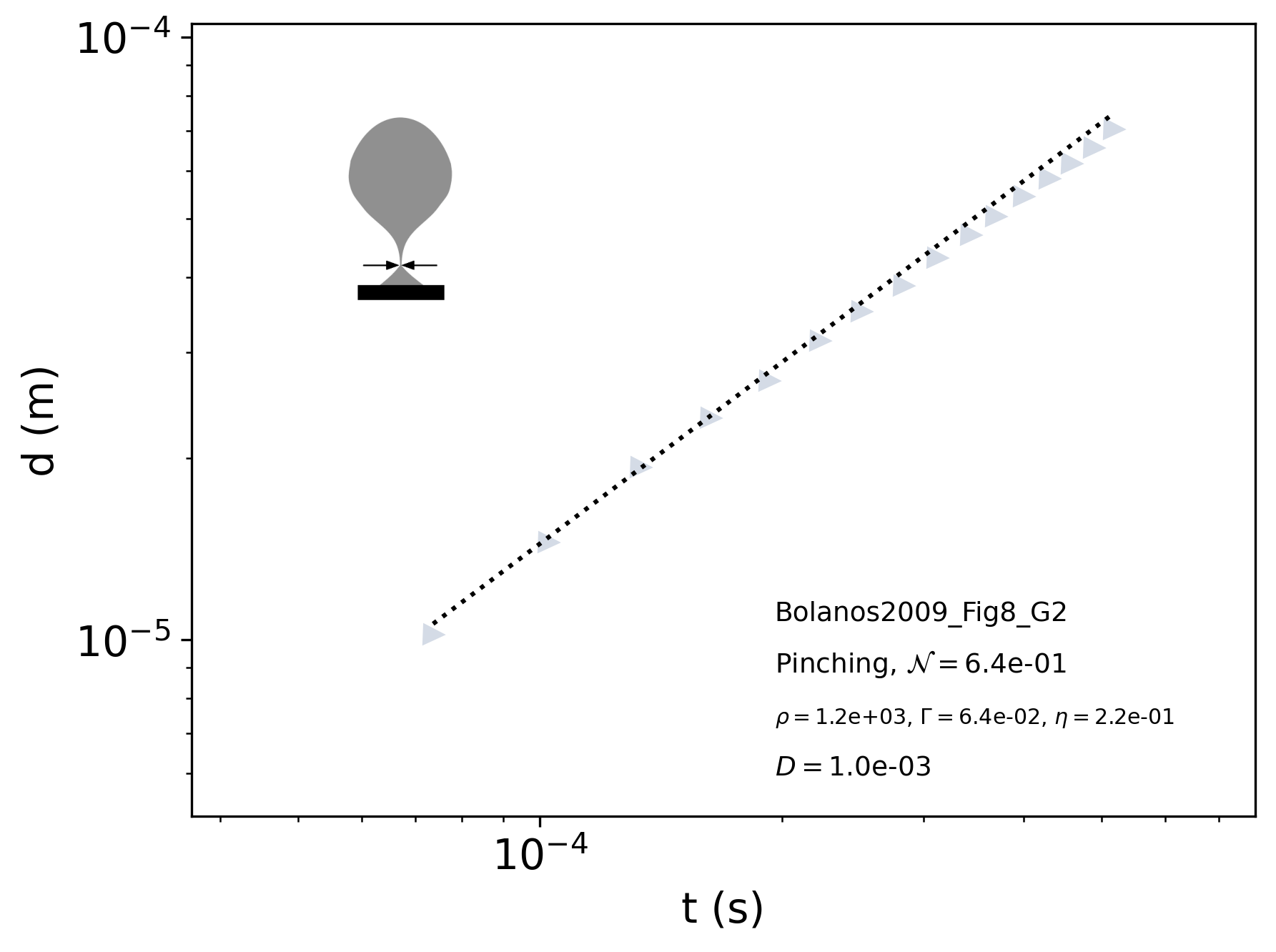} 
 \end{minipage}
 & 
 \begin{minipage}{.5\textwidth} 
   \includegraphics[width=\linewidth]{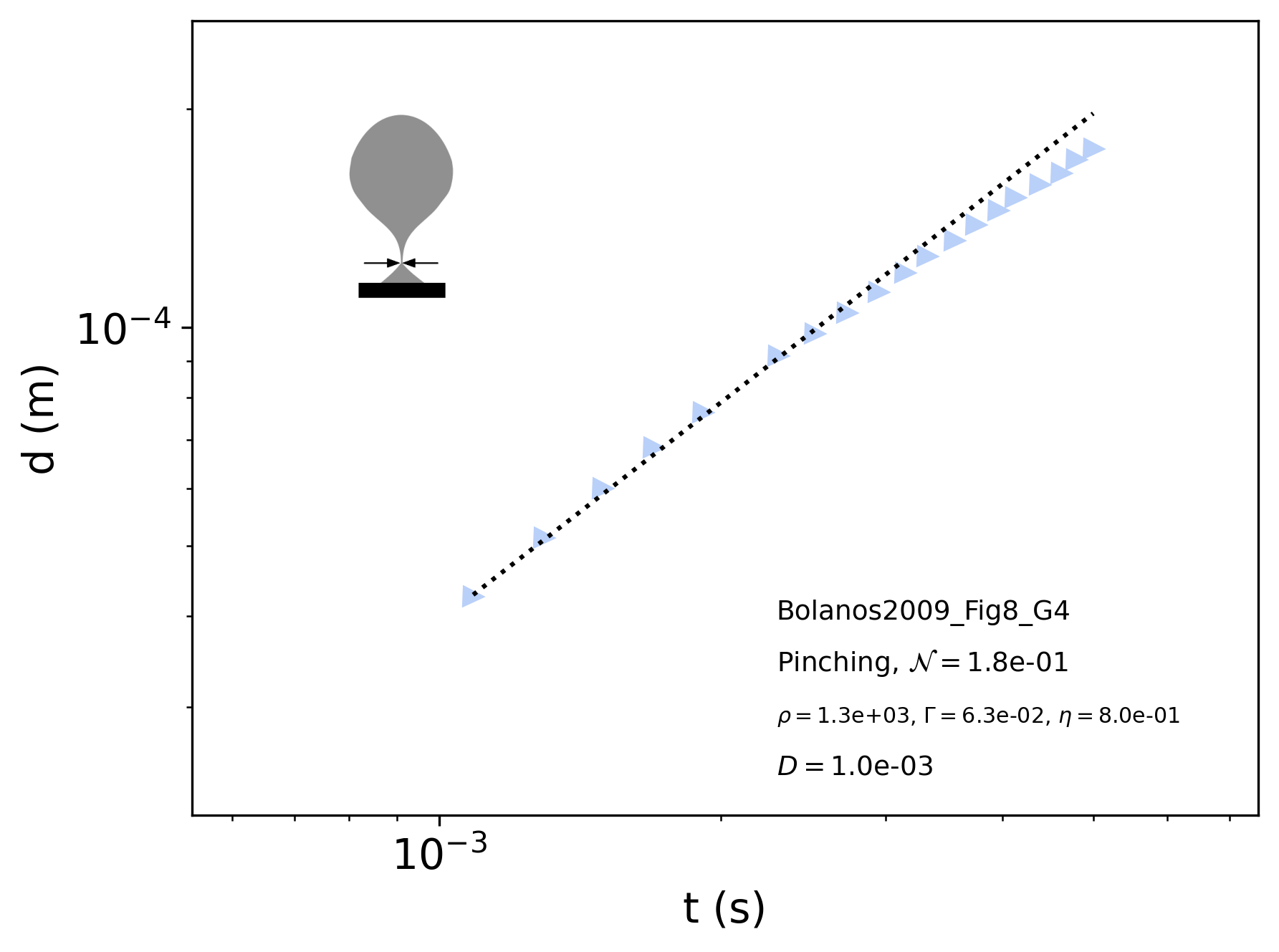} 
 \end{minipage} 
 \\ 
Air bubble in water-glycerol mixture. \newline Density and viscosity are that of the outer fluid. & Air bubble in water-glycerol mixture. \newline Density and viscosity are that of the outer fluid.\\ \hline  \hline 
\textbf{Bolanos2009 Fig9 G5} & \textbf{Bolanos2009 Fig9 G6}  \\ 
 \begin{minipage}{.5\textwidth} 
   \includegraphics[width=\linewidth]{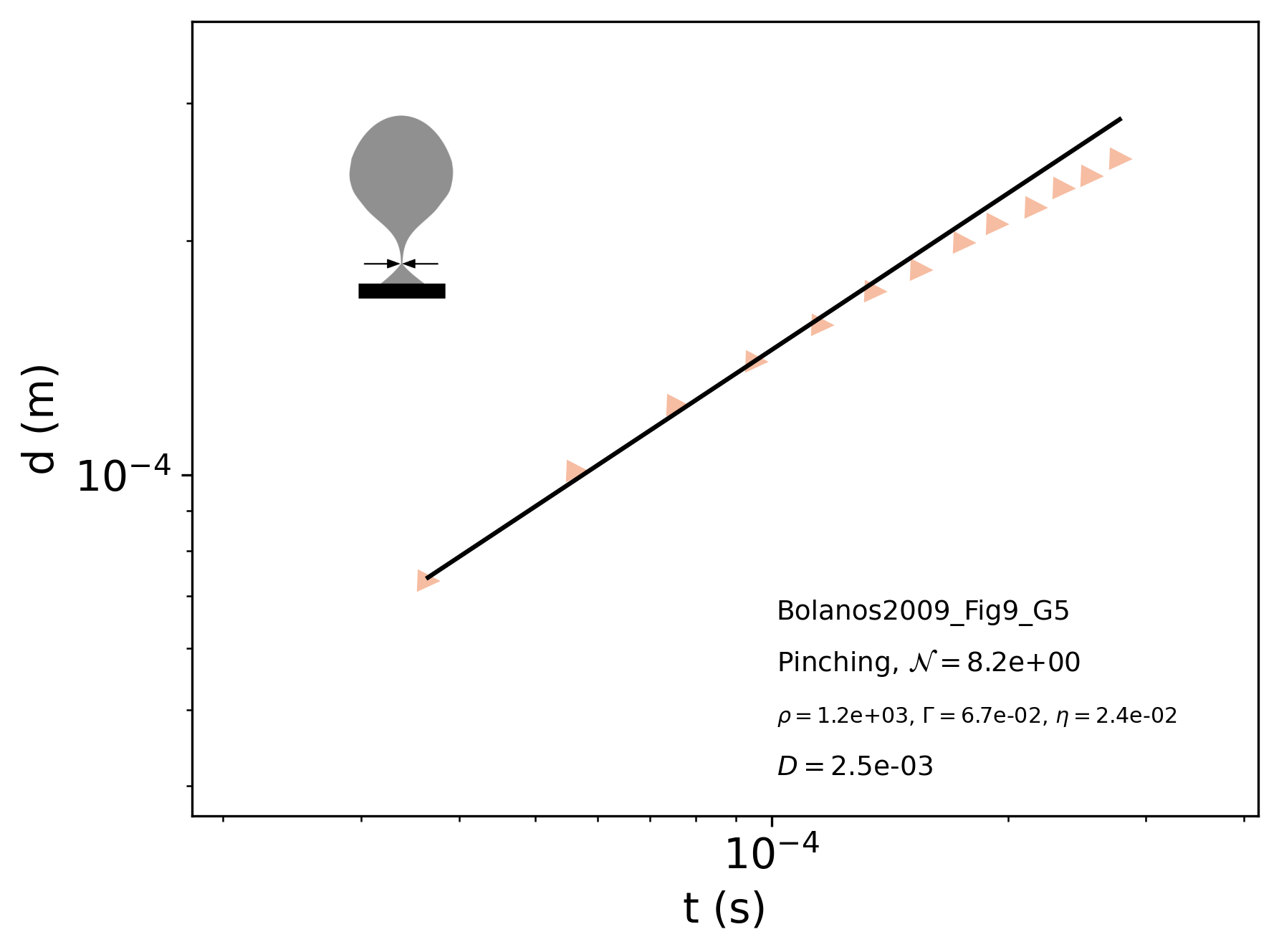} 
 \end{minipage}
 & 
 \begin{minipage}{.5\textwidth} 
   \includegraphics[width=\linewidth]{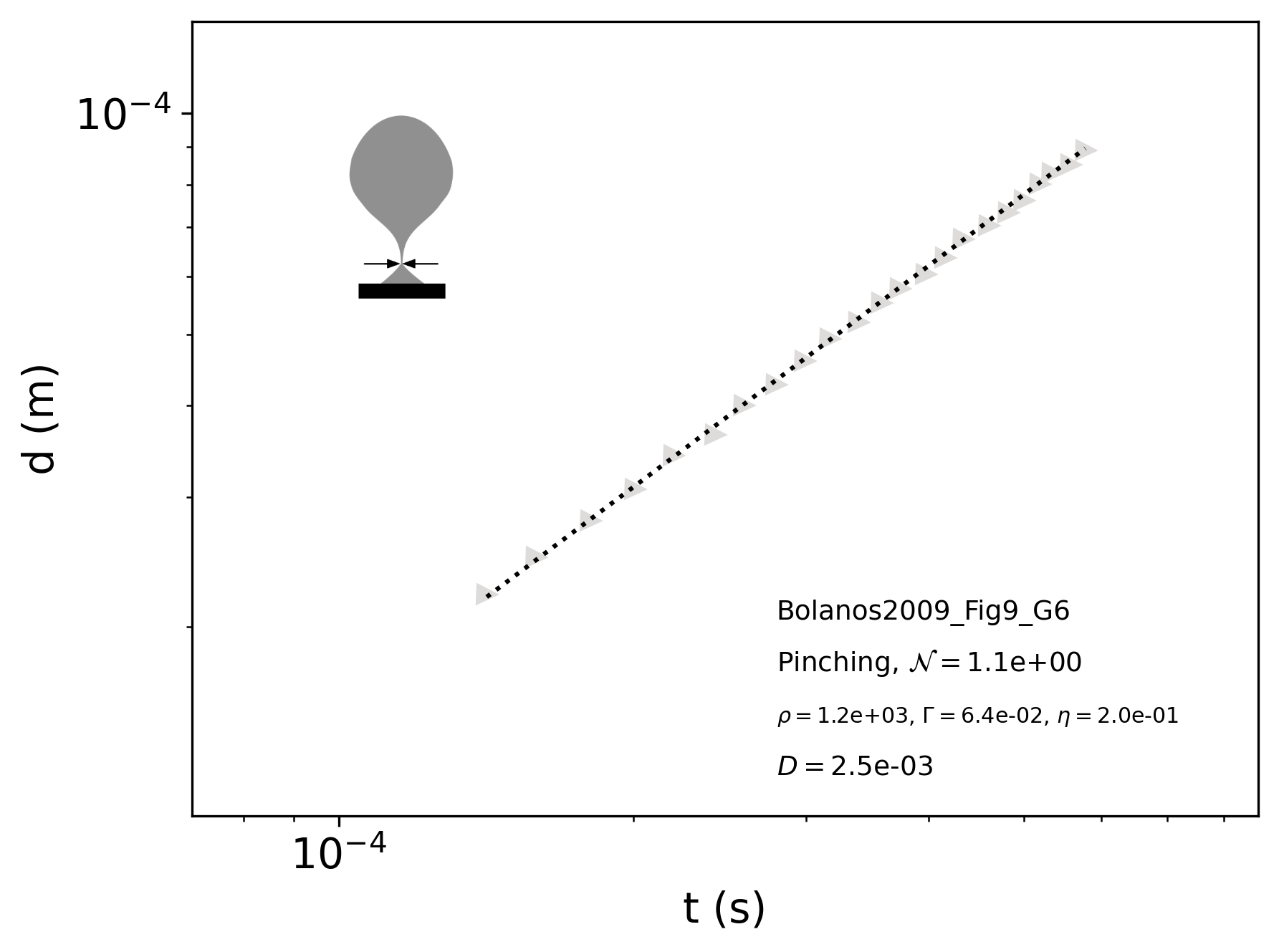} 
 \end{minipage} 
 \\ 
Air bubble in water-glycerol mixture. \newline Density and viscosity are that of the outer fluid. & Air bubble in water-glycerol mixture. \newline Density and viscosity are that of the outer fluid.\\ \hline 
\end{tabular} 
 \end{table}

\begin{table} 
 \centering 
 \begin{tabular}{ | p{9cm} | p{9cm} | } 
 \hline 
 \textbf{Bolanos2009 Fig6 O1} & \textbf{Bolanos2009 Fig6 O3}  \\ 
 \begin{minipage}{.5\textwidth} 
  \includegraphics[width=\linewidth]{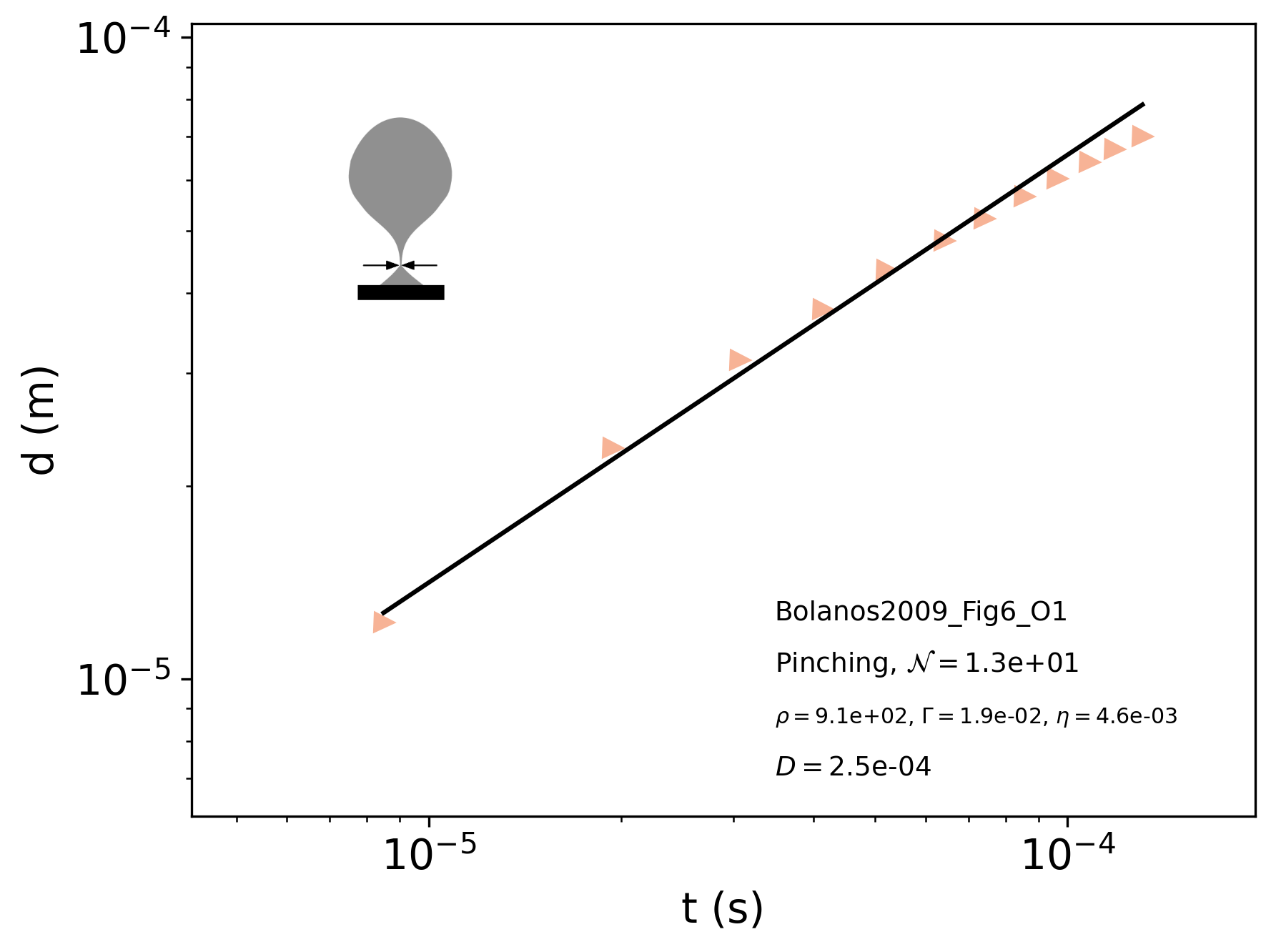}
 \end{minipage}
 & 
 \begin{minipage}{.5\textwidth} 
  \includegraphics[width=\linewidth]{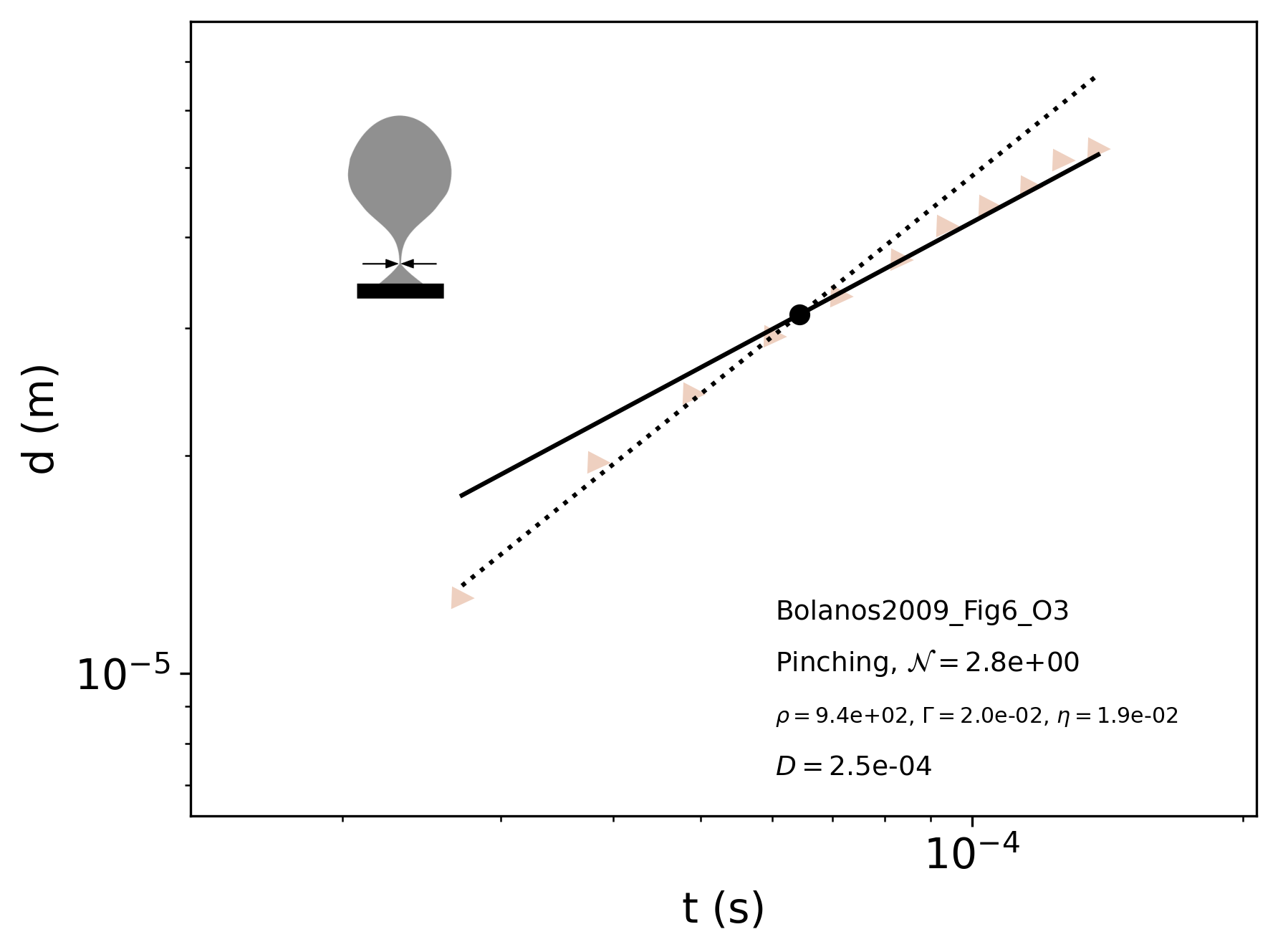} 
 \end{minipage} 
 \\ 
Air bubble in silicone oil. \newline Density and viscosity are that of the outer fluid. & Air bubble in silicone oil. \newline Density and viscosity are that of the outer fluid.\\ \hline  \hline 
\textbf{Bolanos2009 Fig9 G8} & \textbf{Goldstein2010 Fig5}  \\ 
 \begin{minipage}{.5\textwidth} 
   \includegraphics[width=\linewidth]{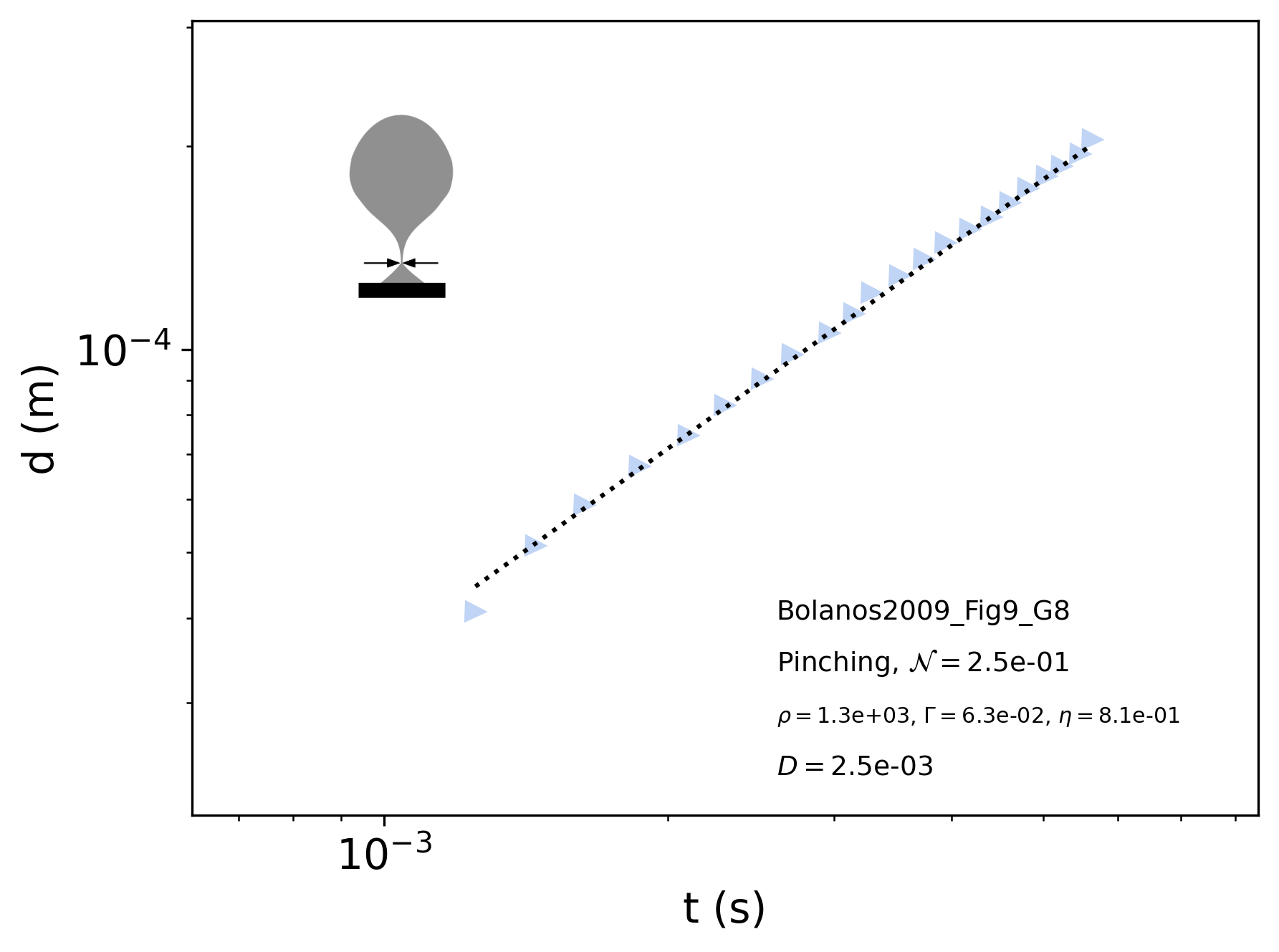} 
 \end{minipage}
 & 
 \begin{minipage}{.5\textwidth} 
   \includegraphics[width=\linewidth]{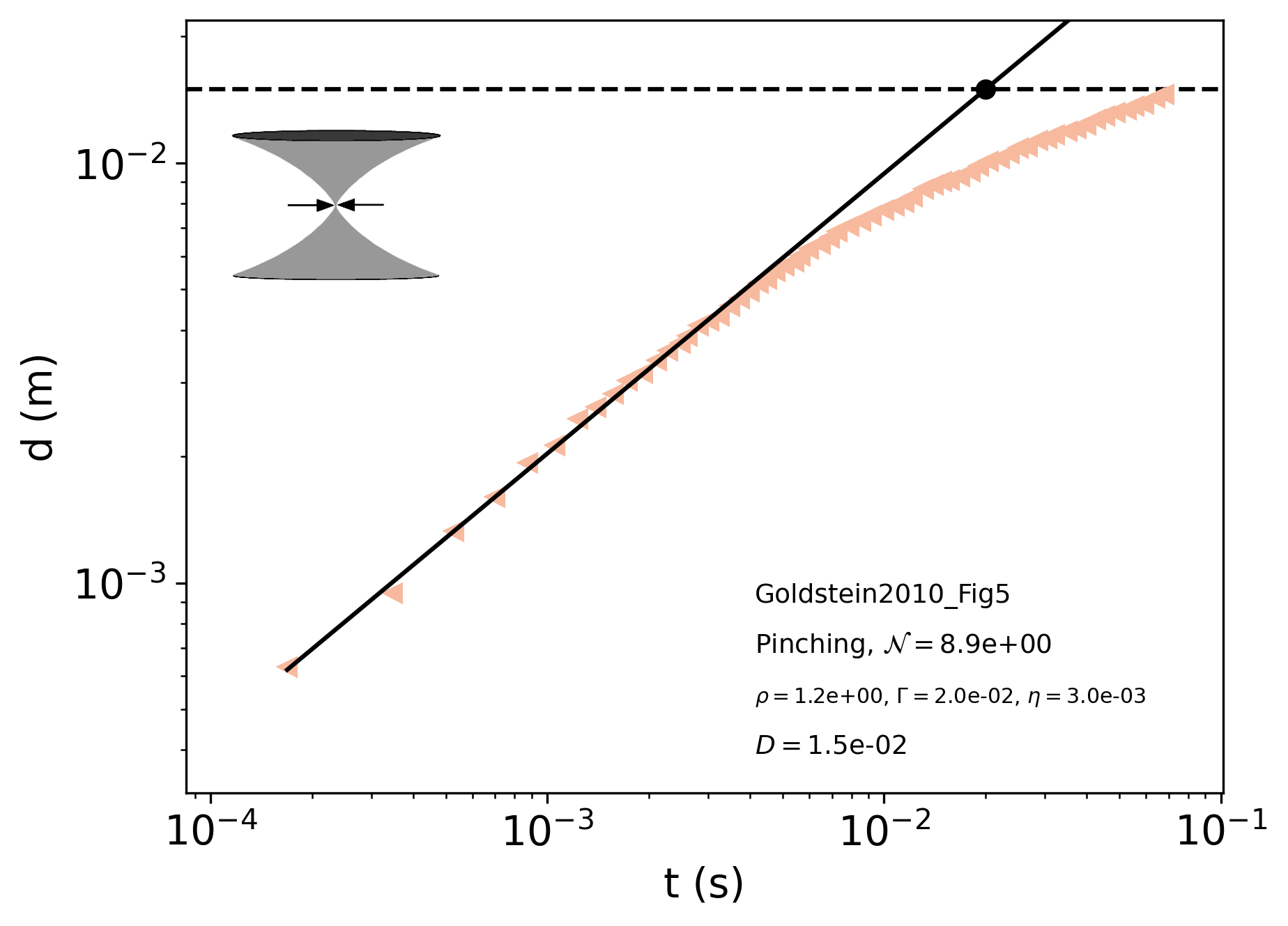} 
 \end{minipage} 
 \\ 
Air bubble in water-glycerol mixture. \newline Density and viscosity are that of the outer fluid. & Soap film on Mobius strip. \newline Values of material parameters estimated since not provided. \newline Density is that of air, viscosity is that of soap film.
\\ \hline 
\end{tabular} 
 \end{table}

\end{widetext}

\end{document}